  \documentclass[12pt,preprint]{aastex}
\newcommand{\ri}{r_{\rm i}}

\newcommand{\cs}{c_{\rm s}}
\newcommand{\css}{c_{\rm s}^2}

\newcommand{\rsun}{{\rm R}_{\sun}}
\newcommand{\etat}{{\eta}_{\rm t}}
\newcommand{\betat}{{\beta}_{\rm t}}

\newcommand{\deltai}{{\delta}_{\rm i}}

\newbox\grsign \setbox\grsign=\hbox{$>$}
\newdimen\grdimen \grdimen=\ht\grsign
\newbox\laxbox \newbox\gaxbox
\setbox\gaxbox=\hbox{\raise.5ex\hbox{$>$}\llap
     {\lower.5ex\hbox{$\sim$}}}\ht1=\grdimen\dp1=0pt
\setbox\laxbox=\hbox{\raise.5ex\hbox{$<$}\llap
     {\lower.5ex\hbox{$\sim$}}}\ht2=\grdimen\dp2=0pt

\shorttitle{Jets from star-disk magnetospheres}
\shortauthors{Ch.~Fendt}

\begin{document}

\title{Formation of protostellar jets as two-component outflows from star-disk magnetospheres}

\author{Christian Fendt}
\affil{ Max Planck Institute for Astronomy, K\"onigstuhl 17, D-69117 Heidelberg, Germany}
\email{fendt@mpia.de}

\begin{abstract}
Axisymmetric magnetohydrodynamic (MHD) simulations have been applied to 
investigate the interrelation of a central stellar magnetosphere and 
stellar wind with a surrounding magnetized disk outflow and how
the overall formation of a large scale jet is affected.
The initial magnetic field distribution applied is a superposition of two 
components - a stellar dipole and a surrounding disk magnetic field, 
in both either parallel or anti-parallel alignment.
Correspondingly, the mass outflow is launched as stellar wind plus a 
disk wind.
Our simulations evolve from an initial state in hydrostatic equilibrium
and an initially force-free magnetic field configuration.
Due to initial differential rotation and induction of a strong 
toroidal magnetic field the stellar dipolar field inflates and is
disrupted on large scale.
Stellar and disk wind may evolve in a pair of collimated outflows.
The existence of a reasonably strong disk wind component is essential
for collimation.
A disk jet as known from previous numerical studies will become 
de-collimated by the stellar wind.
In some simulations we observe the generation of strong flares triggering 
a sudden change in the outflow mass loss rate by a factor of two
and also a re-distribution in the radial profile of momentum flux and 
jet velocity across the jet.
We discuss the hypothesis that these flares may trigger internal shocks
in the asymptotic jets which are observed as knots.
\end{abstract}

\section{Introduction}
Astrophysical jets are launched by magnetohydrodynamic (MHD)
processes in the close vicinity of the central object -- 
an accretion disk surrounding a protostar or a compact object
\citep{blan82,pudr83,came90,pudr07,cabr07}.
However, the principal mechanism which actually launches the
outflow - the transition from accretion to ejection - for a certain 
disk at a certain time is still not completely understood.

During the recent decade, 
numerical simulations of MHD jet formation became more and more 
feasible and substantially helped to improve our understanding of 
how jets emerge.
In general, these simulations may be distinguished in those 
taking into account the time evolution of the disk structure 
(see e.g. 
\citet{good97,mill97,kudo98,kuwa00,roma02,kato02,cass02,kuwa05,zann07})
and others considering the disk surface as a fixed-in-time 
boundary condition for the disk wind or jet
(see e.g. \citet{usty95,ouye97,kras99,fend00,fend02,ouye03,pudr06,fend06}).

The first approach allows to study the launching process directly 
in particular the mechanism lifting matter from the disk plane into the 
outflow.
However, this approach is computationaly very expensive and yet limited 
by spatial and time resolution.
In order to study the jet formation process - the acceleration 
and collimation of a disk/stellar wind -  it is essential to follow
the dynamical evolution of the system for many (several thousands) 
of rotational periods and on a sufficiently large grid with
appropriate resolution. 
For such a goal, the second approach is better suited. 
Since it is computationally less expensive it also allows to perform 
series of simulations for parameter studies. 
It is clear that the prescription of mass flow rate and magnetic flux profile
constrains the result of the simulation more than a consistent simulation of 
the jet-disk evolution which could in fact provide the mass loss rate from the
disk into the jet.
On the other hand, the current status of MHD disk modeling has its own
limitations. 
In particular the magnetic field structure in the disk is a rather open question
unless radiative MHD, global simulations of dynamo-active disk models provide 
fully self-consistent results.
The aim of this paper was to investigate the interaction and interrelation 
between a stellar wind and a disk wind and represents a unique approach 
in that field.
For this first step it is of advantage to govern the simulation by well
understood boundary conditions.
Further, for our goal it is essential to study the long term evolution
of the outflow at considerable distances from the star.
If we would include the disk evolution in the simulation it would be hard
to reach the appropriate time scales. In fact all disk simulations so far
stop at earlier time scales.
Future work should include the disk evolution for the jet launching.

One may further distinguish between the different initial setup for these 
simulations - some of them consider 
a pure stellar dipole 
(see e.g. \citet{uchi84,uchi85,good97,mill97,fend99,fend00,roma02,matt08}),
others a pure disk field
(see e.g. \citet{ouye97,kras99,fend02,cass02,ouye03,fend06}).
The case of a superposed stellar and disk magnetic field is yet rarely 
treated in simulations, however, the first such model configuration was 
discussed already by \citet{uchi81}.
One example is \citet{mill97} who superposed a central dipole with 
an aligned vertical disk field.

In this paper, we study the long-term evolution of a two-component MHD
outflow consisting of a stellar wind launched from a stellar magnetosphere 
and a surrounding disk wind.
It essential to follow the time evolution for very long term in order
to be able to take into account the evolution also of the outer regions 
of the disk magnetosphere as much as possible.
 
\section{Stellar magnetospheres inside disk outflows}
The magnetic field of protostellar jets (and probably also those of micro-quasars) 
most probably consists of two components - a central stellar, probably dipolar field
plus a field component provided by the surrounding accretion disk 
(either generated by a disk dynamo or advected from interstellar space or both).

\subsection{Impact of a stellar magnetosphere on the large-scale outflow}
In the following we qualitatively expose the main aspects of the interaction 
between a central stellar field and the disk magnetic field and how that may
affect the overall jet formation.

{\em Enhanced magnetic flux}.
The stellar field adds magnetic flux to the system.
Assuming a polar field strength $B_{\star}$ and a stellar radius $R_{\star}$,
the dipolar field scales with
\begin{equation}
B_{\rm p,\star} (r) \simeq 40\,{\rm G} \left(\frac{B_0}{1\,{\rm kG}}\right) 
                                       \left(\frac{r}{3\,R_{\star}}\right)^{-3}.
\end{equation}
This has to be compared to the disk poloidal magnetic field provided
either by a disk dynamo or by advection of ambient interstellar field.
The latter can be estimated by equipartition arguments, and is limited
to
\begin{eqnarray}
\label{eq_beq}
B_{\rm p,disk} < B_{\rm eq}(r) = 20\,{\rm G}\,\alpha^{-\frac{1}{2}}
\left(\frac{\dot{M}_{\rm a}}{10^{-6}\,M_{\odot}/{\rm yr}}\right)^{\frac{1}{2}} 
\nonumber \\
\!\left(\frac{M_{\star}}{M_{\odot}}\right)^{\frac{1}{4}}
\!\left(\frac{H/r}{0.1}\right)^{-\frac{1}{2}}
\!\left(\frac{r}{10\,R_{\odot}}\right)^{-\frac{5}{4}}\!\!.
\end{eqnarray}
The stellar magnetic field will not remain closed, but will inflate due
to shear between the disk surface and the star 
(e.g. \citet{uchi84, fend00, uzde02, matt05}).
The additional Poynting flux that threads the disk may support the MHD jet 
launching and may provide an additional energy reservoir for the conversion
of magnetic energy to jet kinetic energy, 
thus, implying a greater asymptotic jet speed
(Michel scaling, see \citet{mich69, fend96, fend04}).

{\em Additional magnetic pressure}.
The central magnetic field provides additional magnetic pressure implicating
possible de-collimation of the overall outflow.
The stellar magnetic field may drive a strong stellar wind which will remove
stellar angular momentum.
This stellar outflow will interact with the surrounding disk wind.
The observed protostellar jets may consist of two components -- a 
stellar wind and the a wind, both with a strength depending on intrinsic
(yet unknown) parameters.

{\em Angular momentum exchange}.
In the scenario of magnetic ``disk locking", the stellar field threading the
disk will re-arrange the global angular momentum budget.
Angular momentum of the star is transfered by the dipolar field and
deposited at the inner disk.
Thus, matter orbiting in this region will be accelerated to slightly
super-Keplerian rotation.
This has two interesting implications.
(i) Due to the super-Keplerian speed this disk material could be 
more easily expelled into the corona by magneto-centrifugal launching
\citep{blan82, ferr97} and forms a disk wind.
(ii) Excess angular momentum in that disk area will slow down accretion 
unless removed by some further process.
The disk outflow launched from this very inner part of the disk
can be an efficient way to do this. This scenario is similar to
the X-wind models \citep{shu94, ferr06}.

{\em Non-axisymmetric effects}.
In addition to the simple picture of an axisymmetric configuration, 
an inclined stellar magnetic dipole will add non-axisymmetric effects.
A moderate non-axisymmetric perturbation may result in warping of the 
inner disk, and, thus, a precession of the outflow launched from this area.
For extreme cases of inclination, jet formation may be completely
prevented. 

A rotating inclined dipole further implies a time-variation of
the magnetic field strength at the inner disk radius.
This may lead to a time-variation in the accretion rate and, also,
the mass outflow  flow rate.
Numerical simulations of the warping process \citep{pfei04} indicate
that the warp could evolve into a steady state precessing rigidly.
Disks may be warped by the magnetic torque that arises from the a slight
misalignment between the disk and star's rotation axis 
\citep{lai99}\footnote{This disk warping mechanism may also operate in
the absence of a stellar magnetosphere as purely induced by the interaction 
between a large-scale magnetic field and the disk electric current and,
thus, may lead to the precession of magnetic jets/outflows \citep{lai03}
}.

\begin{figure}
\centering
\includegraphics[width=7cm]{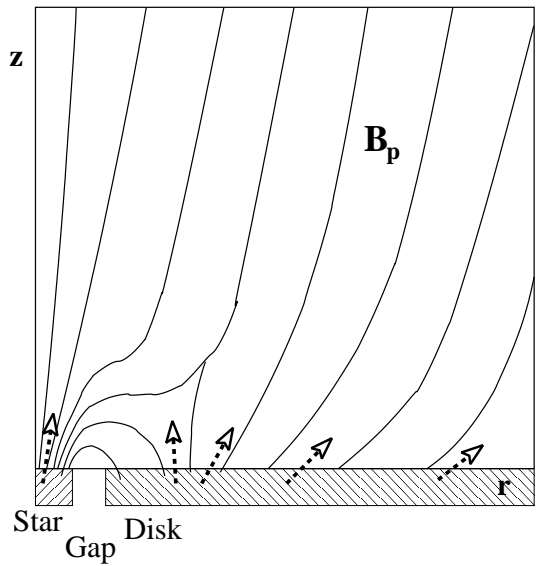}
\caption{Scheme of model setup.
 Along the equatorial plane ($r$-axis) we distinguish jet inflow
 boundary conditions (shaded areas of ghost cells outside active grid)
 from
 the star ($r=0.0, ..., r_{\star} = 0.5$) and 
 the disk region from $r_i = 1.0, ...,  r=r_{\rm out}$).
 The mass flux from star or disk is prescribed by the profile
 for the inflow density and inflow velocity (dashed arrows).
 The latter is typically of about $0.1\%$ of the local Keplerian 
 speed.
 The stellar boundary is in rigid rotation, the disk boundary in Keplerian
 rotation. 
 The disk inner radius is at the co-rotation radius.
 In the gap area between disk and star ($r=0.5, ..., r_i = 1.0$)
 a minimum "floor" mass flux is defined in order to keep the simulation going.
 Thin contours indicate initial poloidal field lines for one of the
 simulation runs. 
\label{fig_model}
}
\end{figure}

\subsection{Simulations of star-disk magnetospheres}
%
The simulation of outflows from star-disk magnetospheres is a difficult numerical 
task due to strong gradients in magnetic field strength, density/pressure, 
and also the velocity field.
Still, the first simulations of a stellar dipole surrounded by a disk have 
been presented in seminal papers as early as 1984 by \citet{uchi84,uchi85}.
Probably due to the success of the disk wind models \citep{blan82, pudr83}
the formation of outflows from stellar magnetospheres became somewhat 
unattended before the topic was re-discovered in the early 90ies with 
several model suggestions \citep{came90, shu94}.
Numerical, stationary-state MHD solutions of the star-disk jet formation 
problem were presented subsequently \citep{fend95,fend96,saut94}.

Time-dependent simulations of dipolar magnetospheres started again later 
\citet{haya96, hiro97, mill97}.
However, the time scale of these simulations including the disk evolution 
was short - a few inner disk rotations.

Goodson et al. \citep{good97, good99, matt02} in a series of 
papers presented simulations runs of dipolar star-disk magnetospheres for 
up to 150 (inner) disk rotations, 
in particular considering the reconnection/flaring behavior of the 
magnetic field close to the inner disk radius. 
As another result, a highly collimated, narrow jet emerged along the
rotation axis for which the authors also presented forbidden emission
line maps.
Simulations by \citet{matt02} followed a similar setup, however, based 
on an increased grid scale of $(0.8 \times 3.2 \rm AU)$ 
and run for times scales of 150 days (stellar case).
Magnetic flares emerging from the inner disk location lead
to episodic mass ejections.
These flares were both triggered and triggering a variation
of the disk mass accretion rate and a corresponding change in the inner 
disk radius.
Reconnection is strongly interrelated to magnetic diffusivity
(resp. electric resistivity), however, such model assumptions were not
particularly specified in these papers.
\citet{matt02} investigated the setup of an ambient vertical field 
aligned with a central dipolar field.
Their scenario is somewhat different to what we consider: 
the surrounding magnetic field is rather weak and vertical and,
thus, not an active outflow-generating disk field as in our case.
The authors identified three classes of field geometries - as in 
our setup depending on the direction of the ambient field.
As a main result, in all three cases the collimation of the central 
outflow was observed, similar to earlier suggestions by \citet{kwan88}.

In comparison, \citet{fend99, fend00} did not treat the disk structure 
in their simulations, allowing them to follow the evolution of the 
the stellar magnetosphere for more than 2000 rotation periods of the
inner disk. 
These simulations showed for the first time that the axial jet feature observed 
in some of the simulations before \citep{mill97, good97, good99} is in fact an 
intermittent feature and disappears on the long term.
The initial axial jet appears essentially during relaxation process of the 
hydrostatic initial condition towards a hydrodynamic steady state.
\citet{fend00} were first to introduce a physical stellar wind boundary 
condition in addition to the disk wind. 
The emerging two component outflow from star and disk remained
un-collimated.
This was understood as caused by the fact that no net electric current was driven 
along the outflow (since the initial dipolar field distribution).
Since the stellar dipolar field becomes very weak along the disk,
a classical magneto-centrifugal wind launching becomes inefficient.
This is another reason for the lack of collimation observed in these
simulations.

Simulations of the disk-star interaction \cite{roma02,kuek03,roma04}
focus on the dipolar accretion process and potential angular
momentum exchange between star and disk.
No evidence was found for persistent outflows and only episodic ejections 
from the inner disk area occurred.
Time-dependent simulations of winds from an initially spherically symmetric 
stellar magnetic field were performed by \citet{kepp99}, evolving exactly
into the classical stationary Weber-Davis solution.
\citet{matt08} derive the angular momentum loss of pure stellar winds
by numerical simulations. 
The setup is similar to the stationary-state models calculated
by \citet{fend95} and \citet{fend96}. 
The outflow angular momentum loss derived from these simulations is
comparable to what can be derived from the stationary 
models\footnote{Sean Matt, private communication. Number values for the
angular momentum flux were not included in \citet{fend95,fend96} }.

While the simulations from dipolar magnetospheres failed to show collimated
outflows, MHD simulations of disk winds did actually proof the self-collimating
characteristics of MHD winds.
After seminal papers by \citet{usty95} and \citet{ouye97} this approach was
further developed taking into account
the time-dependent change of disk field inclination \citep{kras99}, 
turbulent magnetic diffusivity in the jet \citep{fend02}, 
a variation in the disk wind magnetic field and density profile \citep{pudr06,fend06},
or non-axisymmetric effects \citep{ouye03, kigu05, ande06}, 
and also the disk dynamical evolution \citep{kudo98,cass02,kigu05,zann07}.

Recent simulations by \citet{mats08} investigated the 
"topological stability" of two-component (star-disk) self-similar solutions 
derived from stationary MHD. 
In difference to our approach their simulations do not start from an initial
hydro-{\em static} state, 
but from an initial dynamical steady state solution of the MHD equations, 
with analytical extrapolations in the case of radially self-similar 
solutions (which are singular on the axis).
\citet{meli07} presented two-component outflow simulations including the
treatment of the disk evolution. 
However, although the wind dynamics indeed consists of two components,
the magnetic field distribution basically resembled a monotonous field
profile (i.e. no stellar dipole, no stellar outflow involved).

\section{Simulation model setup}
We perform axisymmetric MHD simulations of jet formation for a set
of different magnetic field geometries and mass fluxes.
The general model setup follows \citet{ouye97}, \citet{fend02}
and \citet{fend06}, however, with important modifications.
The original ZEUS-3D ideal MHD code \citep{stone92a,stone92b,hawl95}
extended for physical magnetic resistivity 
(see description and tests in \citet{fend02}) is used.
For the purpose of this paper the magnetic diffusivity was set to 
such a low level that it does not affect the overall collimation of 
the outflow. 
However, resistivity/diffusivity is essential for our simulations, 
as it allows for magnetospheric reconnection phenomena. 

The set of MHD equations considered is the following,
\begin{equation}
{\frac{\partial \rho}{\partial t}} + \nabla \cdot (\rho \vec{v} ) = 0,\quad
\nabla \cdot\vec{B} = 0,\quad
\frac{4\pi}{c} \vec{j} = \nabla \times \vec{B},
\end{equation}
\begin{equation}
\rho \left[ {\frac{\partial\vec{u} }{\partial t}}
+ \left(\vec{v} \cdot \nabla\right)\vec{v} \right]
+ \nabla (p+p_A) +
 \rho\nabla\Phi - \frac{\vec{j} \times \vec{B}}{c} = 0,
\end{equation}
\begin{equation}
{\frac{\partial\vec{B} }{\partial t}}
- \nabla \times \left(\vec{v} \times \vec{B}
-{\frac{4\pi}{c}} \eta\vec{j}\right)= 0,
\end{equation}
\begin{equation}
\label{eq_energ}
\rho \left[ {\frac{\partial e}{\partial t}}
+ \left(\vec{v} \cdot \nabla\right)e \right]
+ p (\nabla \cdot\vec{v} )
- {\frac{4\pi}{c^2}}\eta \vec{j}^2= 0,
\end{equation}
with the usual notation for the variables
(see \citet{fend00,fend06}).
The magnetic diffusivity is space and time dependent and
is denoted by $\eta(r,z)$.

We apply a polytropic equation of state for the gas with the polytropic 
index $\gamma =5/3$. 
As \citep{ouye97} we have added turbulent Alfv\'enic pressure in order
to allow to keep the corona "cool".
In fact, we do not solve the energy equation \ref{eq_energ} and apply the internal 
energy of the gas reduced to $ e=p/(\gamma-1)$.
Two major reasons to do this are both computational speed and numerical 
stability.
Some of our long-term simulations did last already more than two months
of  CPU time
and it would be impossible to reach the desired evolutionary steps 
if the energy equation would have been solved.
Following \citet{ouye97} this approach also allows to combine gas 
pressure forces and gravity under the same derivative in the code.
Thus, instead of subtracting gradients, we apply the gradient of the
difference, which results in perfect stability of the initial state.
Note, however, that recent work by Clarke and collaborators seems to
indicate that relaxation of the polytropy assumption may affect the 
dynamical evolution in certain domains of the jet, in particular regions
with shocks or contact discontinuities \citep{clar04}.

The magnetic diffusivity can be considered as turbulent and, thus, be
related to the Alfv\'{e}nic turbulent pressure $p_{\rm A}$ if we assume
that it is primarily due to
the turbulent Alfv\'{e}nic waves which are responsible for the turbulent 
Alfv\'{e}nic pressure applied in the simulations.
In our previous work \citep{fend02} we have derived a toy parameterization
relating both effects by parameterizing the {\em turbulent magnetic 
diffusivity} similar to the Shakura-Sunyaev parameterization of turbulent 
viscosity, $\etat = \alpha_{\rm m} v l$, where $\alpha_{\rm m}\leq\!1$
and $l$ and $v$ are the characteristic dynamical length scale and velocity
of the system, respectively. 
With that we obtain $\etat = \alpha_{\rm m} v_{\rm t} l$ and with
$\betat = ( \cs / v_{\rm t})^2$ and $\css = \gamma\,p/\rho$, 
it follows that 
$ v_{\rm t}^2 = \frac{\gamma}{\betat} \frac{p}{\rho} $, 
or, normalized, 
${v'}_{\rm t}^2 = \frac{\gamma}{\deltai\betat} \rho'^{\gamma-1}$.
For the chosen polytropic index this implies a magnetic diffusivity 
$\eta\sim \rho^{1/3}$ if $l$ is constant (see also the discussion in \citet{fend06}).
We again stress the point that in the present paper the magnetic diffusivity
is on a such a low level (much below the critical level found in
\citet{fend02}) that it does not affect the dynamics, but allows for
reconnection.

For the numerical grid, we use the ``scaled grid'' option by ZEUS with the 
element size decreasing inwards by a factor of 0.99.
The size of the cylindrical grid is $(256\times 256)$ elements resulting 
in a physical grid size for all simulations is 
$(r\times z) = (80\times 80)\ri$
corresponding to $(4\times 4)$\,AU for $\ri \simeq 10\,\rsun$.
Thus, the disk gap ($r<1$) is resolved with 11 grid elements.
Time is measured in rotation periods (Keplerian orbits) at the inner 
disk radius.

In summary, compared to our previous studies 
\citep{fend00,fend02,fend06},
the main new feature included in the present approach is that
the initial magnetic field distribution consists of two components -
a stellar dipolar field and a disk field. 
In difference to recent studies by \citep{meli07} we included both
outflow components and magnetic field components in the simulation box,
in particular treating the star-gap-disk boundary.

\subsection{Boundary conditions}
Along the $r$-boundary we distinguish between 
the star extending from $r=0.0$ to $r=r_{\star} = 0.5$,
the gap region extending from $r=0.5$ to $r=r_i = 1.0$, and 
the disk region from $r=1.0$ to $r=r_{\rm out}$, 
see Fig.\ref{fig_model}.
A Keplerian disk is taken as a (fixed in time and space) boundary
condition for the mass inflow from the disk surface into the corona
and the magnetic flux.
The stellar surface is approximated by a rigidly rotating "disk" in
cylindrical coordinates.
The stellar rotation is chosen such that the inner disk radius is 
located at the co-rotation radius.

The initial magnetic field is purely poloidal. 
Magnetic field lines are anchored in the disk and the rotating star
and are in co-rotation with their respective foot point.
The poloidal magnetic field profile along the $r$-boundary remains fixed in 
time and is,
hence, determined by the choice of the initial magnetic field distribution.

The disk region governs the mass inflow from the disk surface into corona
(denoted as "disk wind").
In addition to that we prescribe a stellar wind with different mass load.
The hydrodynamic boundary conditions are ``inflow'' along the $r$-axis for
the disk and stellar region and either "inflow" (very light mass flow) or 
"reflecting" for the gap region,
``reflecting'' along the symmetry axis and ``outflow'' along the outer 
boundaries.
Matter is ``injected'' from the disk and the star into the corona parallel 
to the poloidal magnetic field lines with very low velocity
$\vec{v}_{\rm inj}(r,0) = \nu_{\rm i} v_{\rm K}(r) \vec{B}_{\rm P}/B_{\rm P}$
and with a density $\rho_{\rm inj}(r,0) = \eta_{\rm i}\,\rho(r,0)$.
The proportionality constants are typically $\nu_{\rm i} \simeq 10^{-3}$ and
$\eta_{\rm i} \simeq 100$ for both, stellar and disk wind (but usually different for both
components). 
Along the gap we impose a floor value for the density $\eta_{\rm i} \simeq 10^{-3}$
and a similarly low velocity.
For the injection velocity, the assumption is that the initial disk wind speed is
in the range of the sound speed in the disk,
$v_{\rm inj}(r) \simeq c_{\rm s}(r) \simeq v_{\rm kep} \sim r^{-1/2}$.

These mass loss rates from star and disk, respectively,
are our other main parameters besides the respective magnetic flux (see Tab.~\ref{tab_all}).

\begin{figure}
\centering
\includegraphics[width=6.5cm]{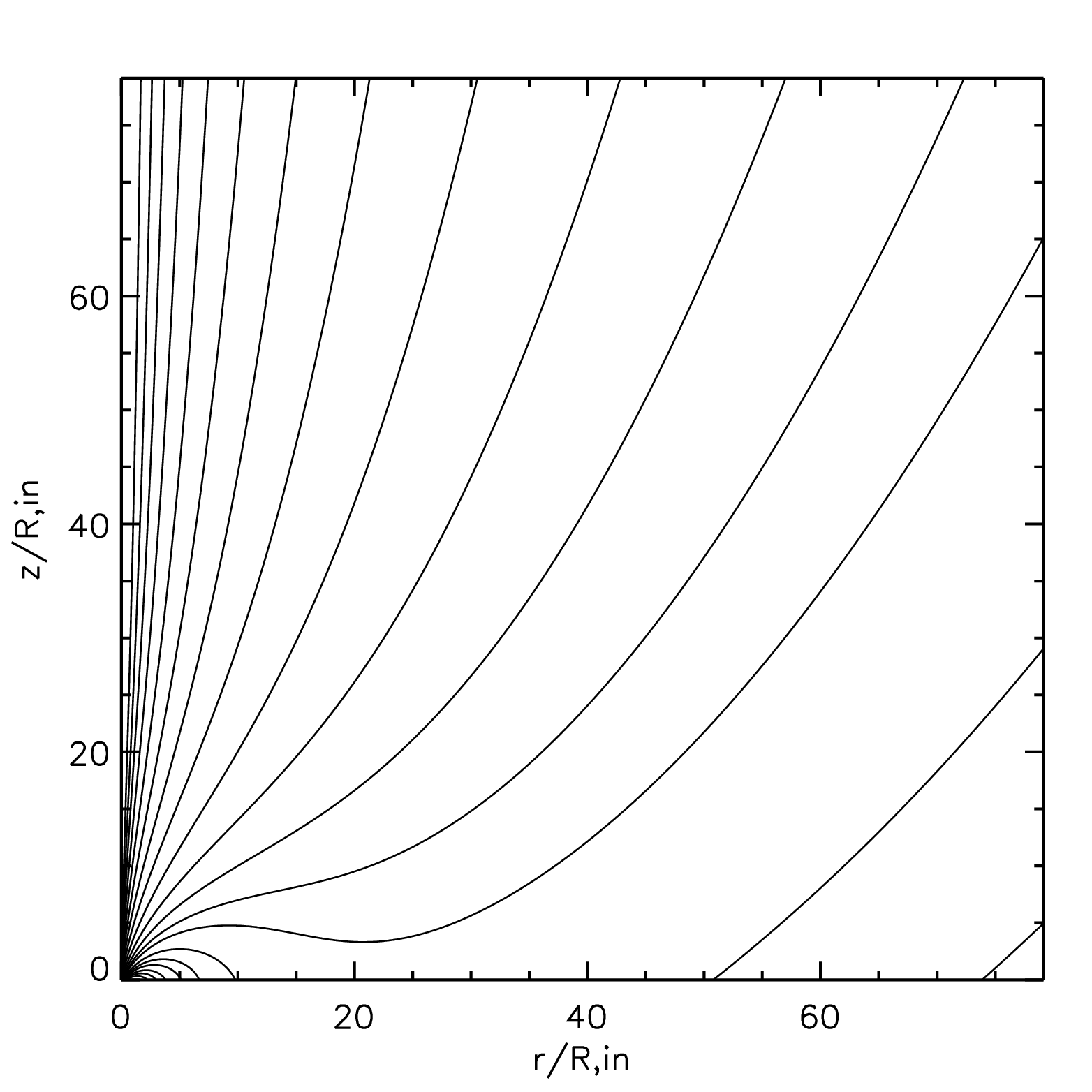}
\includegraphics[width=6.5cm]{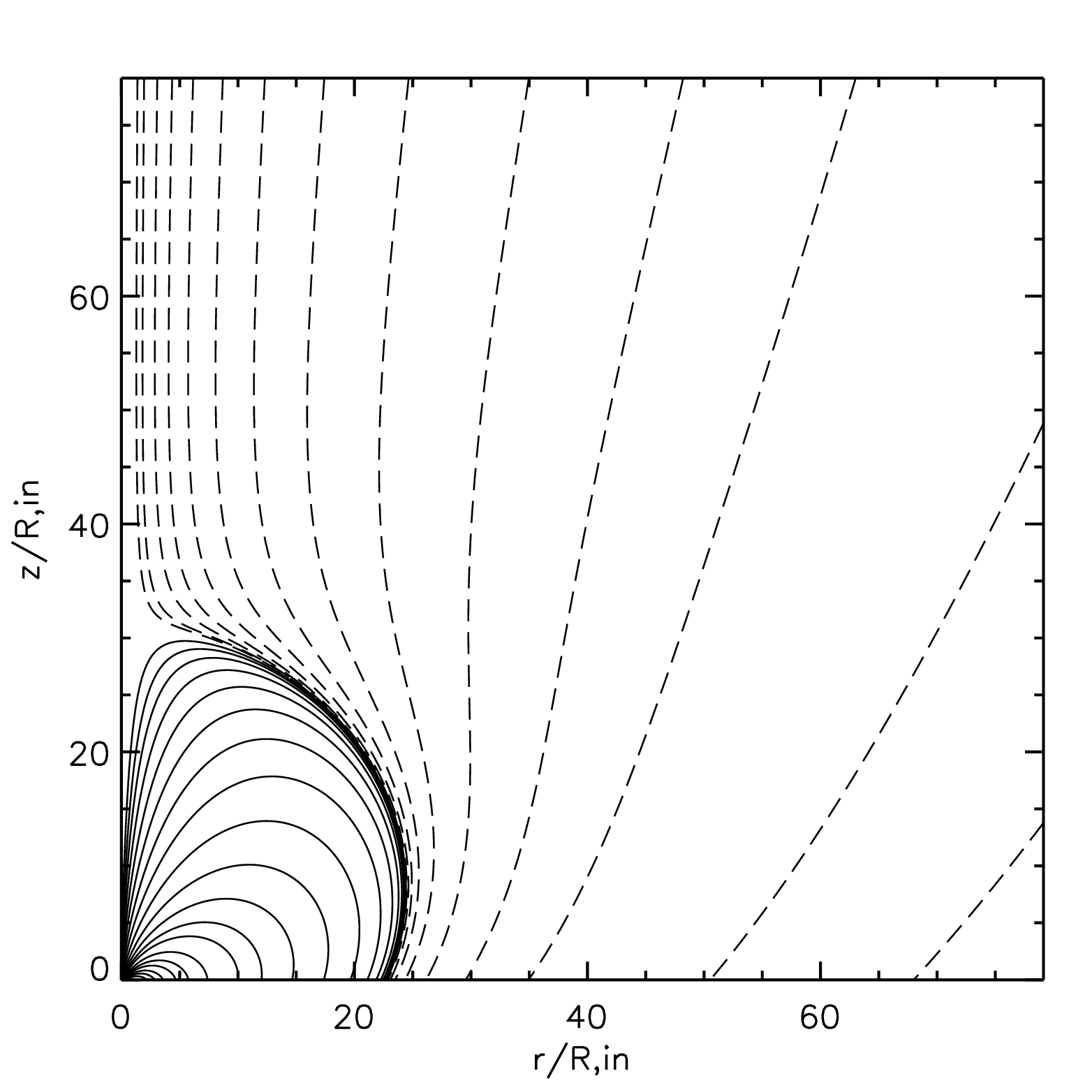}
\includegraphics[width=6.5cm]{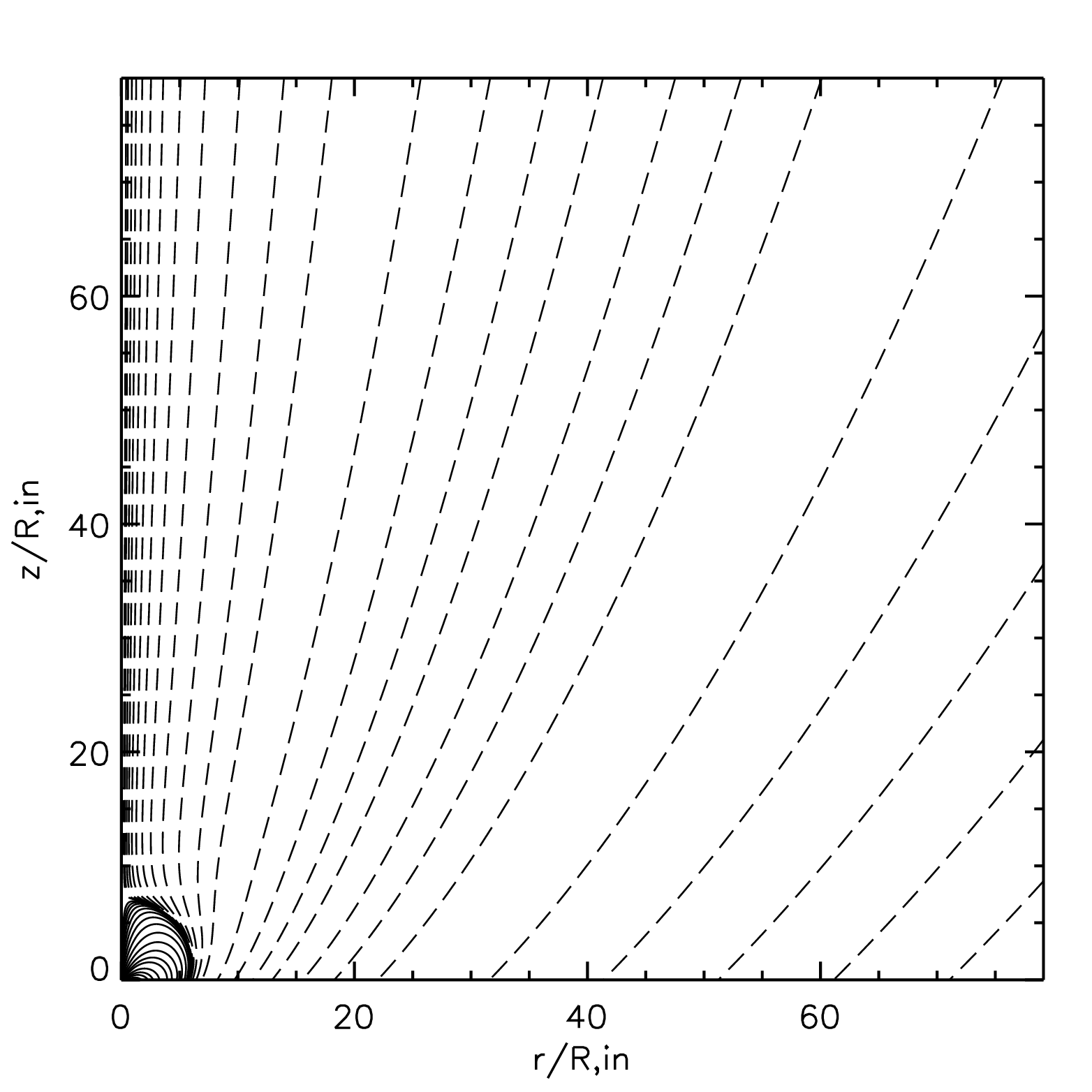}
\caption{Example initial magnetic field distributions.
    Shown are poloidal magnetic field lines. Full and dashed lines indicate 
    the direction of magnetic flux. 
    The plots show different options for the strength and orientation of the 
    superposed stellar and disk magnetic field components
    $\Psi_{0,\rm d} = 0.01, -0.01, -0.1, $ resp. 
    $\Psi_{0, \star} = 5.0, 5.0, 3.0$
    (from top to bottom; runs A2, A3, A4a).
\label{fig_ini}
}
\end{figure}

\subsection{Initial conditions}
As initial state we prescribe a force-free magnetic field along with a 
gas distribution in hydrostatic equilibrium.
Both is essential in order to avoid artificial relaxation processes 
caused by a non-equilibrium initial condition.
The initial density distribution is $\rho(r,z, t=0) = (r^2 + z^2)^{-3/4}$.
As initial magnetic field distribution we superpose a dipolar stellar field 
and a disk potential field. 
For the disk field component we apply the model of \citet{ouye97} and
\citet{fend02}.  
We prescribe the magnetic field distribution as derivative of 
the magnetic flux distribution 
$\Psi(r,z) \equiv \int \vec{B}_{\rm p} d\vec{A}$ 
(i.e. the $\phi$-component of the vector potential).
For the superposed magnetic flux from star and disk we have
\begin{eqnarray}
\Psi(r,z) =
\Psi_{0,\rm d}\,\frac{1}{r}\left(\sqrt{r^2+(z_{\rm d}+z)^2}-(z_{\rm d}+z)\right) 
\nonumber \\
 + \Psi_{0,\star}\,\frac{r^2}{\left(r^2+(z_{\rm d}+z)^2\right)^{3/2}}, 
\end{eqnarray}
with the stellar and disk magnetic flux $\Psi_{0,\star}$ and $\Psi_{0,\rm d}$.
The poloidal magnetic field follows from the derivatives 
$B_{\rm r} = -(1/r) (\partial \Psi/(\partial z)$ and 
$B_{\rm z} = (1/r)(\partial \Psi/(\partial r) $,
properly calulated in the staggered mesh in order to obtain a numerically
divergence-free and force-free initial field structure.
The dimensionless disk thickness $z_{\rm d}$ with $(z_{\rm d}+z)>0$ for
$z<0$ is introduced in order to avoid kinks in the field distribution.
Several combinations of both field components were investigated,
parameterized by combinations of $\Psi_{0,\rm d}$ and $\Psi_{0,\star}$.
Figure \ref{fig_ini} shows three examples for the initial field configuration 
for different strength and alignment of the field components.
Similar field geometries have been discussed already by \citet{uchi81}. 
\citet{matt02} did consider similar configurations, however, superimposing
a stellar dipole with a weak vertical disk field.

In order to allow for a clear comparison between all our runs, the respective magnetic
field component {\em profiles} and wind density component {\em profiles} are identical.

\begin{figure*}
\centering
\includegraphics[width=5.3cm]{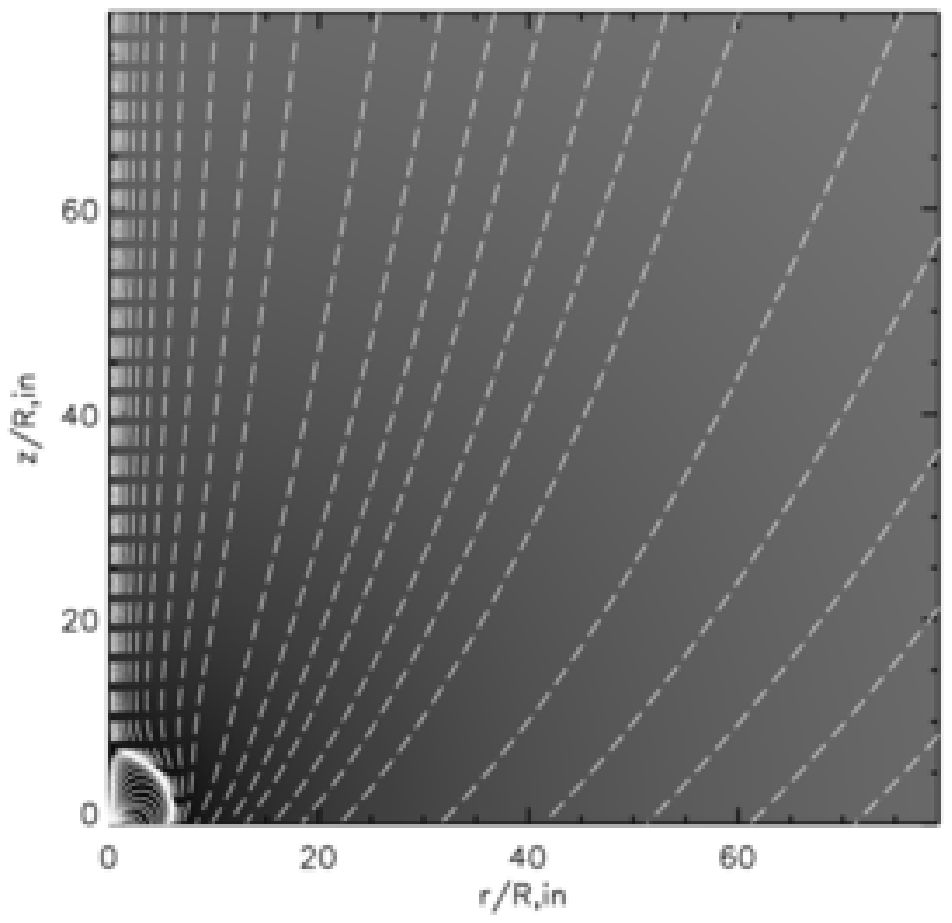}
\includegraphics[width=5.3cm]{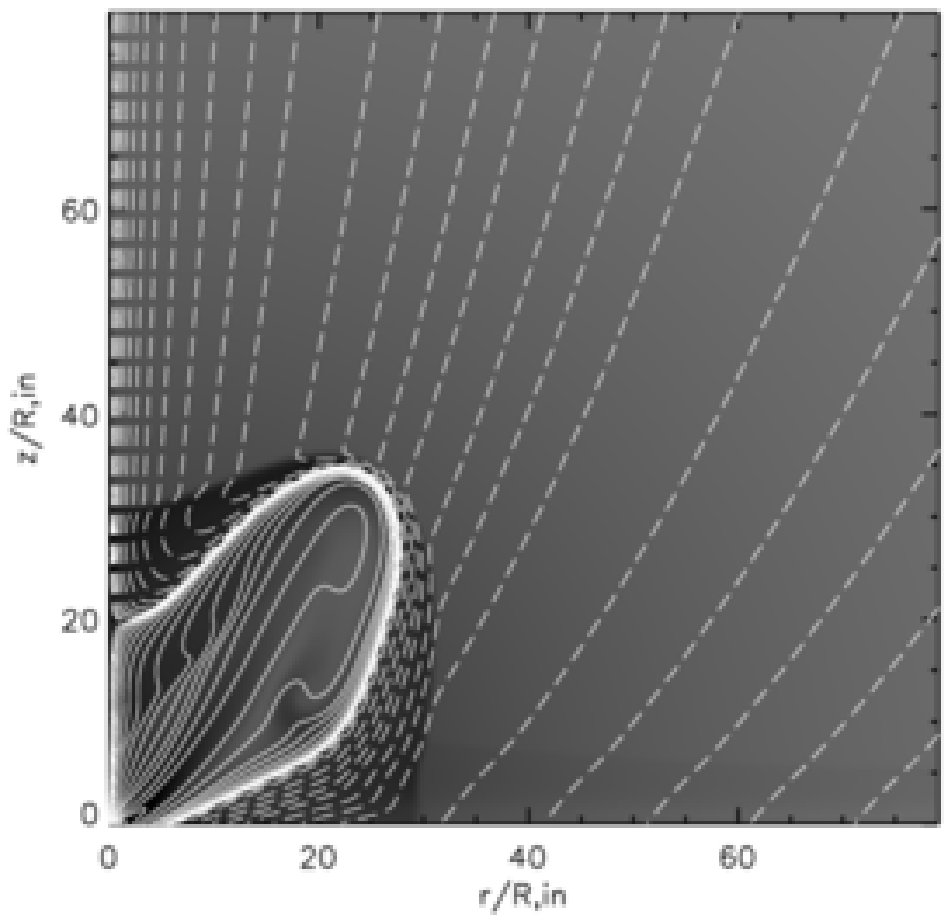}
\includegraphics[width=5.3cm]{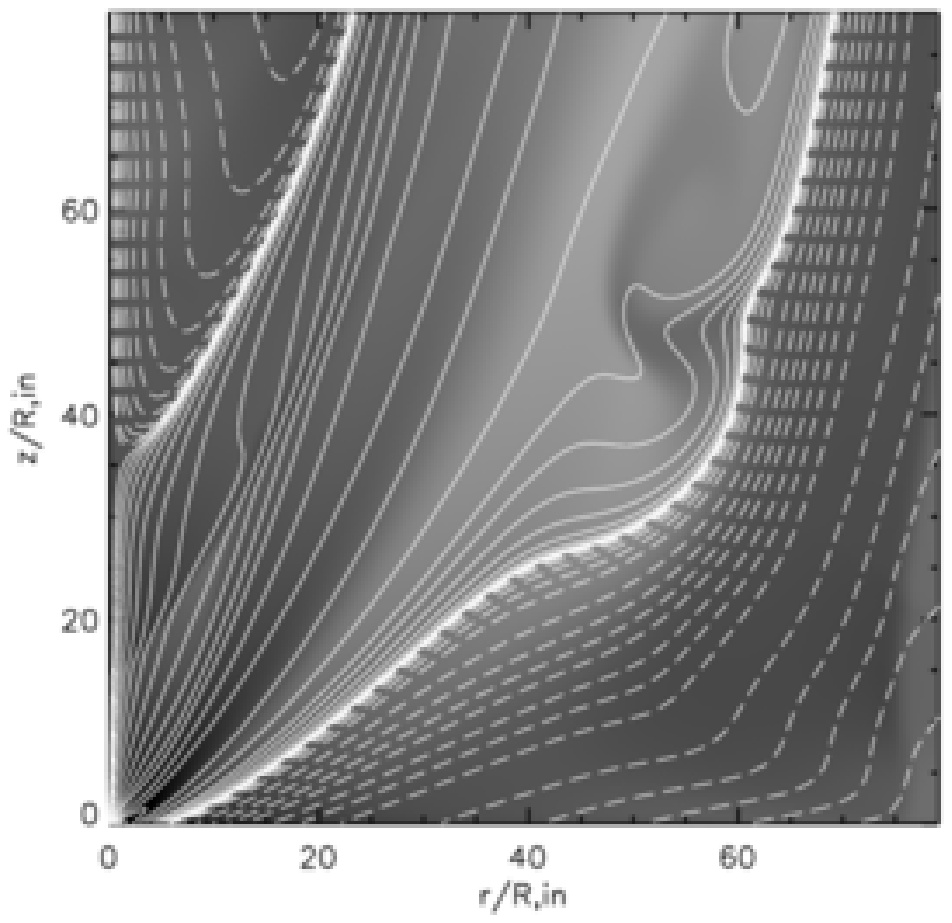}

\includegraphics[width=5.3cm]{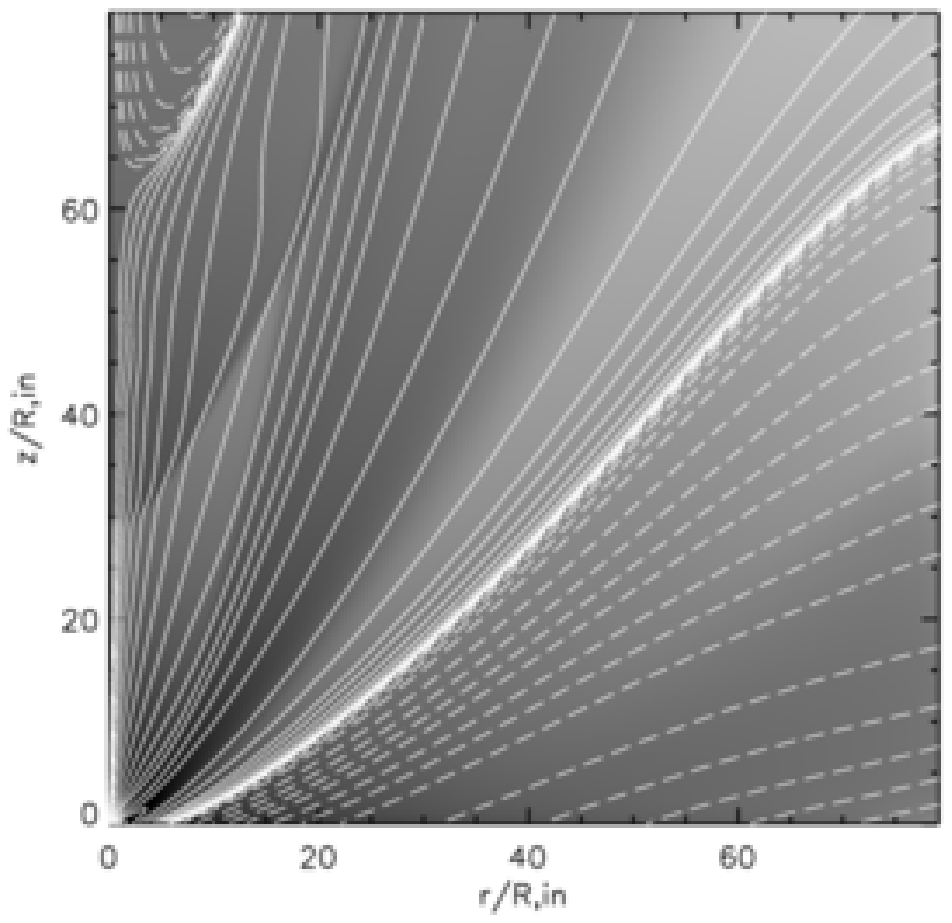}
\includegraphics[width=5.3cm]{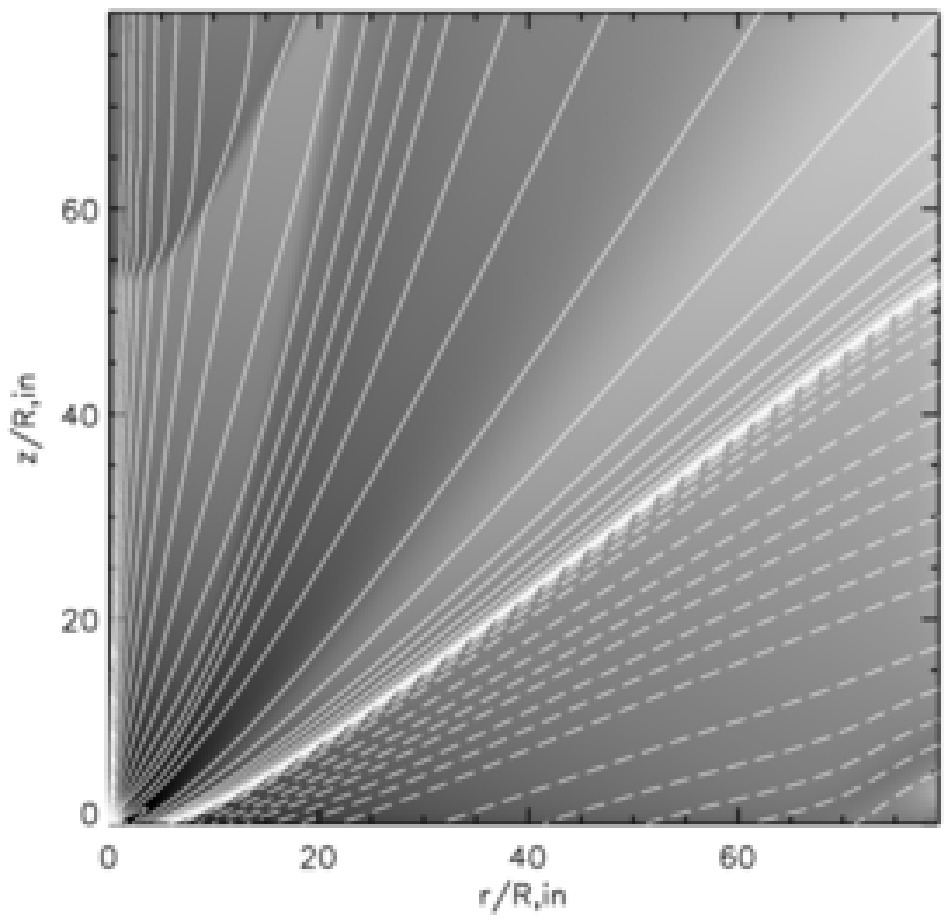}
\includegraphics[width=5.3cm]{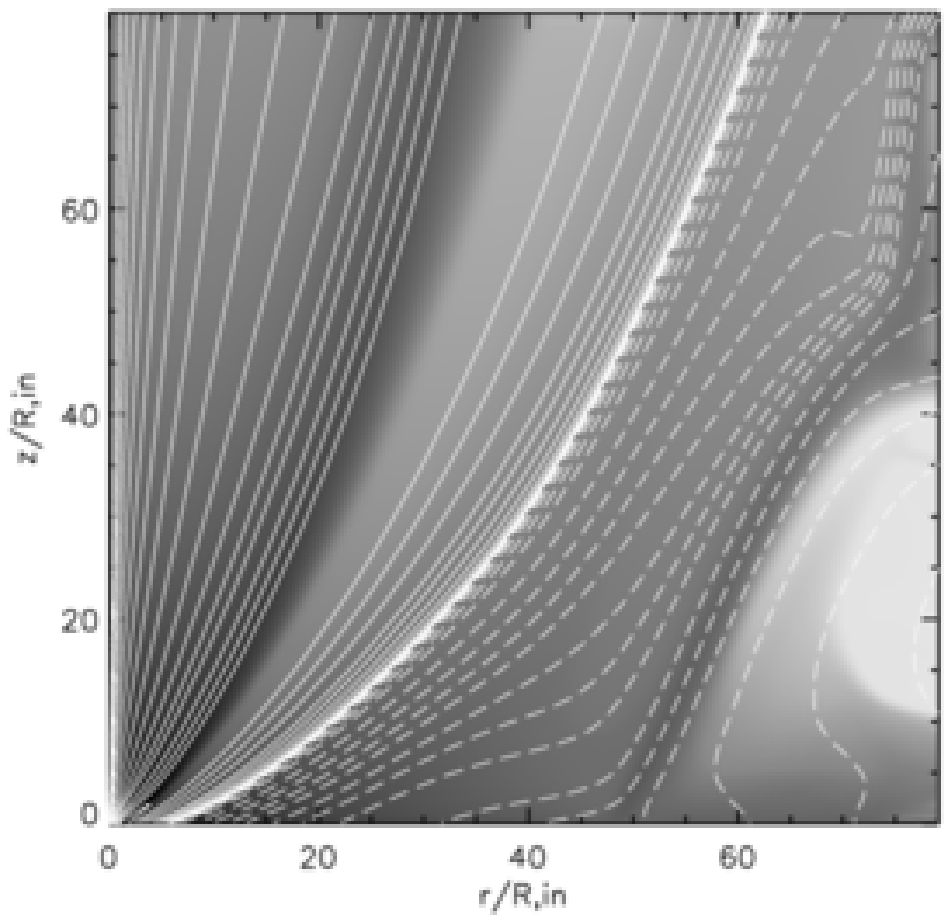}

\includegraphics[width=5.3cm]{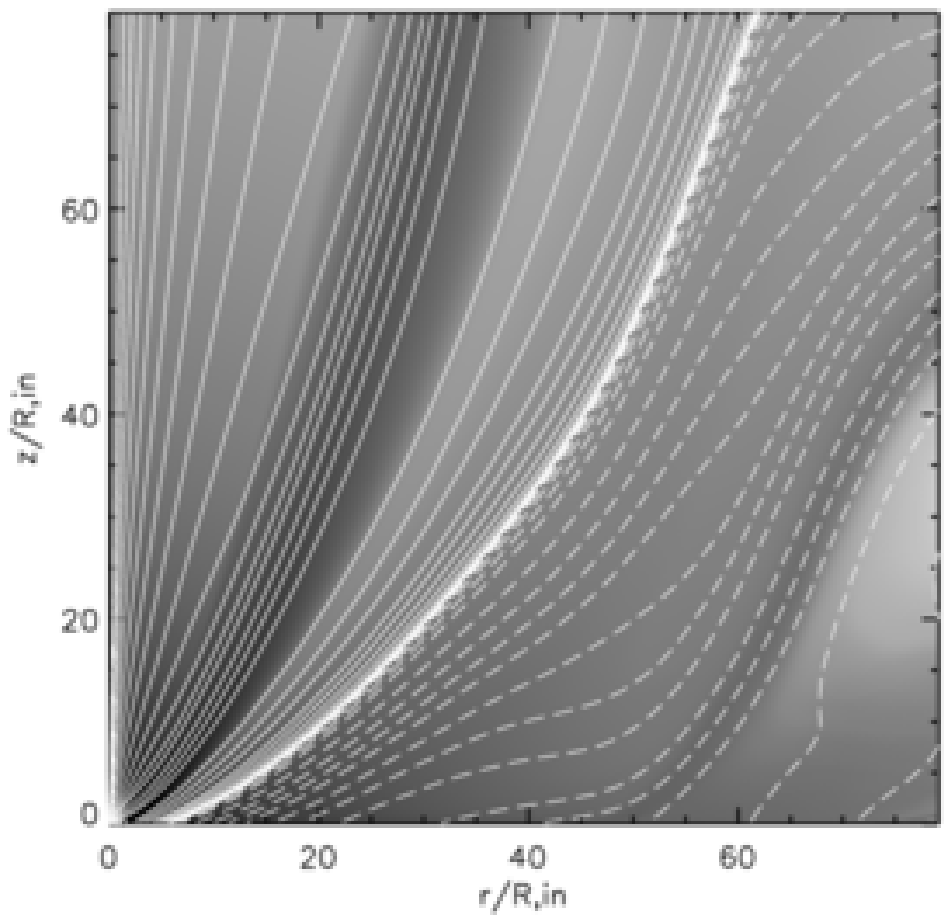}
\includegraphics[width=5.3cm]{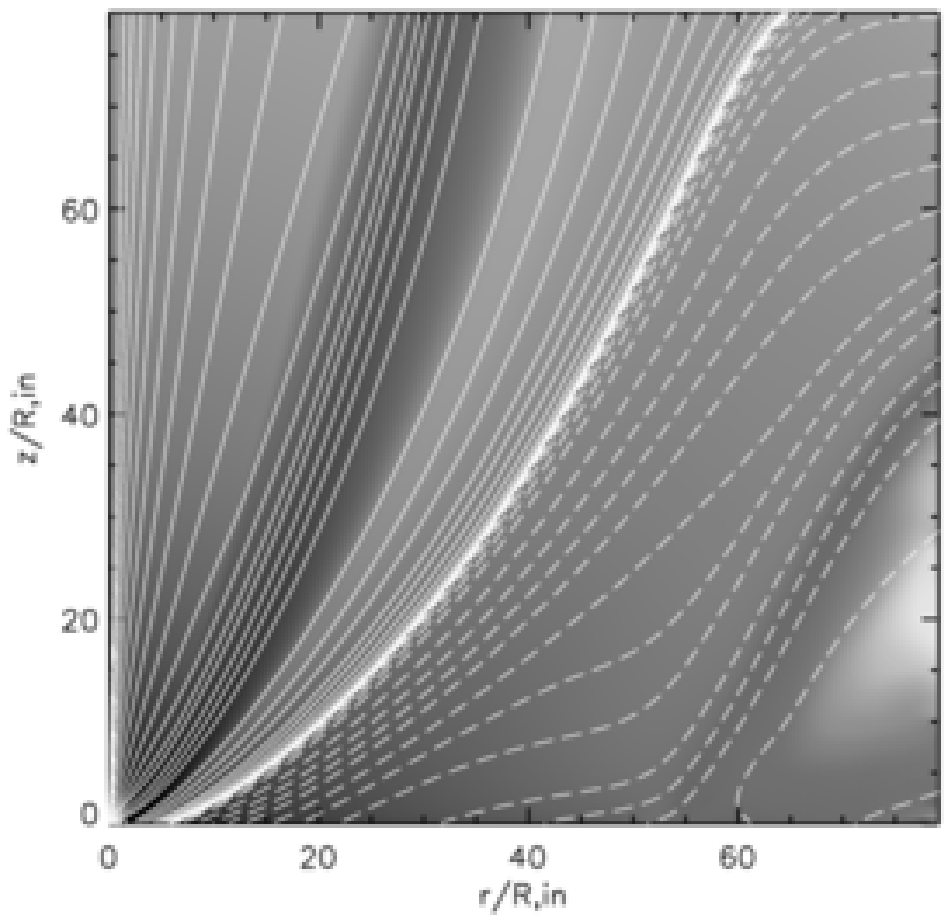}
\includegraphics[width=5.3cm]{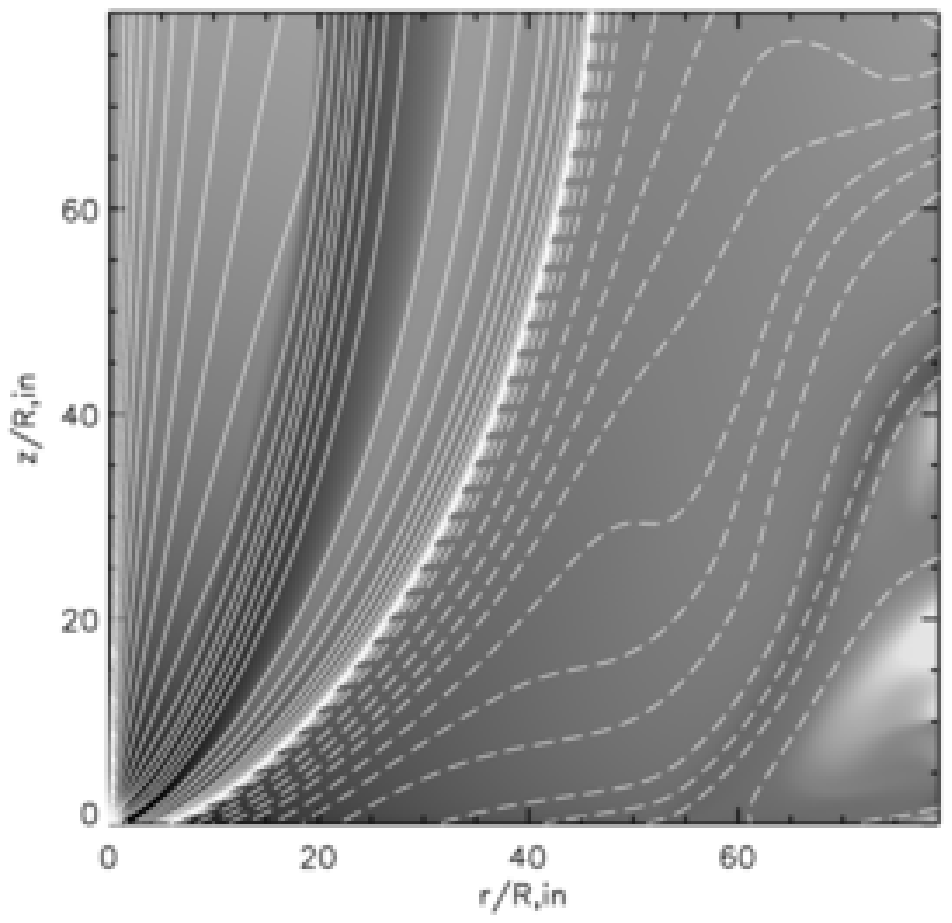}

\includegraphics[width=8.0cm]{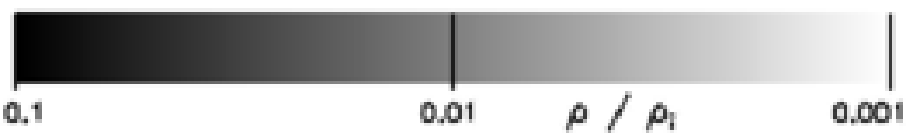}

\caption{Time evolution of simulation run A4a. 
  Poloidal magnetic field distribution at time steps
  $t=0, 50, 200, 500, 1000, 2000, 2500, 3000, 3600$
    (from top left to bottom right).
 Density gray scale as indicated.
 Poloidal field lines as contour levels of the magnetic flux, 
 $-\Psi(r,z) = $ 3.0 ,2.0 ,1.6 ,1.3, 1.0, 0.8, 0.6, 0.4, 0.2,
        0.12, 0.064, 0.032, 0.016, 0.008, 0.004, 0.002,
      0.001, 0.0005, 0.0002, 0.0001, (dashed),
 $ \Psi(r,z) = $ 0.0002, 0.0005, 0.001, 0.002, 0.004, 0.008, 0.016,
     0.032, 0.064, 0.12, 0.2, 0.3, 0.4, 0.6, 0.8, 1.0, 1.3,
      1.6, 2.0 (solid).
\label{fig_evol1}
}
\end{figure*}

\begin{figure}
\centering
\includegraphics[width=8.0cm]{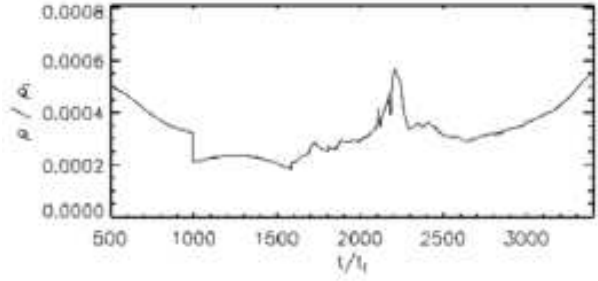}

\caption{Time evolution of the density at $z=44, r=44$ for 
  simulation run A4a. 
\label{fig_quasi}
}
\end{figure}

\begin{figure*}
\centering
\includegraphics[width=8.0cm]{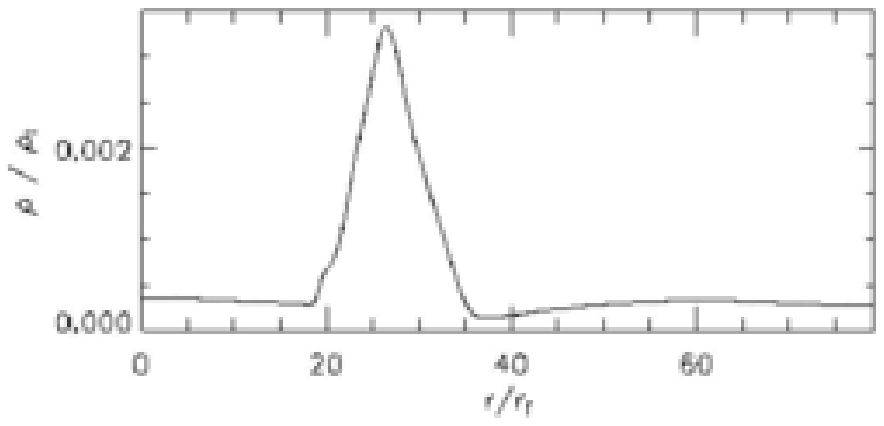}
\includegraphics[width=8.0cm]{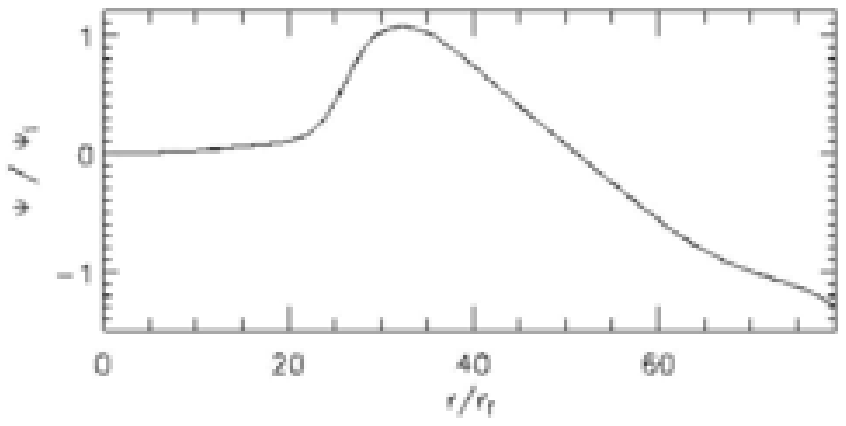}
\includegraphics[width=8.0cm]{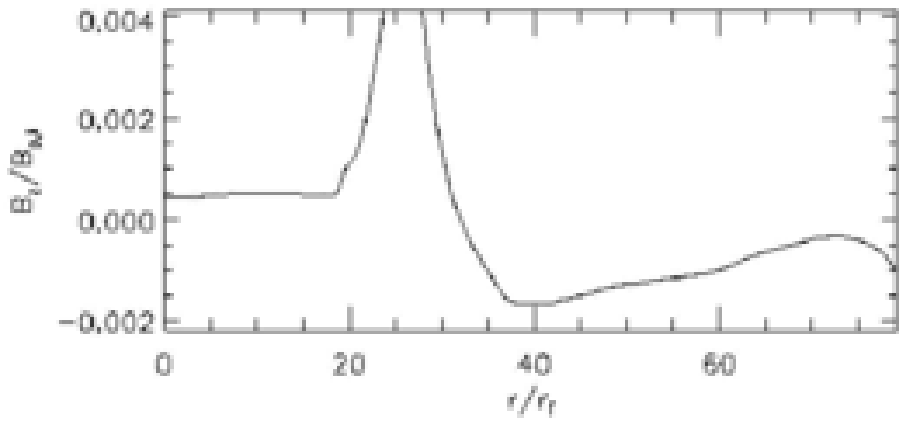}
\includegraphics[width=8.0cm]{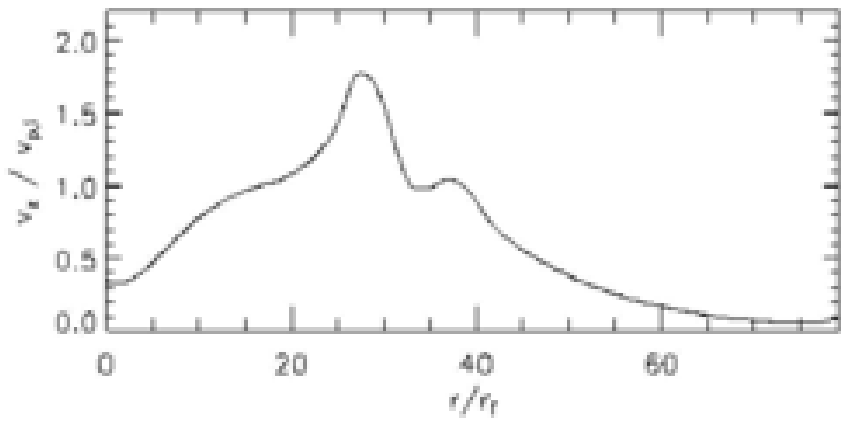}
\includegraphics[width=8.0cm]{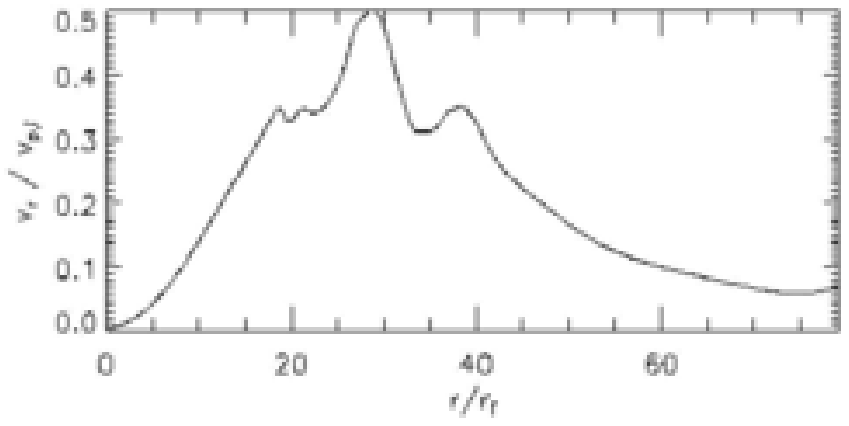}

\caption{Radial profiles of density, magnetic flux, 
 axial magnetic field, axial velocity, and radial velocity
 (from top to bottom)
 at $z=55$ for simulation run A4a at $t=3000$. 
 Note the neutral field line (magnetic field reversal) at 
 $r\simeq 33$ and the magnetic flux reversal at $r\simeq 53$.
\label{fig_reversal}
}
\end{figure*}

\begin{figure*}
\centering
\includegraphics[width=5cm]{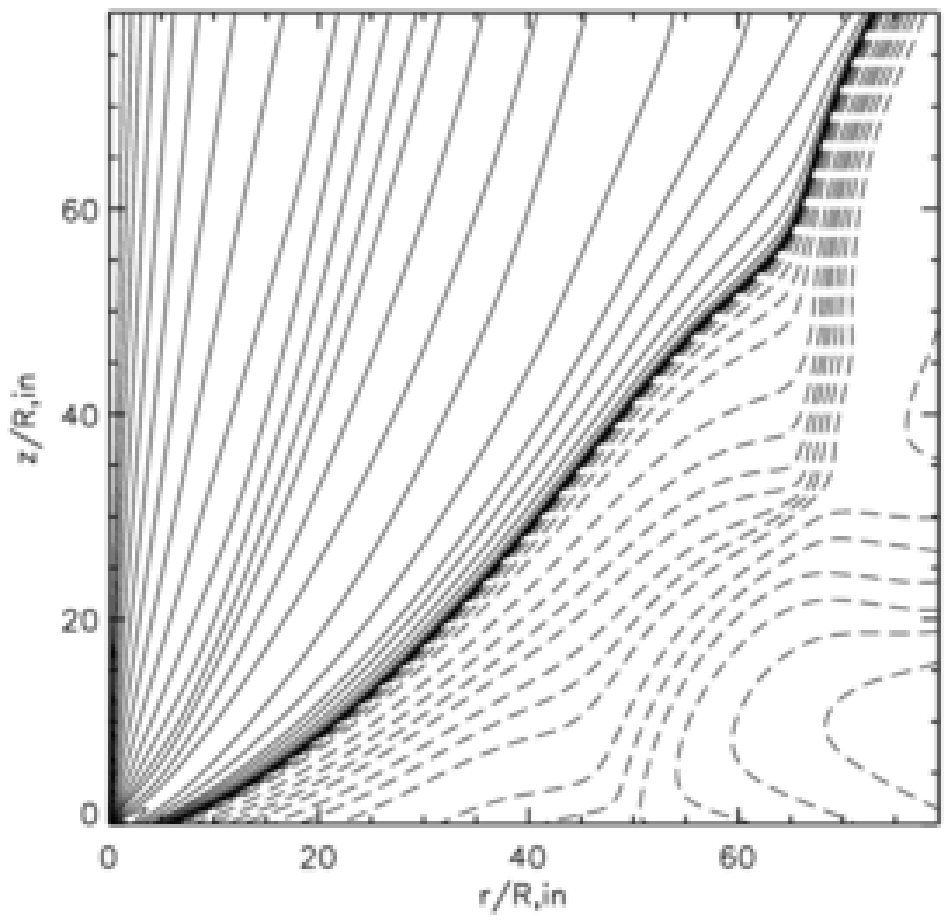}
\includegraphics[width=5cm]{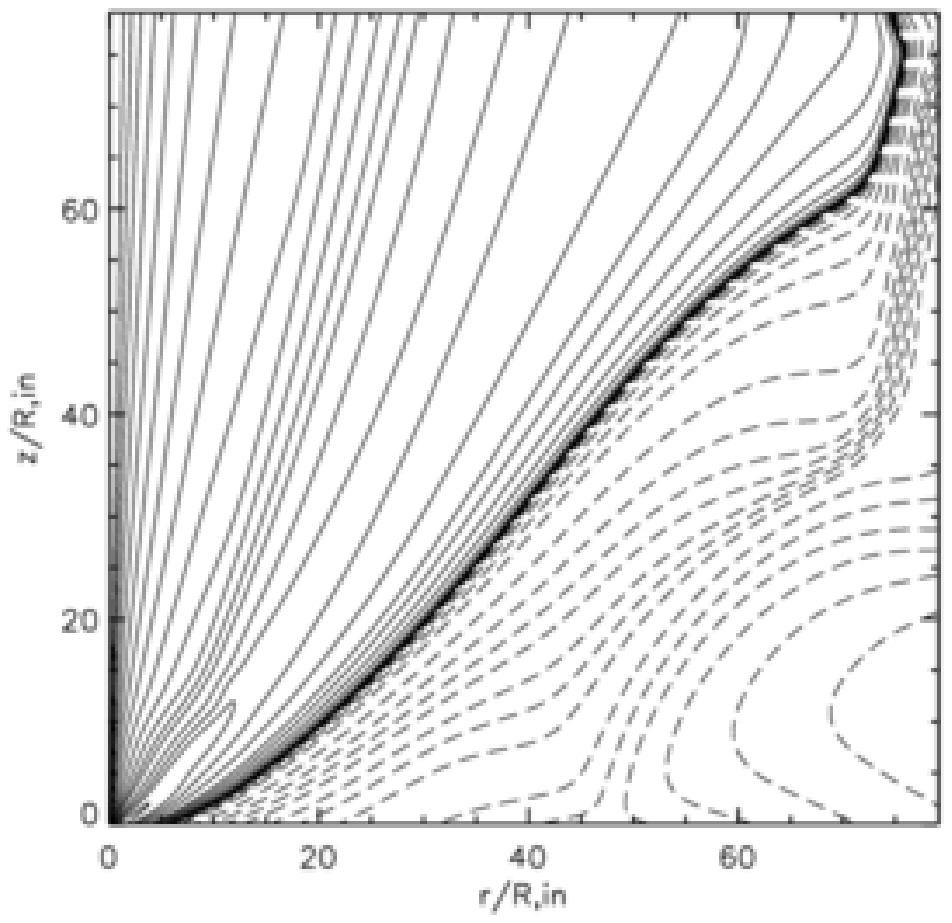}
\includegraphics[width=5cm]{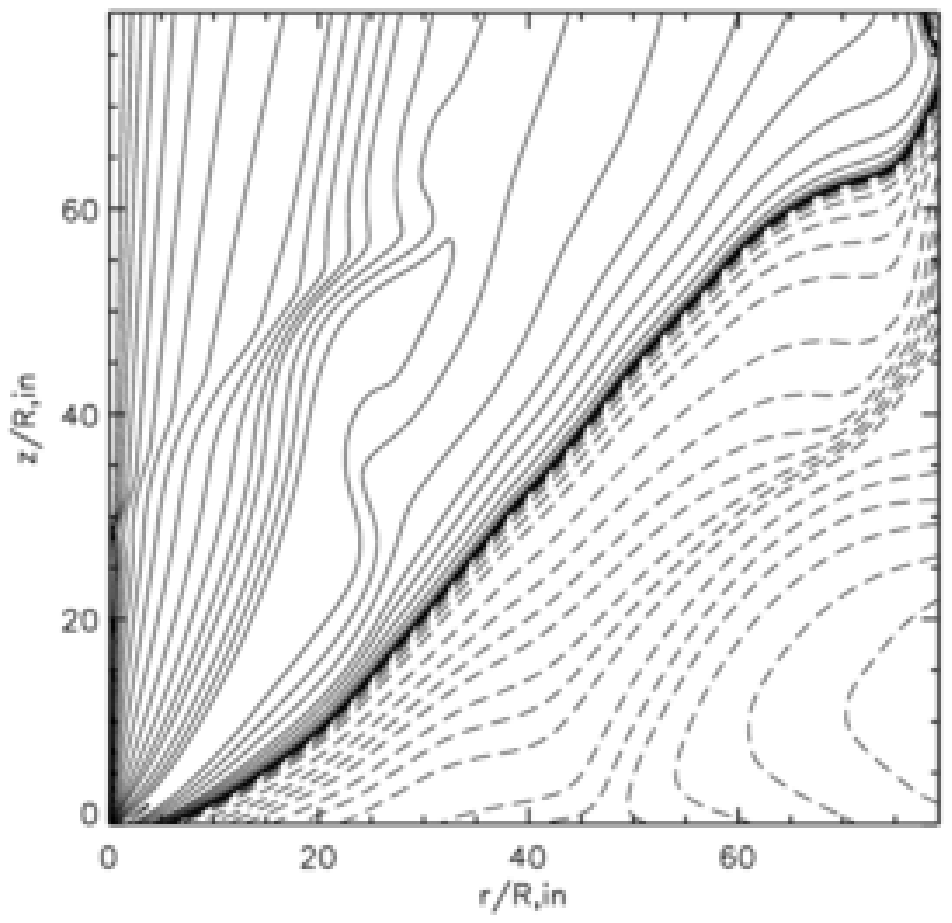}

\includegraphics[width=5cm]{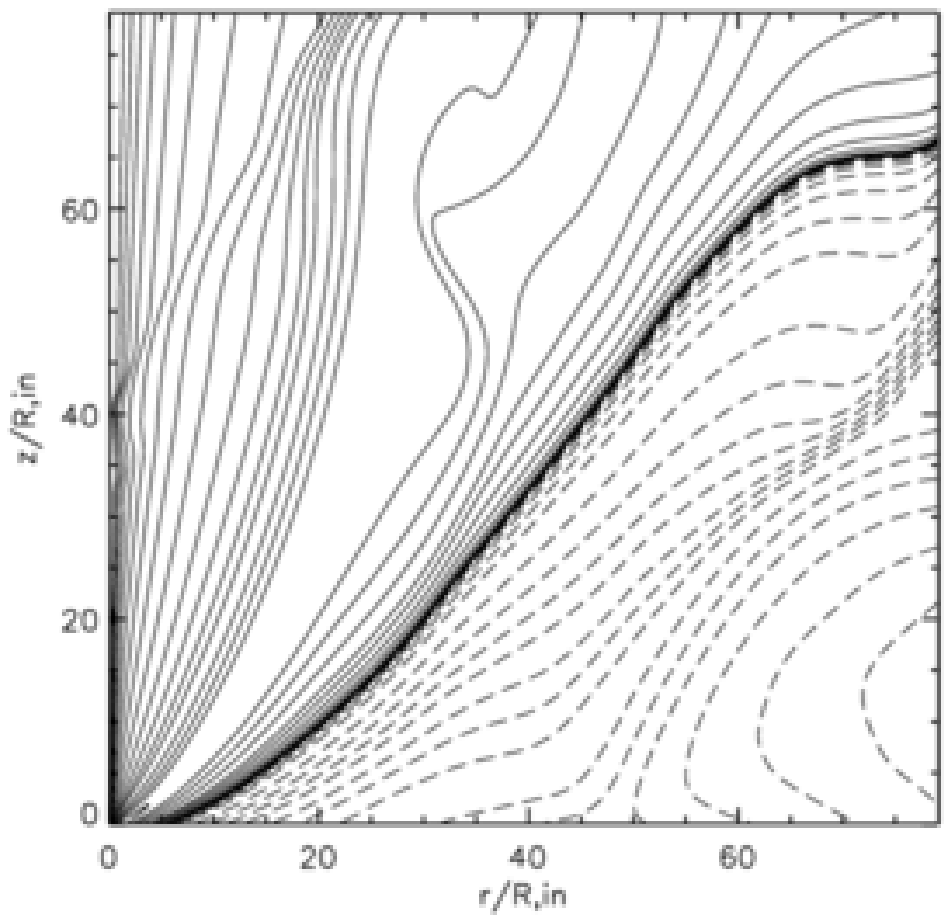}
\includegraphics[width=5cm]{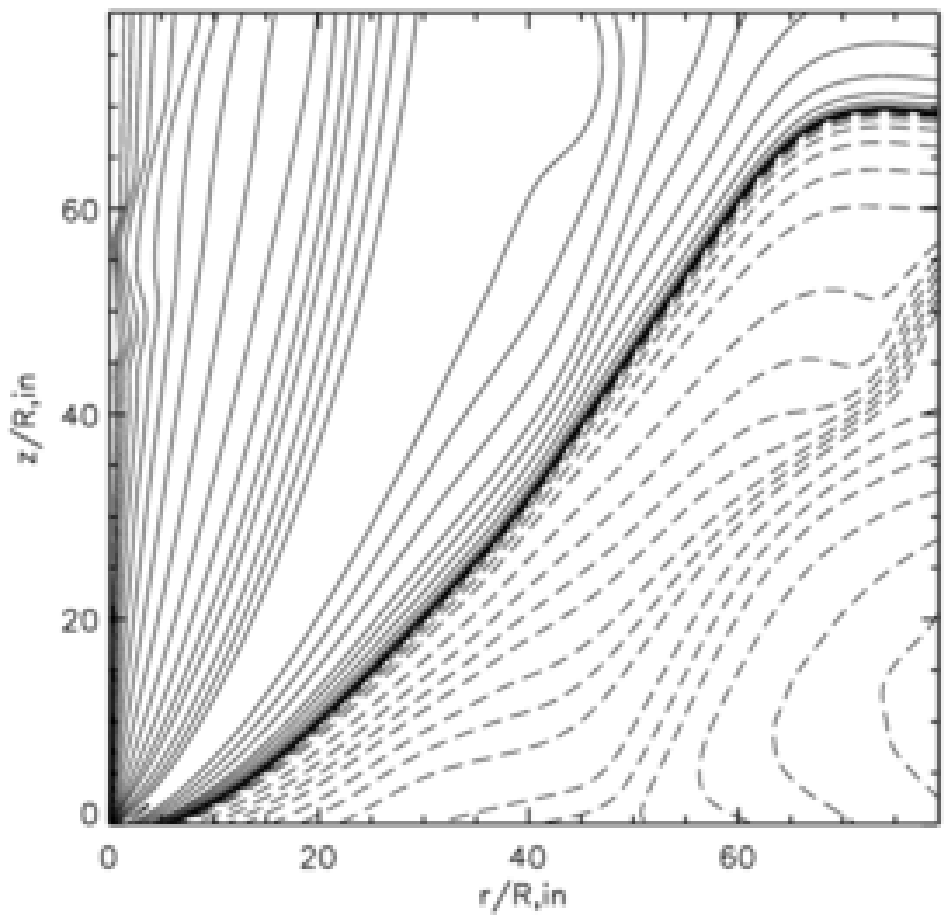}
\includegraphics[width=5cm]{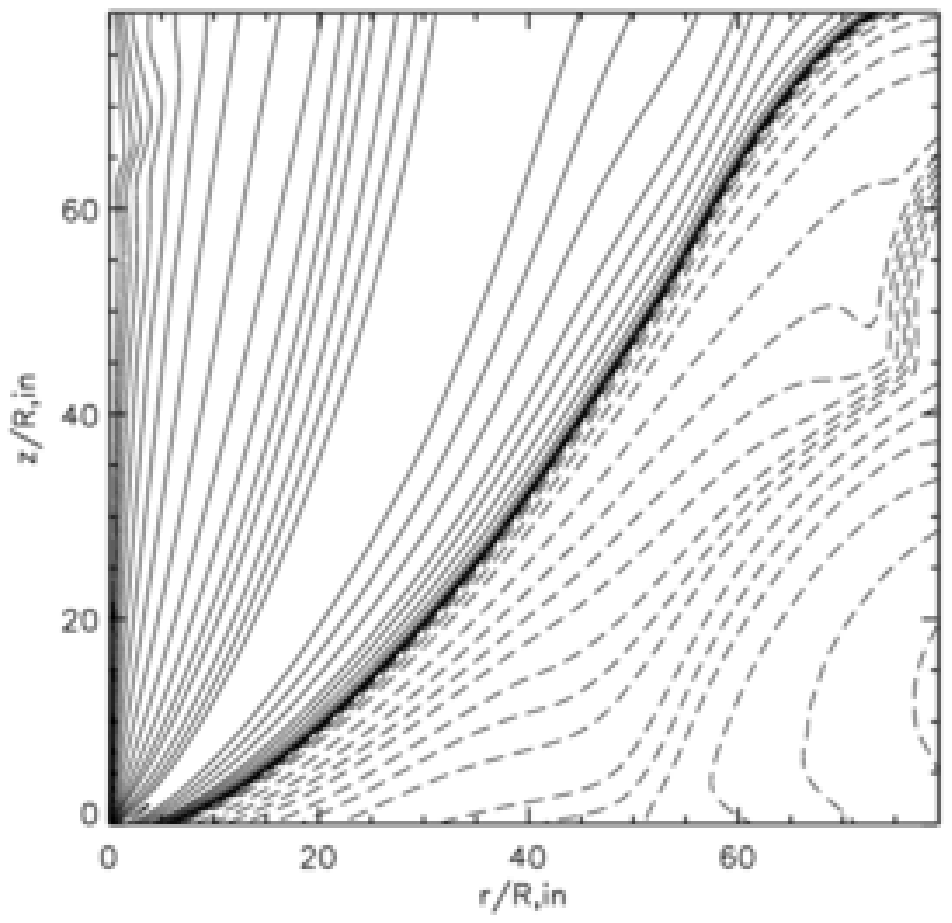}

\includegraphics[width=5cm]{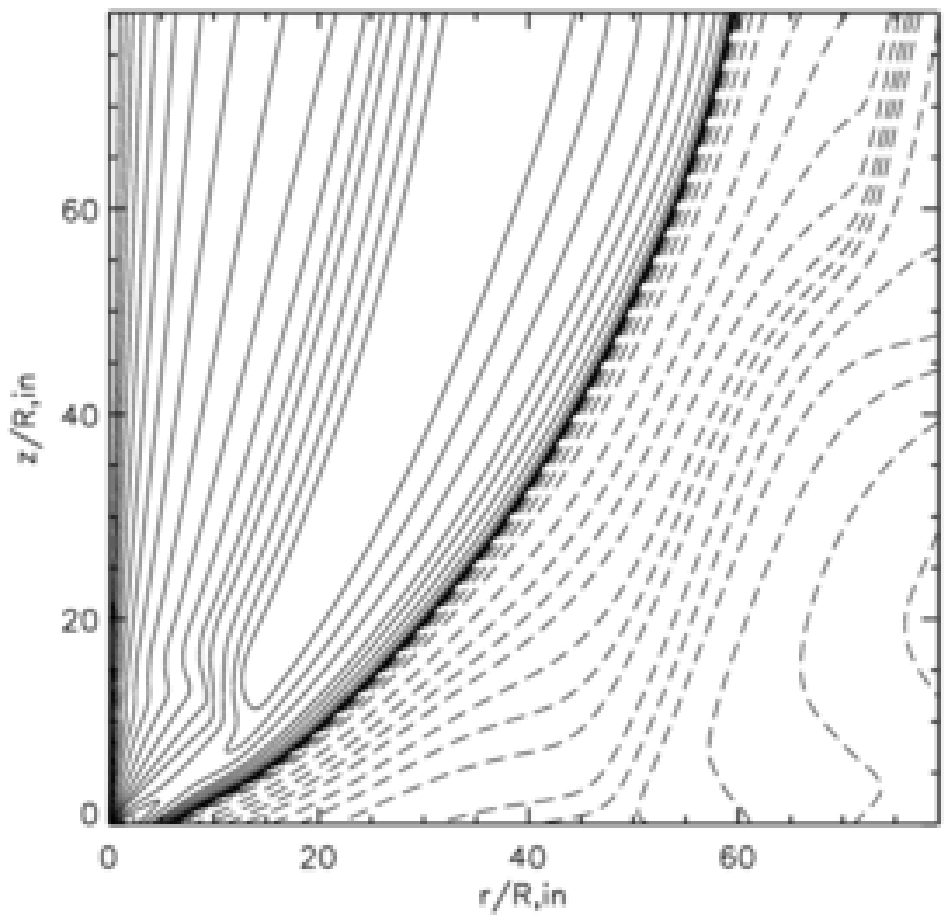}
\includegraphics[width=5cm]{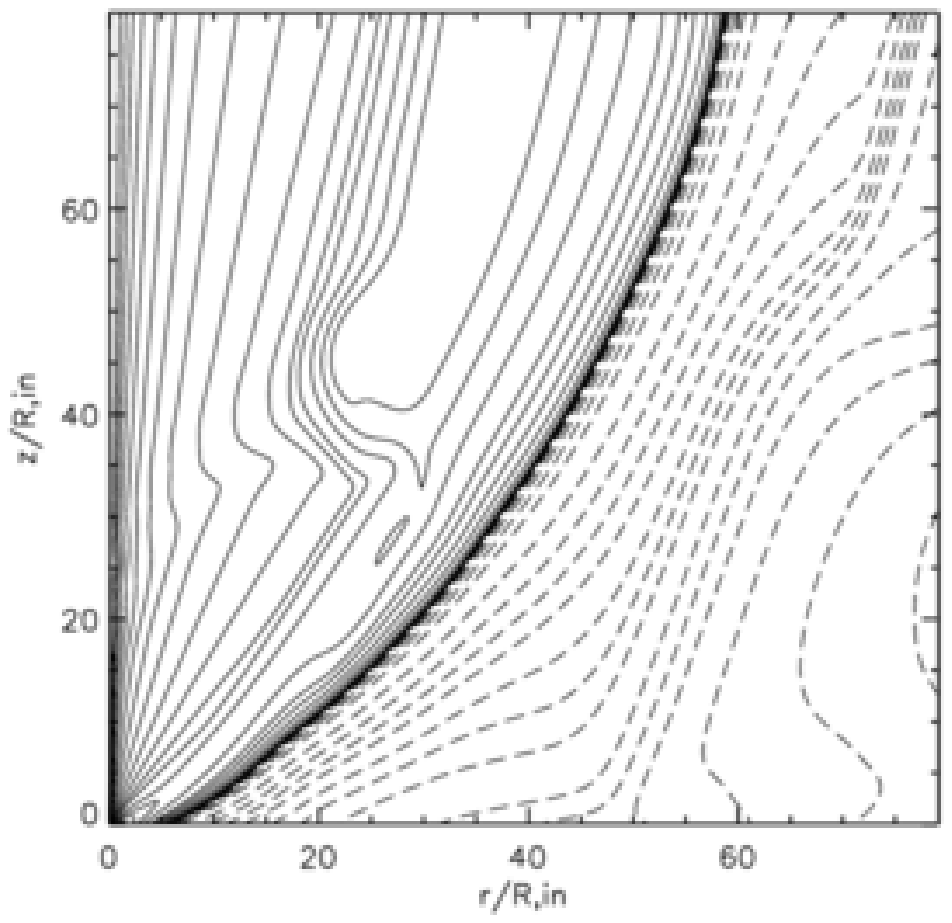}
\includegraphics[width=5cm]{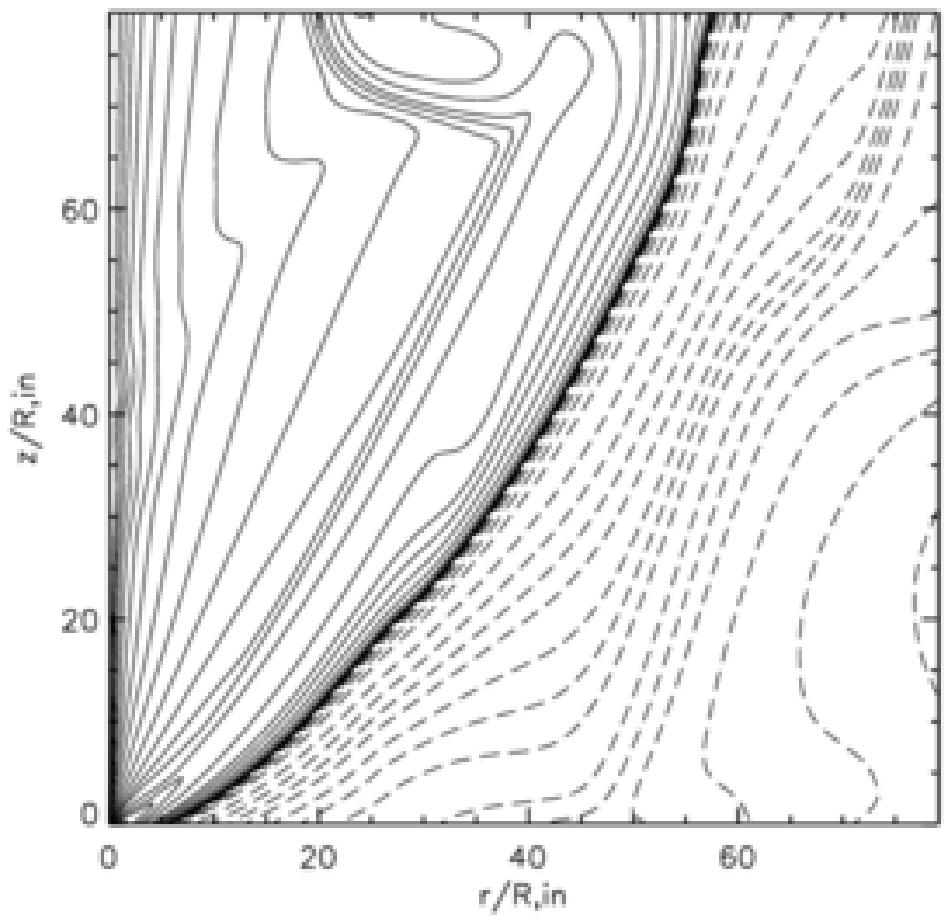}

\includegraphics[width=5cm]{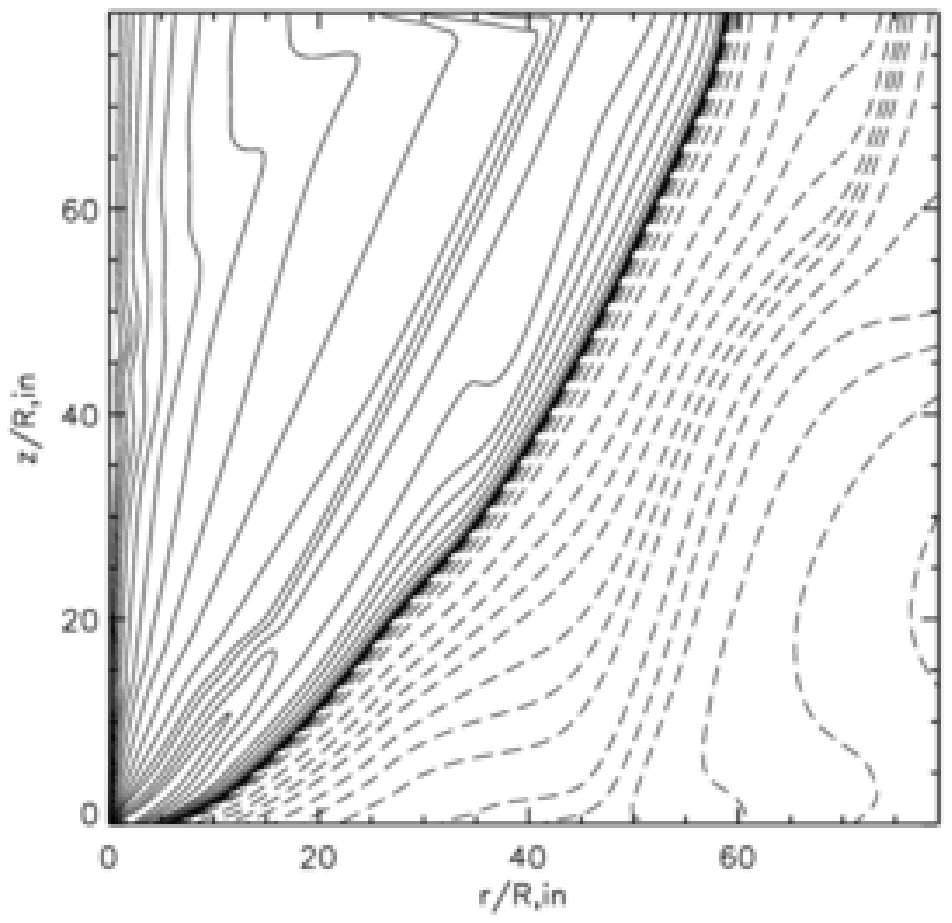}
\includegraphics[width=5cm]{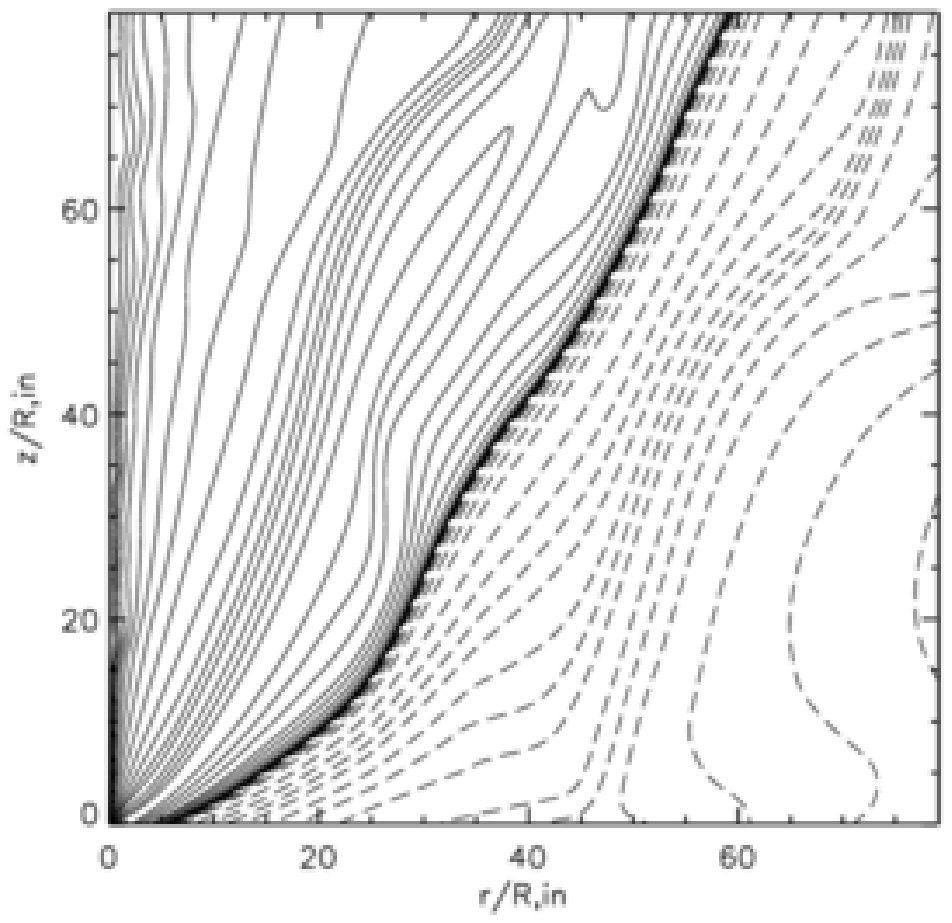}
\includegraphics[width=5cm]{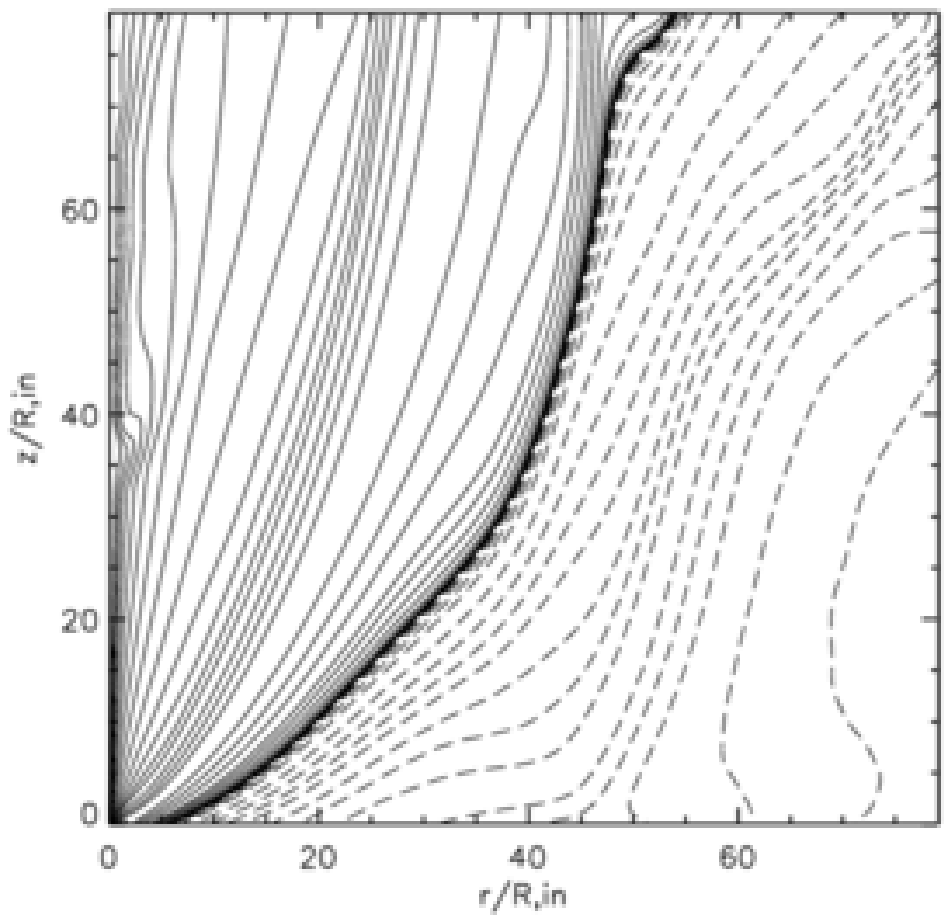}

\caption{Poloidal magnetic field evolution during two flares around $t=176$ and $t=210$.
         Solid and dashed lines indicate the direction of total magnetic flux of the
         superposed dipolar and disk magnetic field components.
         Shown are time steps: 1700, 1760, 1790, 1810, 1840, 1890, 2080, 2100, 2130, 2140, 2180, 2250 
         (from top left to bottom right) of simulation A4a.
         Contour levels as in Fig.~\ref{fig_evol1}.
\label{fig_evol2}
}
\end{figure*}

\section{Results and discussion}
%
We now discuss a number of MHD jet formation simulations covering a 
wide parameter range (see Tab.~\ref{tab_all}) concerning
both the magnitude of the disk and stellar wind mass load and magnetic flux, respectively.
The results presented here are preliminary in the sense that not all simulation runs could be
performed over time scales sufficiently long enough for the MHD flow {\em as a whole} to reach 
the grid boundaries or to establish a stationary state.
This is due to the comparatively large physical grid size in combination with steep gradients 
in the disk wind parameters which in general allow only for a weak outflow from large disk radii.
In particular the steep decline of the stellar field in combination with a reasonable
disk mass loss rate is numerically problematic (see below).
Therefore, for a true comparison between different runs it is essential to also take into account 
the dynamical state of the outflow.

In order to check for resolution issues, we have also run a set of simulations with twice the 
resolution, without seeing significant differences (runs A4b, A5, A6).
Thus, since we are interested in many parameter runs evolving for a long
time, we mainly concentrate on the lower resolution simulations.
We describe first the general evolution of jet formation in our simulations
(see also \citep{ouye97,fend02,fend06}) and later discuss specific features.

\subsection{Overall evolution of an outflow}
Figure~\ref{fig_evol1} shows how the field structure evolves in time 
for the simulation run A4a. 
In this case the stellar dipolar field and the disk field are aligned.
Thus, the X-point is initially along the rotational axis.
We have stopped this simulation after 3600 Keplerian rotations at the 
inner disk radius.
This corresponds to 6 rotations at the outer disk radius and
makes it clear that the outer parts of the disk wind have not yet 
completely evolved into a quasi-stationary state.
This is a generic problem of all disk-jet simulations published so far 
as soon as large disk radii are considered.

The evolution during the very first time steps from the initial 
shows that above the emerging outflow the initial steady state corona
is still present.
The X-point which was initially located at $z<10$ along the rotational 
axis, moves upwards to 
$z\simeq20$ (for $t=50$) and 
$z\simeq40$ (for $t=200$) and is later swept out of the computational domain
for $t>600$ (see top of Fig.~\ref{fig_evol1} and discussion end of Sect.4.2).

Since the initial condition is still kept in steady state, artificial dynamical
re-configuration is prevented. 
This demonstrates that the numerical resolution is sufficient for our
problem also in the out parts of the grid.

The next dominant feature is observed still at early stages of about 
ten rotations.
The initial dipolar field breaks up due to the magnetic pressure of
the toroidal magnetic field induced by differential rotation between
star and disk.
The magnetic pressure gradient drives this outflow. 
Furthermore, an intermediate axial jet is launched due to the re-arrangement
of the initially hydrostatic corona to a new dynamical equilibrium state.
In fact, as the magnetic field along the axis is squeezed by lateral dynamical
pressure, the material is accelerated along the rotational axis.
After the dipole is broken up, there is few direct magnetic connection between
star and disk anymore and the differential rotation induced toroidal field 
decreases.

The outflows from star and disk continue to grow gaining higher kinetic 
energy and momentum.
MHD self-induction of toroidal magnetic field leads to collimation and
magnetic acceleration.
Compared to pure disk winds \citep{ouye97, fend02, fend06} the outflow 
is clearly less collimated as being de-collimated by the central stellar wind.

At intermediate time scales quasi-stationary states may emerge. 
This is demonstrated for example in simulation A4a in Fig.~\ref{fig_quasi} 
by showing the time evolution of the density at point $(z=44, r=44)$.
Two plateaus are clearly seen at different density levels indicating
two quasi-stationary states of the flow evolution.
One is from $t\simeq 1000$ till $t\simeq 1500$, the other from 
$t\simeq 2300-3000$ when the outflow is slowly re-adjusting from
the flaring events (see below).
Note, however, that at this time the outer disk has rotated only about a fifth of
an orbit.
Thus, the field and flow above the outer disk will further evolve in
time and again disturb the enclosed structure in quasi-steady state.
We observe that over even longer time scale such quasi-stationary states 
may be reached (and be disturbed) again and probably again and again.
We believe that this feature is due to the still ongoing evolution of the 
outer disk wind.
The final states of some representative simulation runs are shown in 
Fig.~\ref{fig_final1} and \ref{fig_final2}.
We will discuss them in Sect.\ref{sect_final} below.

As seen already from the initial field configuration, the simulations of 
the anti-aligned field configuration (as e.g. A4a) reveal a change of sign
in the magnetic flux distribution from star and disk (negative magnetic
flux is indicated by dashed contours in Fig.~\ref{fig_reversal}).
The location of flux reversal is accompanied with a concentration of 
contour lines in our figures which may be confused with the existence
of a shock.
This is however an artifact of the choice of contour levels for the 
magnetic flux. In fact, it makes sense to follow the same field lines 
emerging from the stellar surface also back into the disk surface.
Since the strong gradient of the stellar dipolar field, magnetic flux
levels aroung $\Psi=0$ are concentrated.
Figure \ref{fig_reversal} shows that the radial profiles of density,
velocity and poloidal field accross the flux reversal.
Note that the location of magnetic field reversal (the neutral field
line) is at a smaller radius $r=33$ compared to the magnetic flux 
reversal at about $r=53$.
Within the magnetic field reversal magnetic flux is accumulated.
Beyond the field reversal the magnetic flux decreases and then
becomes negative.

\subsection{Reconnection and large scale flares}
Our version of ZEUS code has physical magnetic diffusivity implemented
(see \citet{fend02} for explanation and tests). 
This allows to consider reconnection processes.

We observe reconnection flares along some of the outflows (see Fig.~\ref{fig_evol2}).
These flares are similar to coronal mass ejection.
They rapidly evolve and propagate along the neutral field line.
Once formed, reconnection islands (or rather "tori" in our axisymmetric setup)
propagate across the jet magnetosphere 
within a few rotation times and leave the computational domain.
The flares typically expand and reconnect within 70 time units equivalent
to 70 orbital periods of the inner disk (and 70 stellar rotations).
We may also naively measure the flare propagation speed by the proper
motion of the field lines observed in the simulation.
This gives an average "flare propagation speed" of about unity, as they
travel 80 spatial units in 70 time units - the Keplerian speed at their
foot-point.
The flares propagate along the neutral field line, and leave the physical 
grid at a radius corresponding to about 7-10 AU from the axis.

\begin{figure}
\centering
\includegraphics[width=8cm]{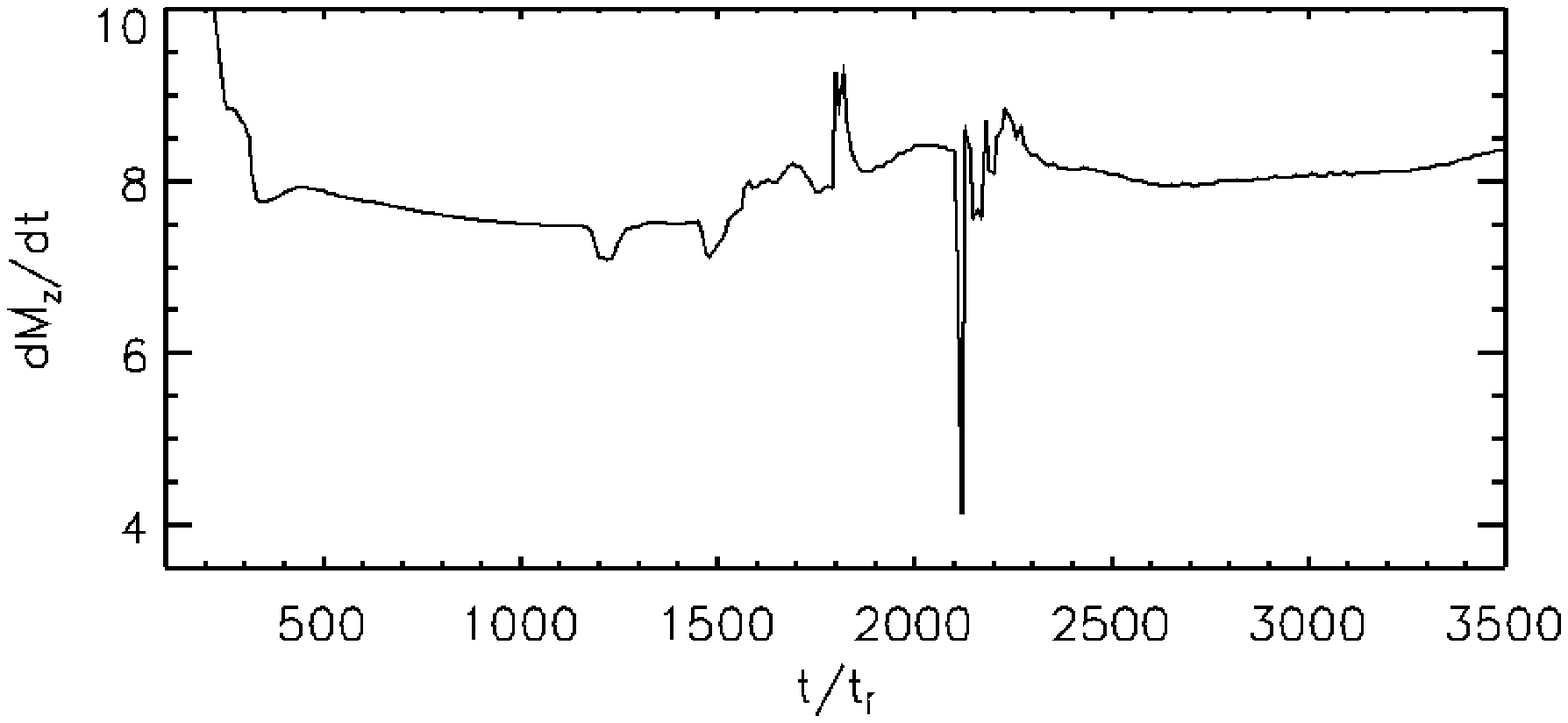}
\includegraphics[width=8cm]{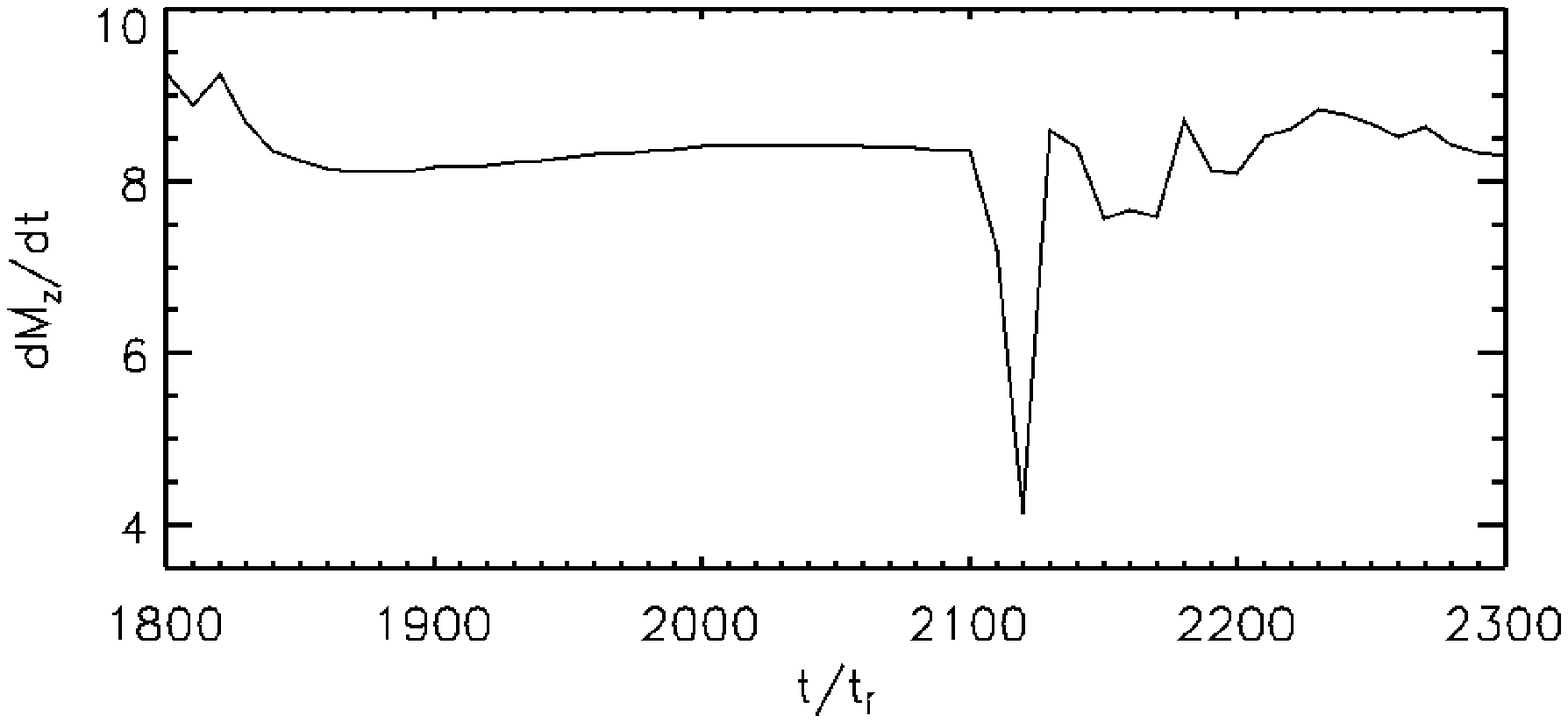}

\caption{Integrated mass flux in axial direction across the upper $z$-boundary
versus time. Note the change of mass flux of about $10-50\% $ during 
the flare events. The high mass flux for $ t<500 $ indicates sweeping off of
the initial hydrostatic corona.
\label{fig_mflux_i}
}
\end{figure}

\begin{figure*}
\centering
\includegraphics[width=8cm]{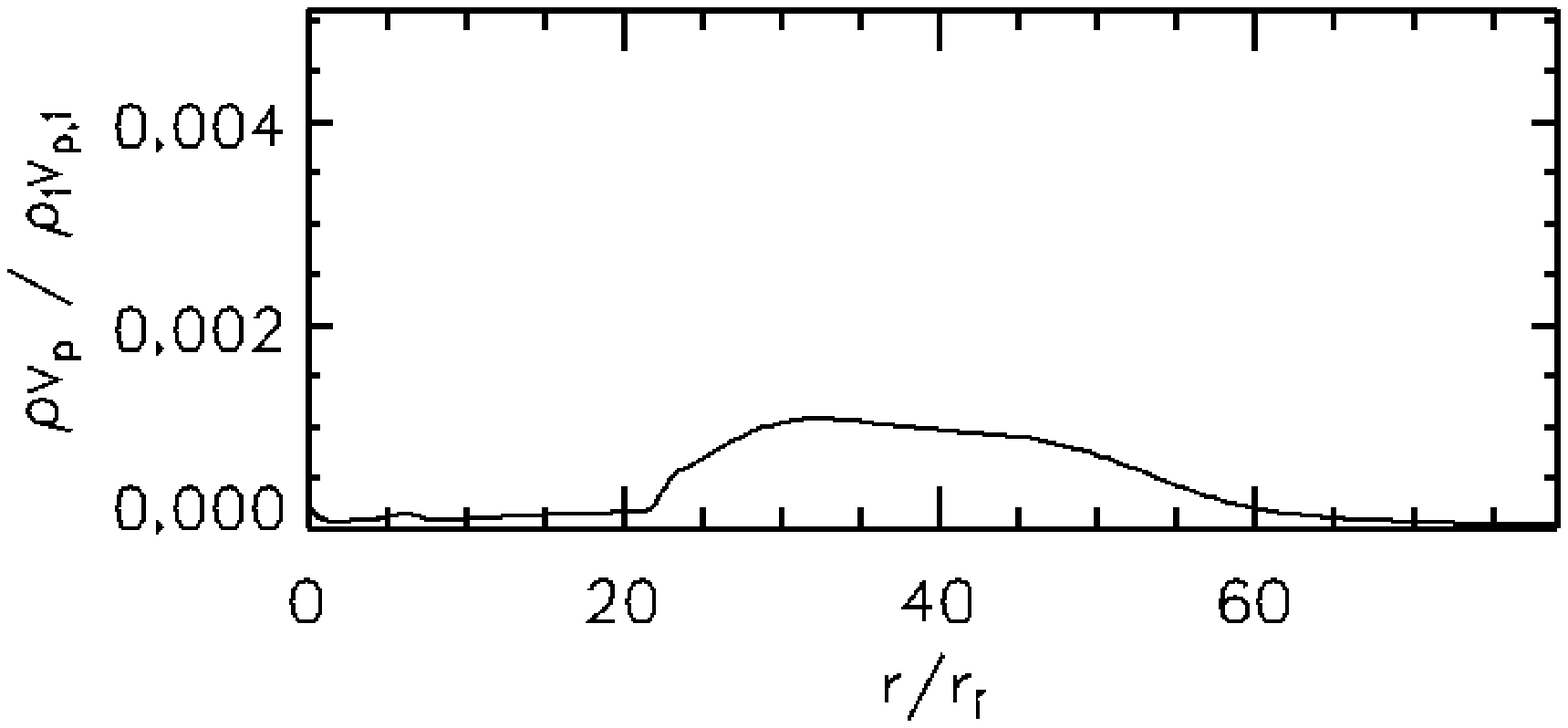}
\includegraphics[width=8cm]{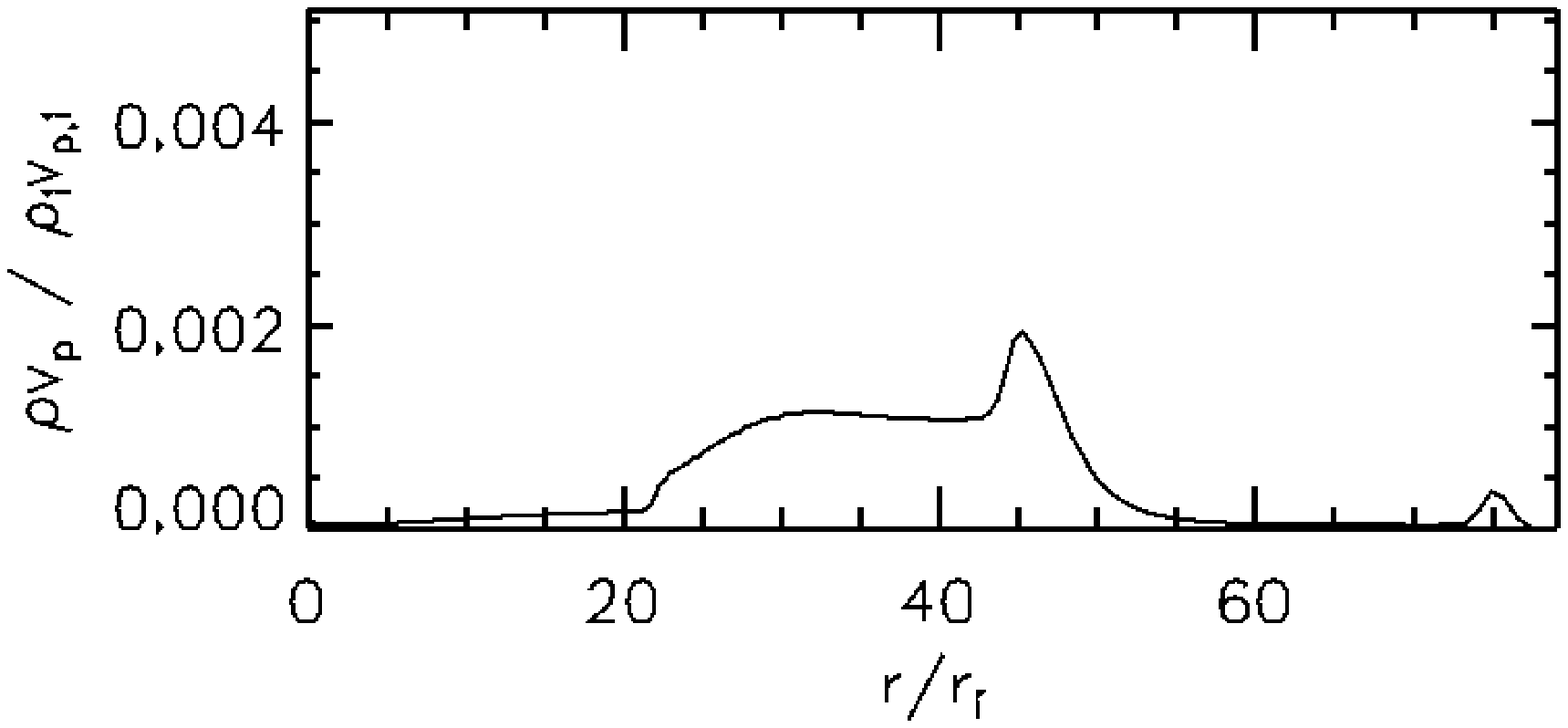}

\includegraphics[width=8cm]{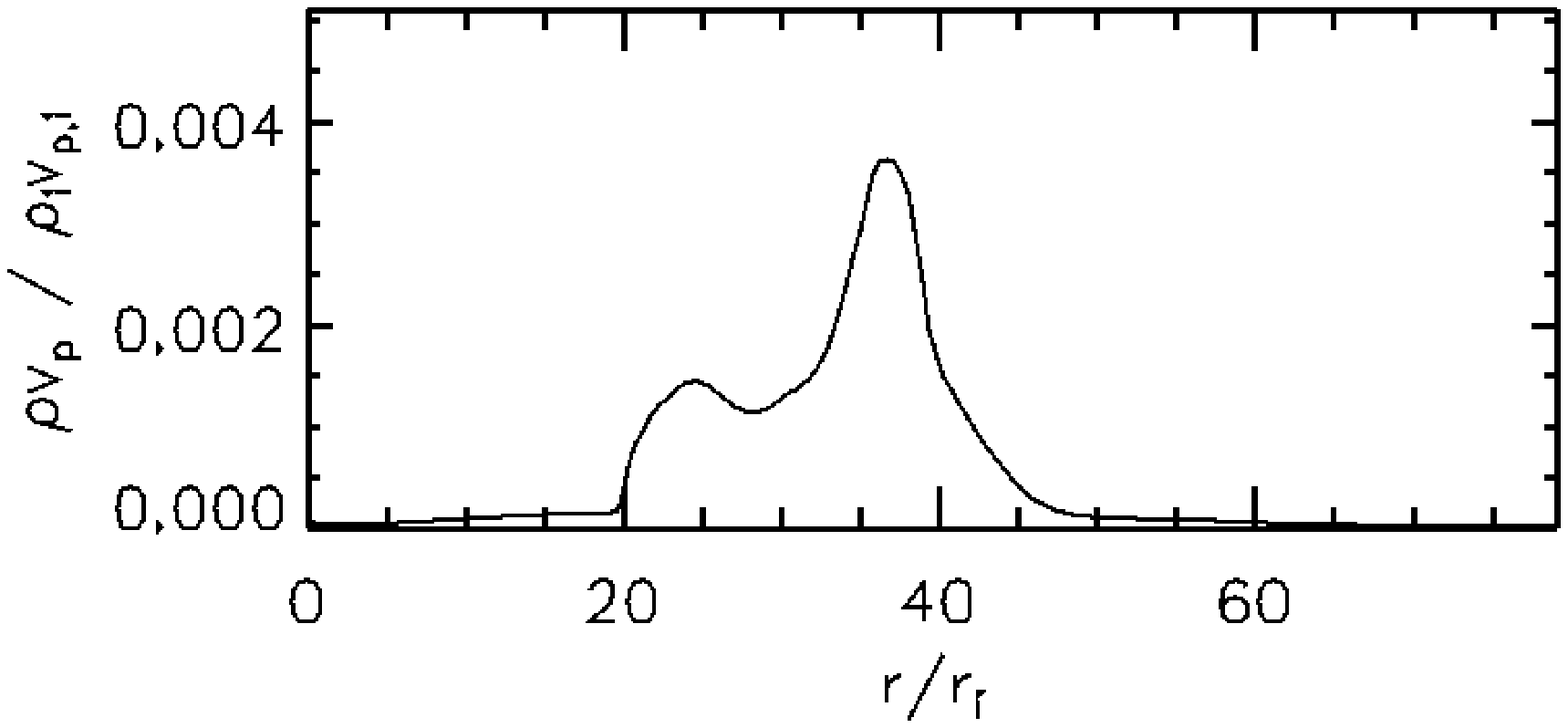}
\includegraphics[width=8cm]{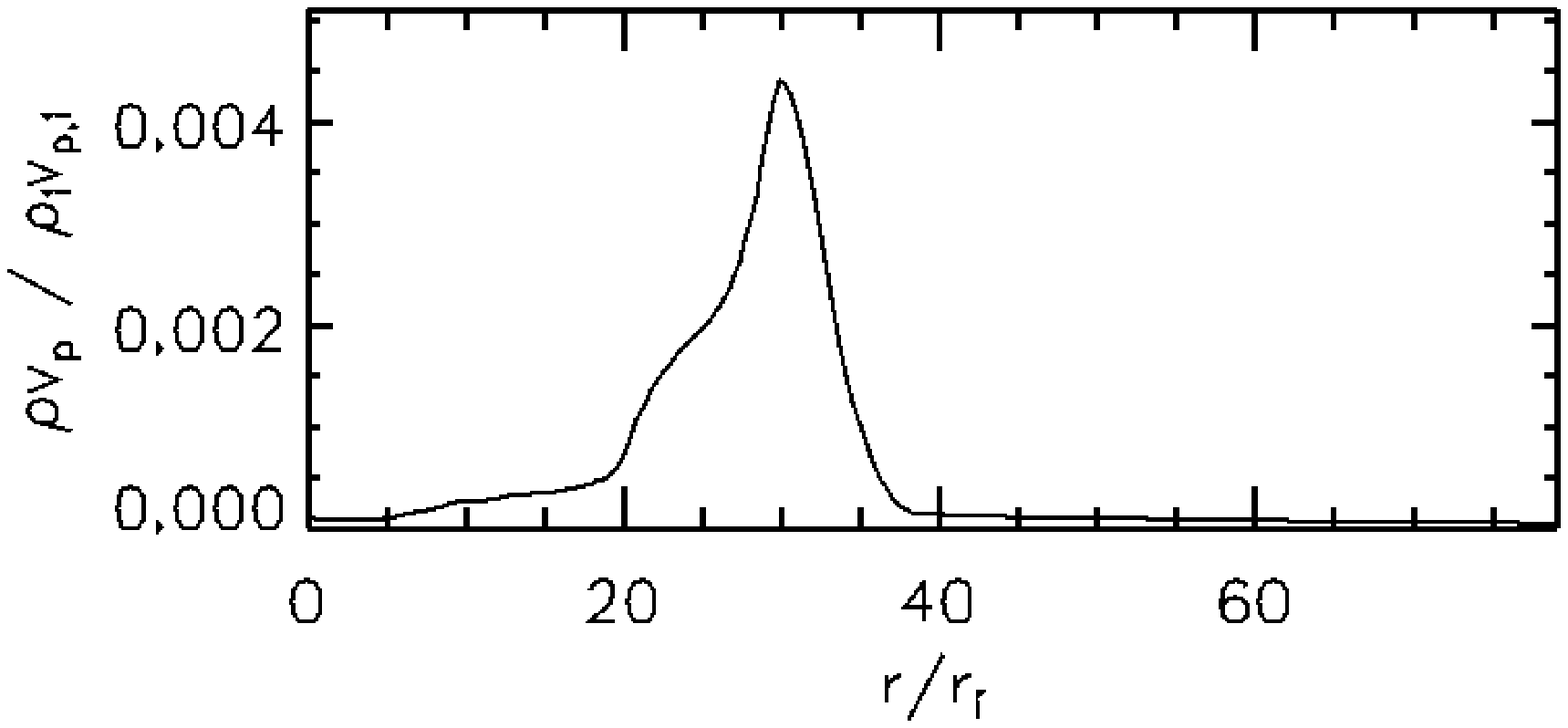}

\includegraphics[width=8cm]{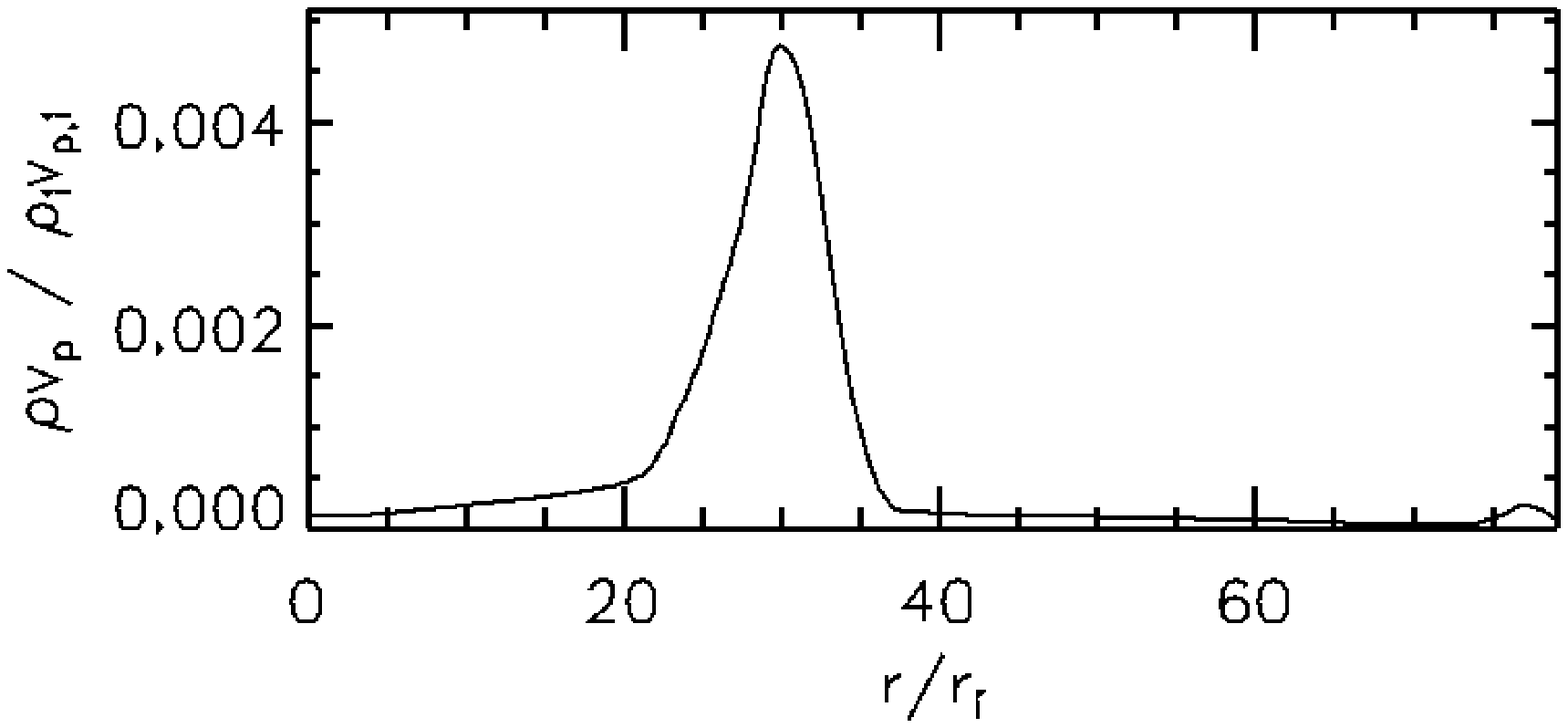}
\includegraphics[width=8cm]{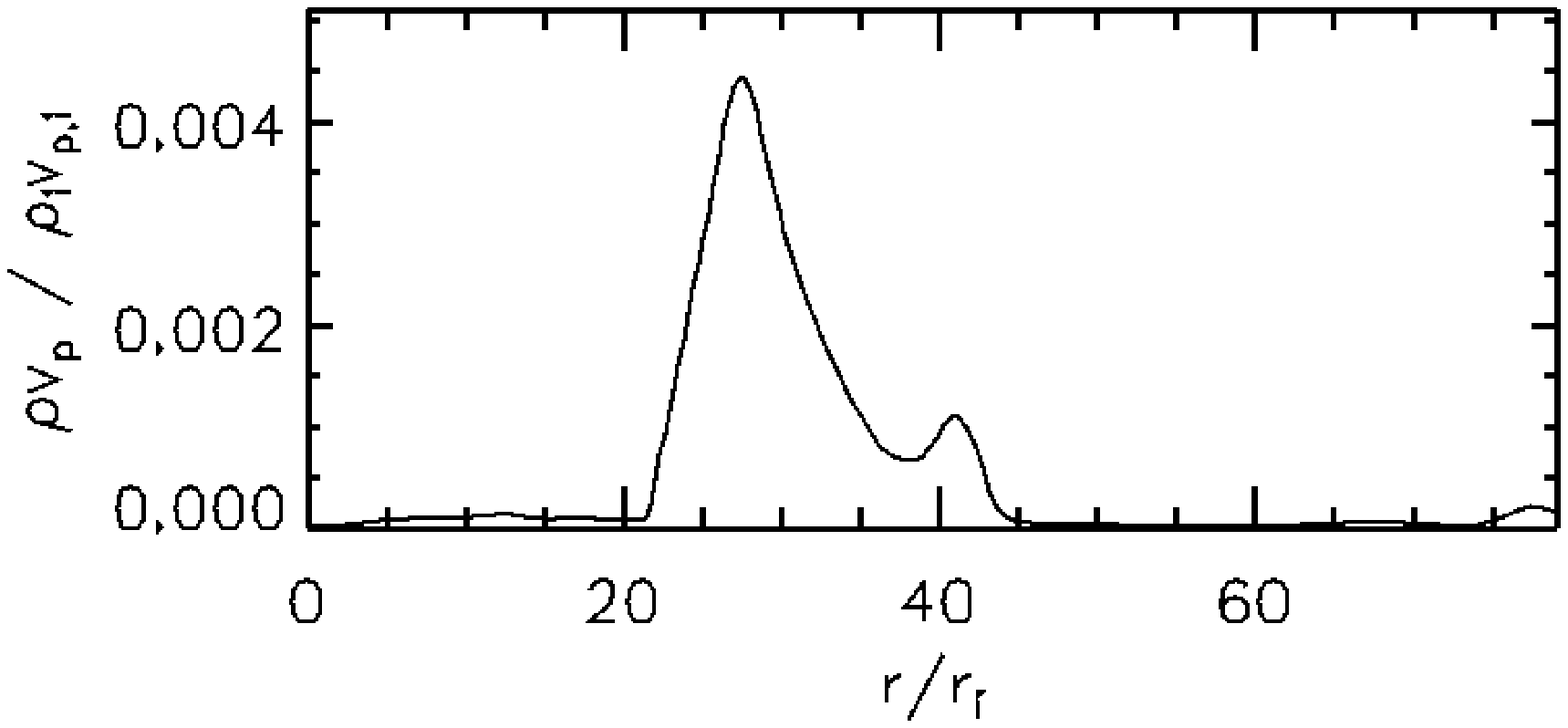}

\includegraphics[width=8cm]{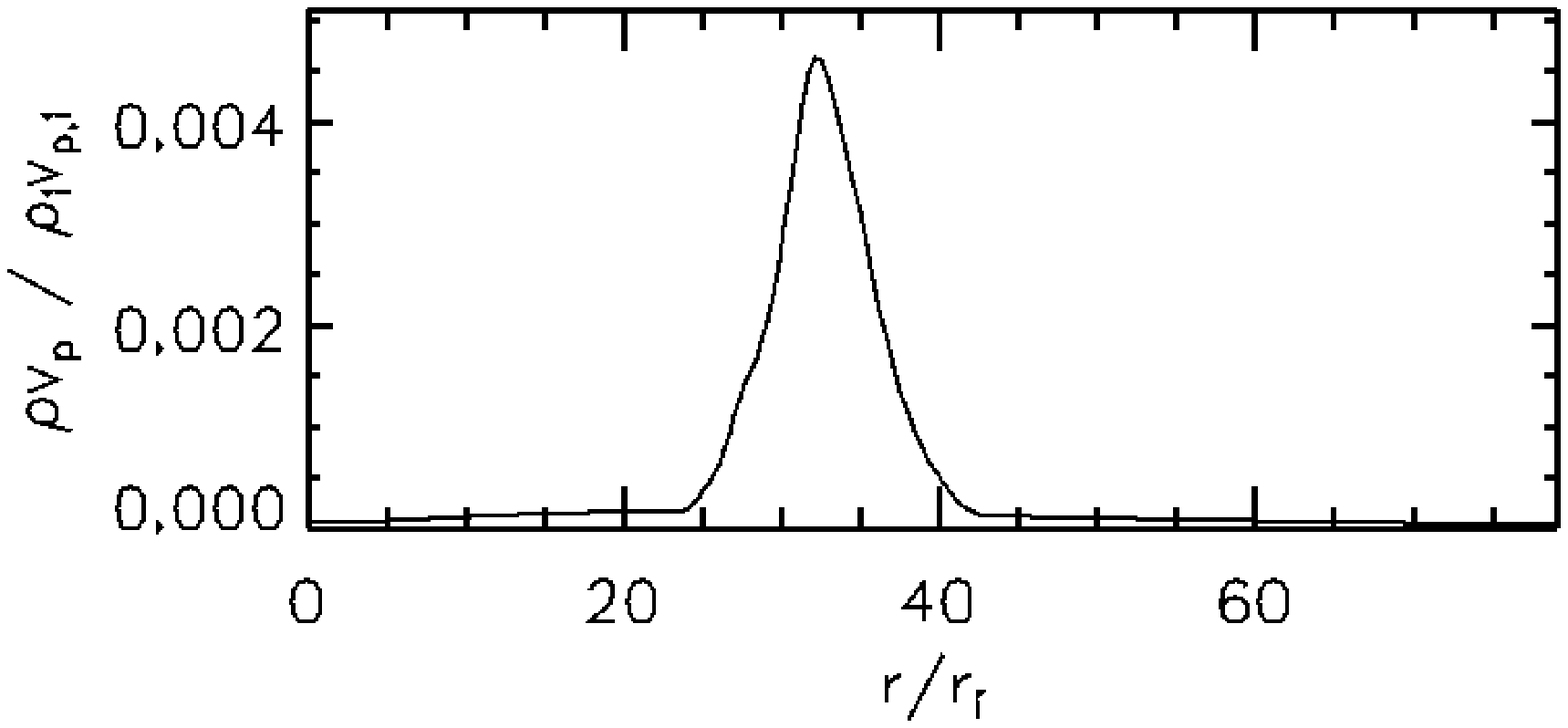}
\includegraphics[width=8cm]{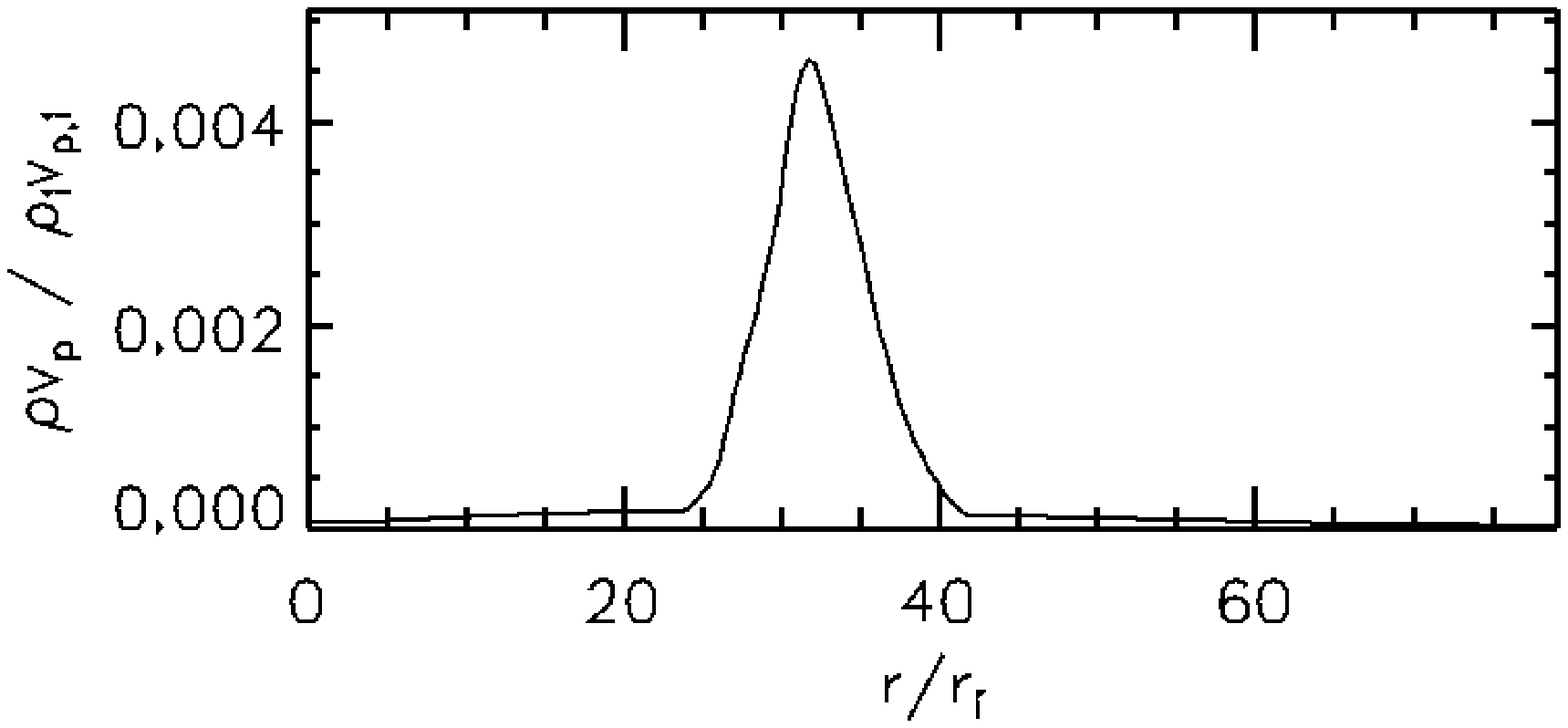}

\caption{Profile of the outflow momentum along the z-boundary
at time $t=1400, 1600, 1800, 1900, 2100, 2200, 2500, 3000$.
Note the re-distribution of the main mass flow channel from
radius $r=60$ to $r=40$.
\label{fig_mflux_mv}
}
\end{figure*}

\begin{figure*}
\centering
\includegraphics[width=8cm]{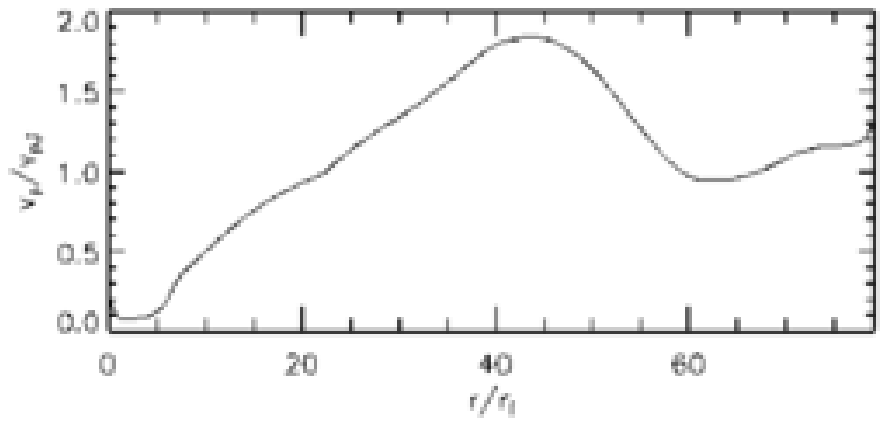}
\includegraphics[width=8cm]{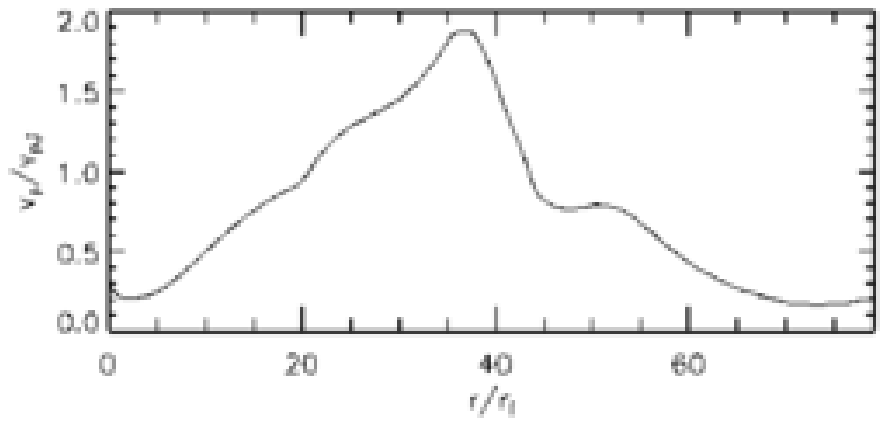}
\includegraphics[width=8cm]{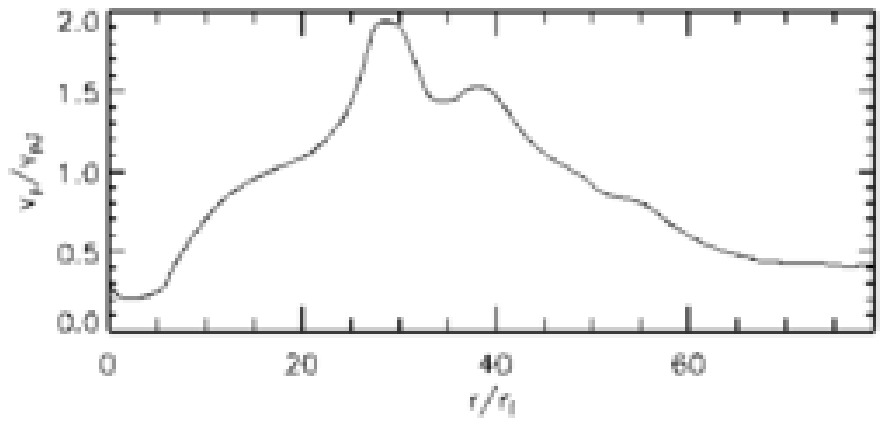}
\includegraphics[width=8cm]{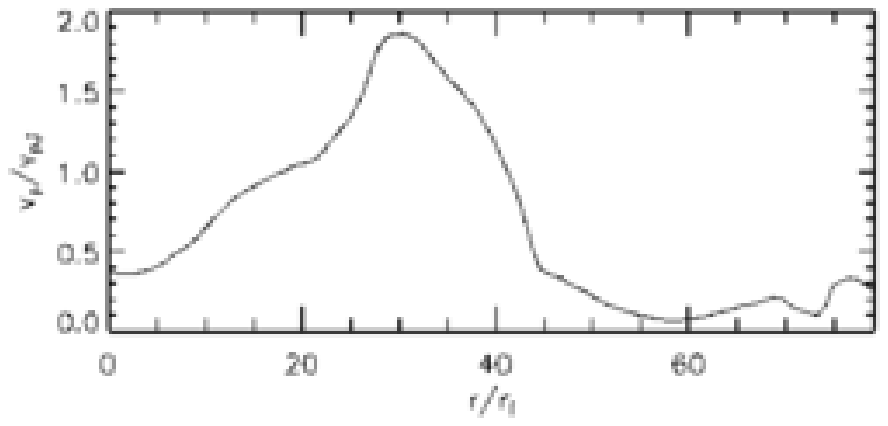}
\includegraphics[width=8cm]{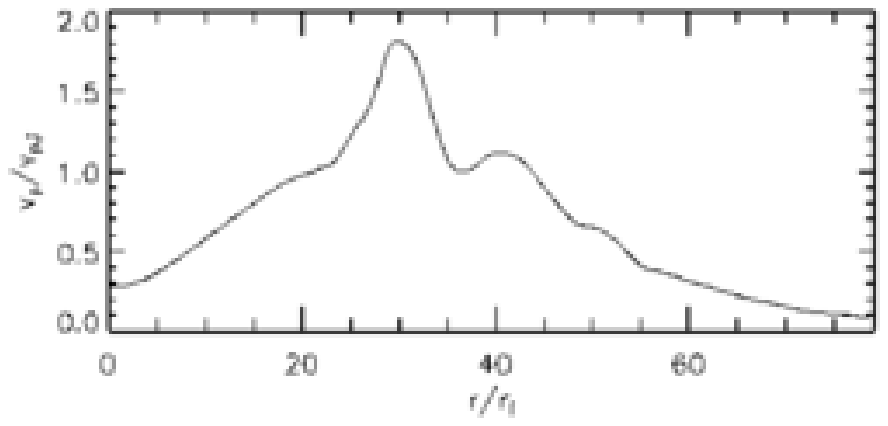}
\includegraphics[width=8cm]{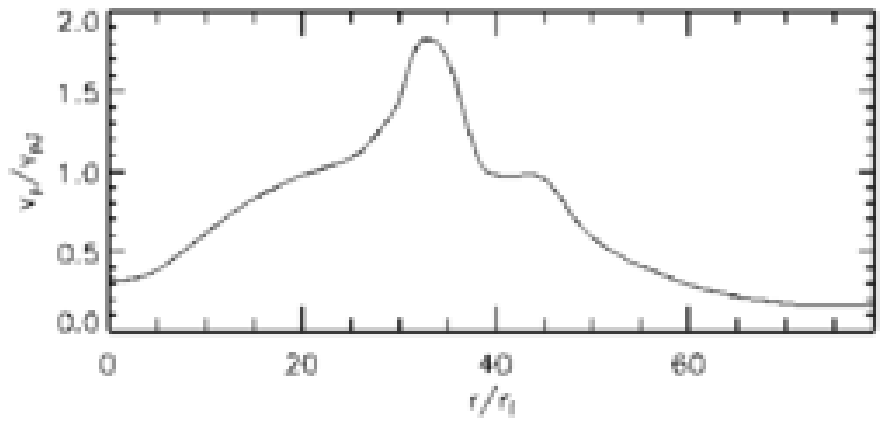}

\caption{Profile of the axial outflow velocity along
the z-boundary at time $t=1400, 1800, 1850, 2200, 2300, 3000$.
\label{fig_mflux_v}
}
\end{figure*}

\begin{figure*}
\centering

\includegraphics[width=7cm]{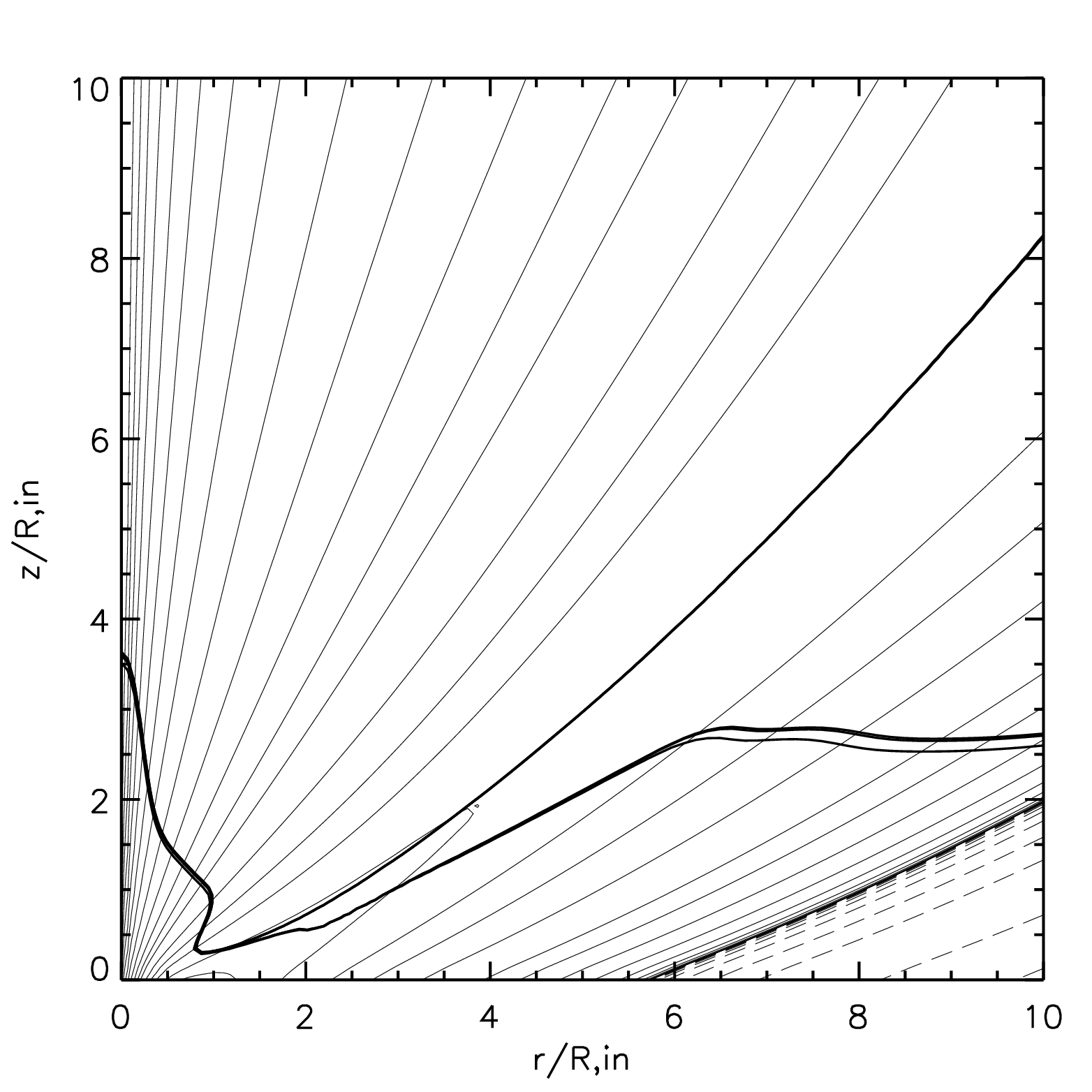}
\includegraphics[width=7cm]{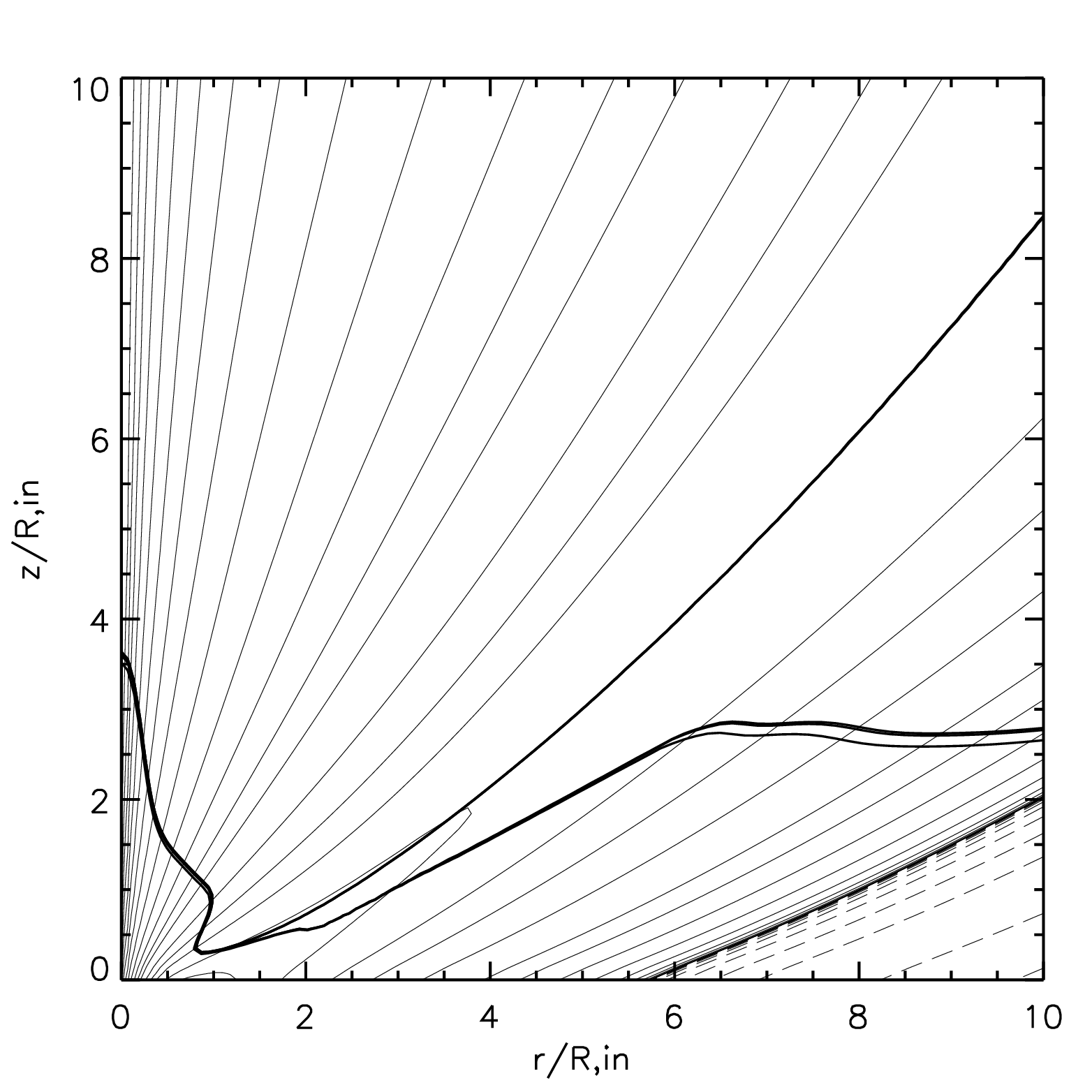}

\includegraphics[width=7cm]{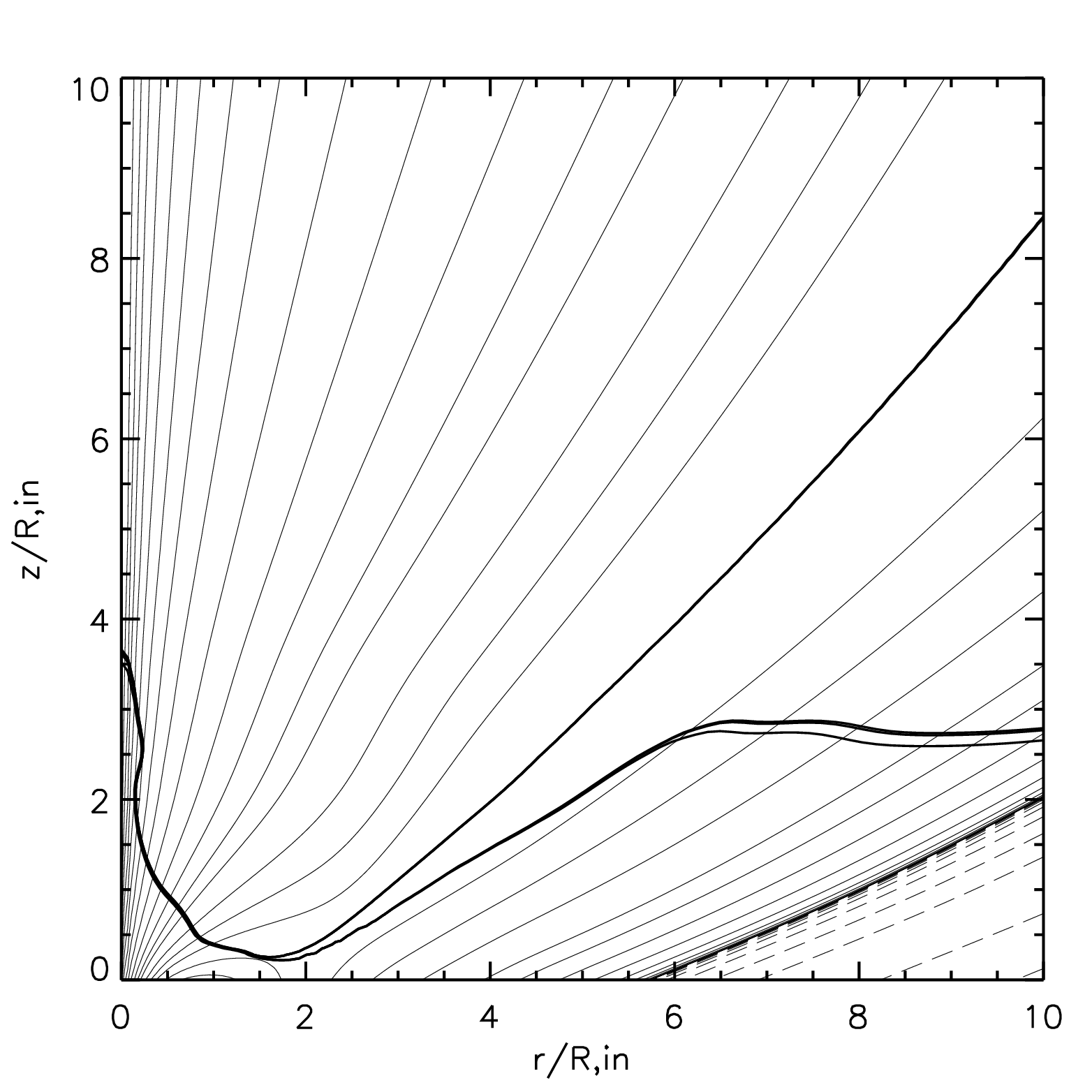}
\includegraphics[width=7cm]{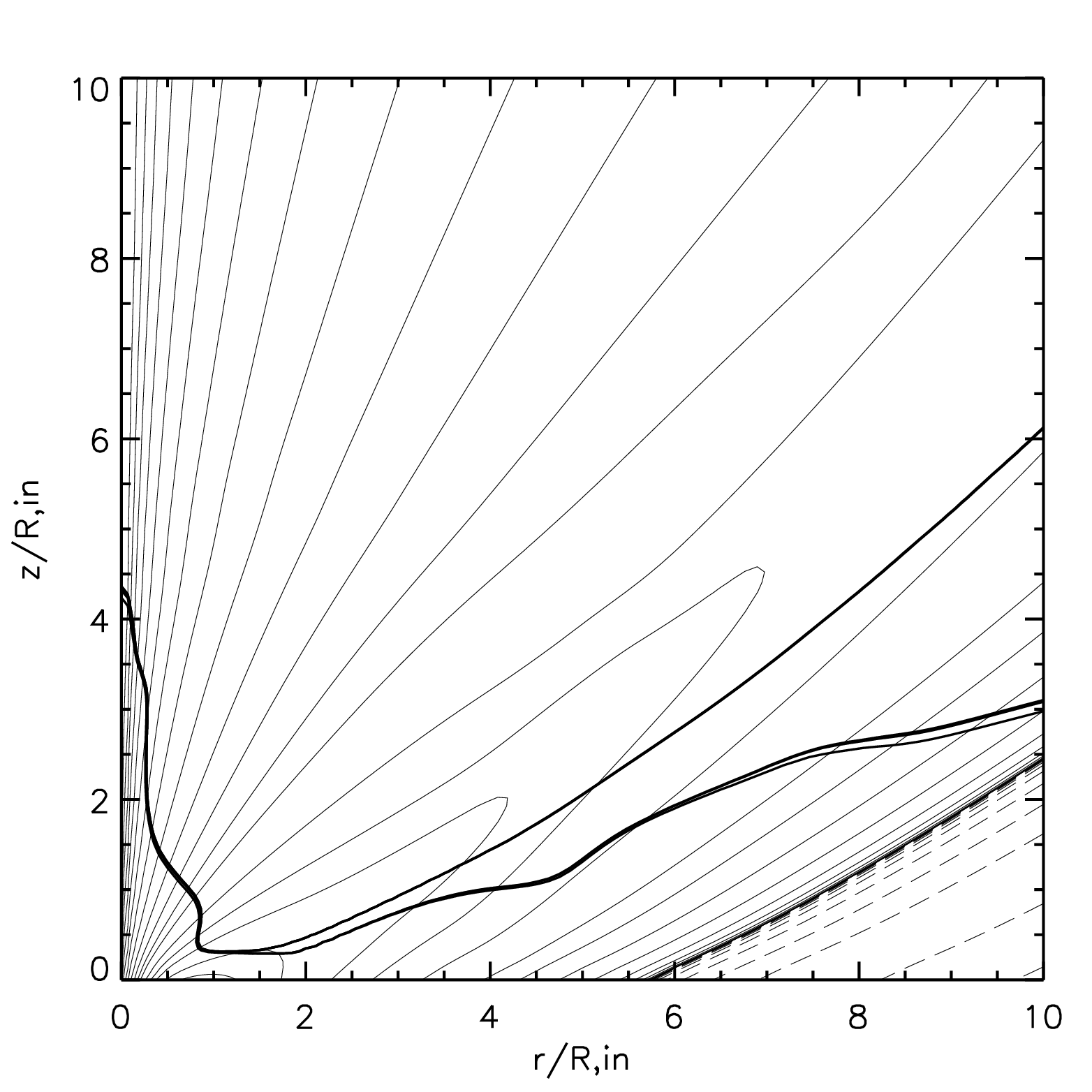}

\includegraphics[width=7cm]{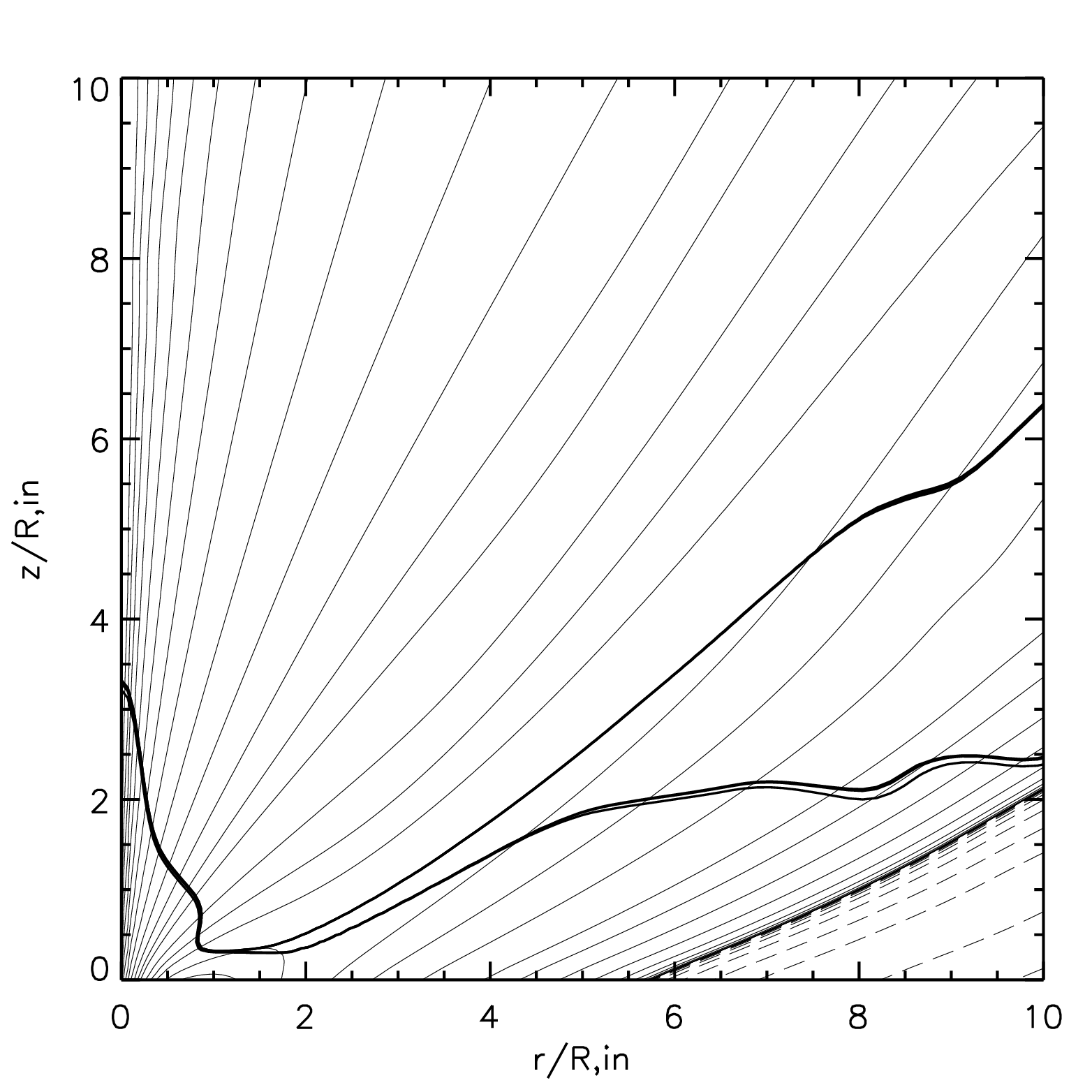}
\includegraphics[width=7cm]{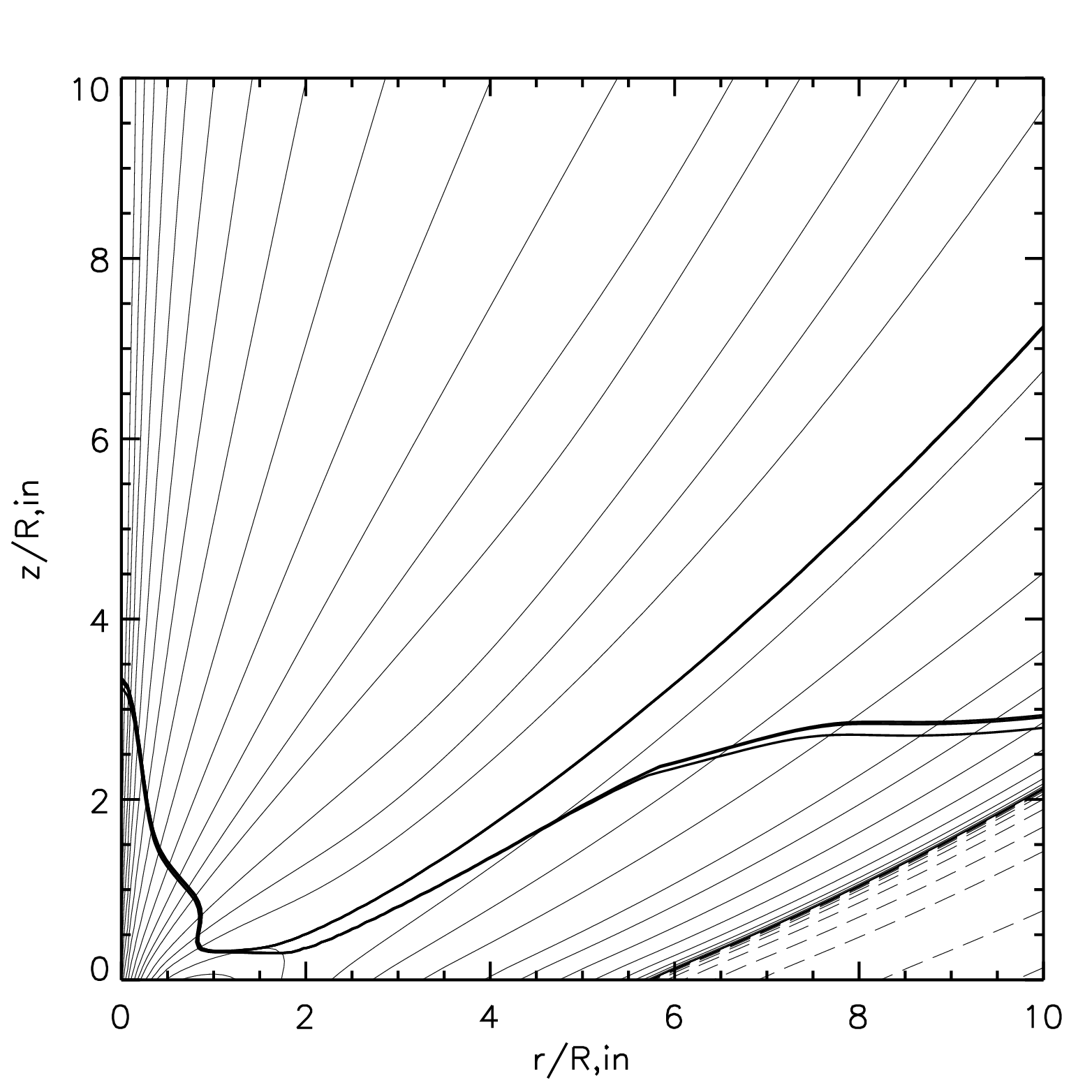}

\caption{Poloidal magnetic field distribution ({\em thin lines})
         during flare evolution (run A04).
    Contour levels as in Fig.~\ref{fig_evol1}.
        Simulation time steps
    t=2000, 2060, 2070, 2130, 2140, 3000.
    (from top left to bottom right). Alfv\'en and fast surface are
    shown by {\em thick lines}. 
    Note the super-Alfvenic /super-fast stellar wind.
\label{fig_small1}
}
\end{figure*}

In the following we mainly concentrate on simulation A4a.
Figure \ref{fig_small1} is a magnification of the inner area of 
Fig.\ref{fig_evol2} and demonstrates that the flares 
are actually from a Y-point above the disk, 
located at ($z\simeq2,r\simeq4$)
Note that the stellar magnetosphere remains closed also for foot-points
along the disk with $r<3$ (at $t\simeq1700$), 
respectively $r<2$ ($t\simeq1700$), beyond the co-rotation radius.
In this case the flare eruptions are launched close to the inner disk area, 
but {\em not} at the inner disk radius.

\citet{good99} also observed flares or reconnection events in their simulations 
of a dipolar magnetosphere connected top a surrounding accretion disk.
Since the evolution of the disk structure was treated in their simulation, 
they were able to observe time-dependent accretion along the equatorial plane.
The frequent ejection events of AU-sized knots along the rotational axis
were correlated with a time-variation of the accretion rate.
This is different in our case, where the reconnection/flares seem to be
triggered by the evolution of the outer disk wind.
As mentioned, during the run-time of our simulations (up to 3600 inner disk rotations)
the corona (or disk wind respectively) above the outer disk has not yet reached
a steady dynamical state\footnote{However, the same argument holds for the
Goodson et al. work}.
Thus, in our case, as the outer disk outflow evolves, the cross-jet pressure equilibrium
is changed and forces the inner magnetic field configuration to
adjust accordingly.
Since Goodson et al did not specify the distribution and magnitude of 
the magnetic diffusivity in their
simulations, it is difficult to compare the results in detail.
Another difference is the duration of the simulation.  
While our physical grid size is only marginally larger, the time scale of our simulations is
substantially longer (3600 inner disk rotations corresponding to about 36,000 days 
compared to 130 days).
This is particularly important when considering the dynamical evolution
above the outer part of the disk and the long-term behavior.

The flare events observed in our simulations are accompanied by a temporal change 
in outflow mass flux, velocity, or momentum, respectively.
Figure \ref{fig_mflux_i} shows the mass loss rate in axial direction integrated
across the upper r-boundary versus time.
During the first flare we see a 10\%-variation in the mass flux followed by
a sudden decrease of mass flux by a factor of two during flare two. 
Figures \ref{fig_mflux_mv} and \ref{fig_mflux_v} show the profiles of jet momentum and 
poloidal velocity across the upper r-boundary.
We see that during flaring these profile indicate a re-arrangement of momentum and
velocity profile across the jet.
Before the flare, the radial jet momentum profile is broad 
(see Fig.~\ref{fig_mflux_mv} upper left for $t=1400$).
In fact, the profile (Fig.~\ref{fig_mflux_mv}, $t=1400$) remains very similar for 
several 100 rotations before until the flare event starts.
After the flare has passed the grid, the jet momentum profile is concentrated 
within a cylindrical sheet of radius
$r\simeq 35$ and thickness $\Delta r \simeq 5$ 
(Fig.~\ref{fig_mflux_mv} lower right at $t=2300$).
Similar to the situation before the flare, the jet momentum profile for time steps
after the flare ($t>2300$) look almost identical 
(see bottom sub-figures for $t=2600$ and $t=3000$).

\begin{figure*}
\centering

\includegraphics[width=7cm]{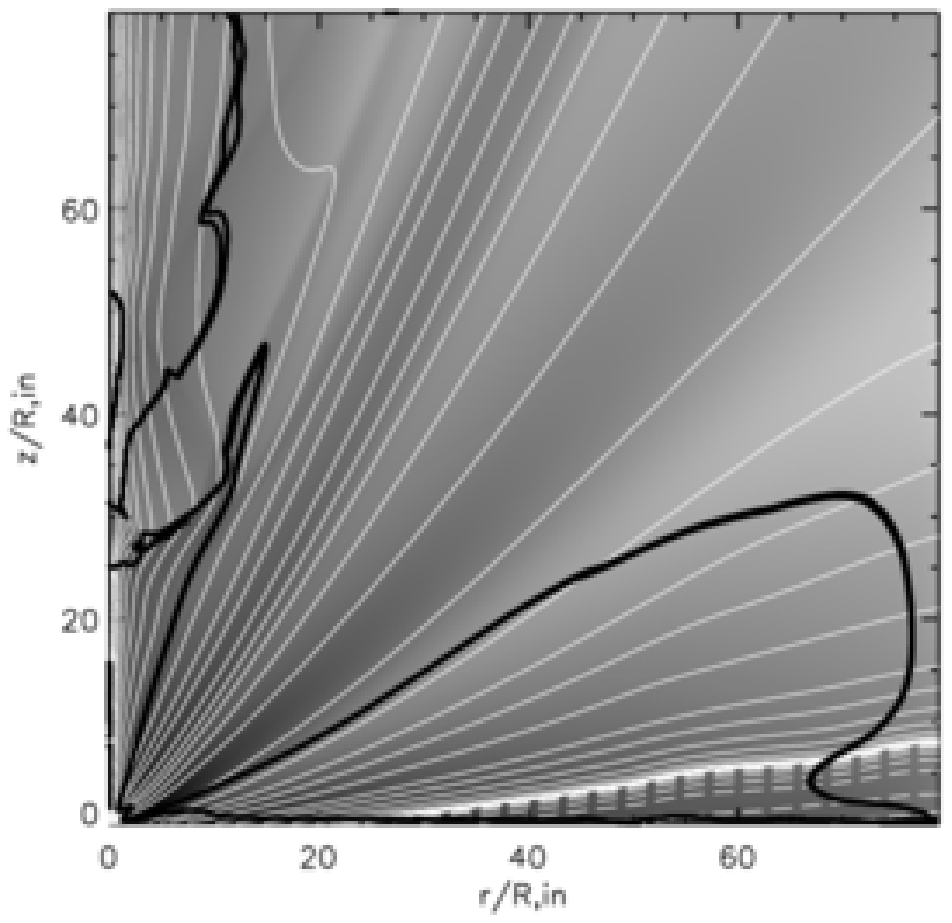}
\includegraphics[width=7cm]{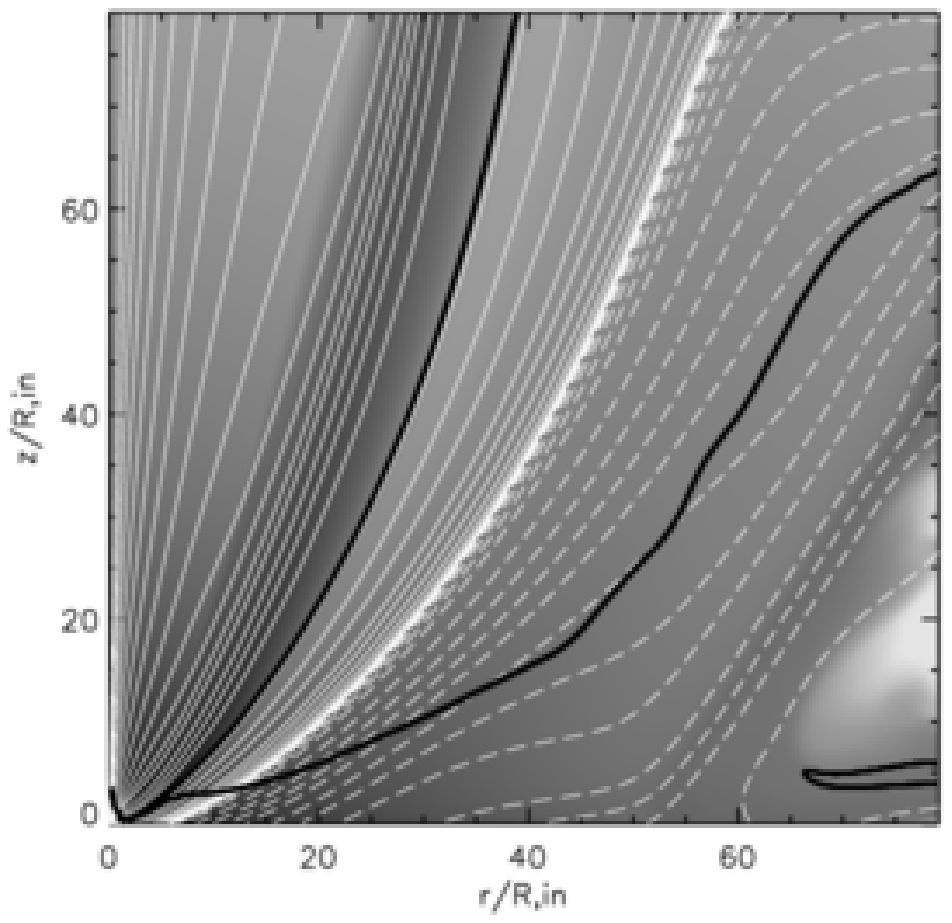}

\includegraphics[width=7cm]{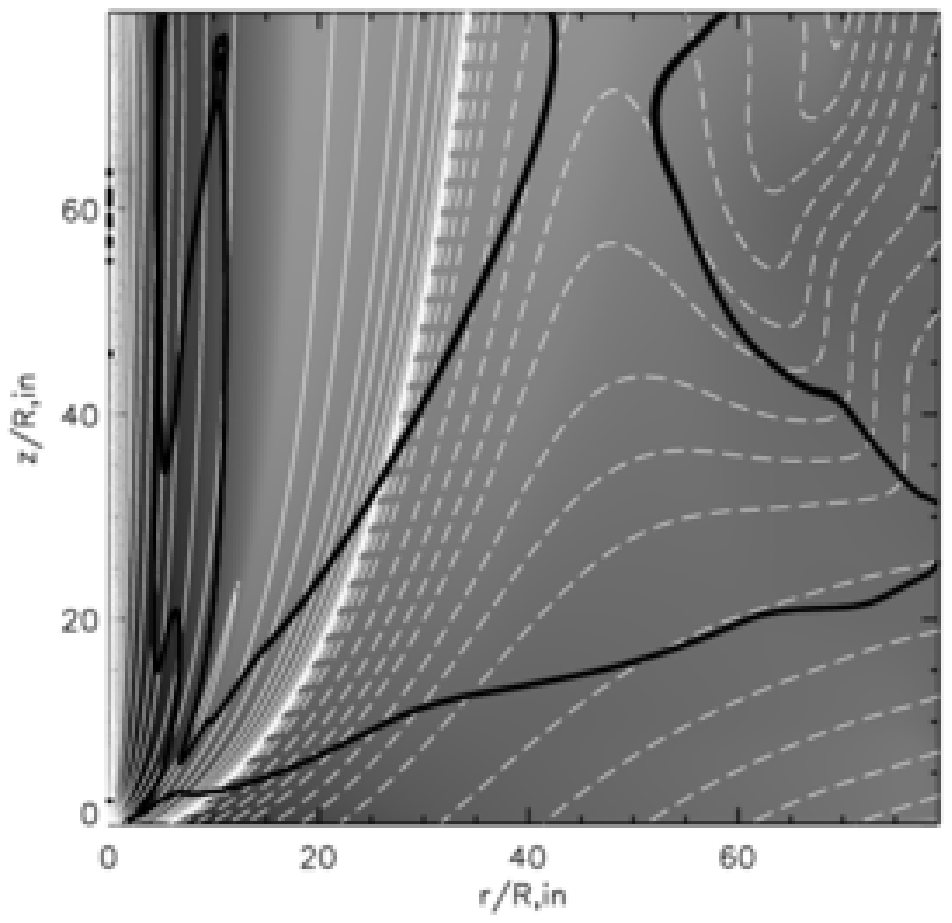}
\includegraphics[width=7cm]{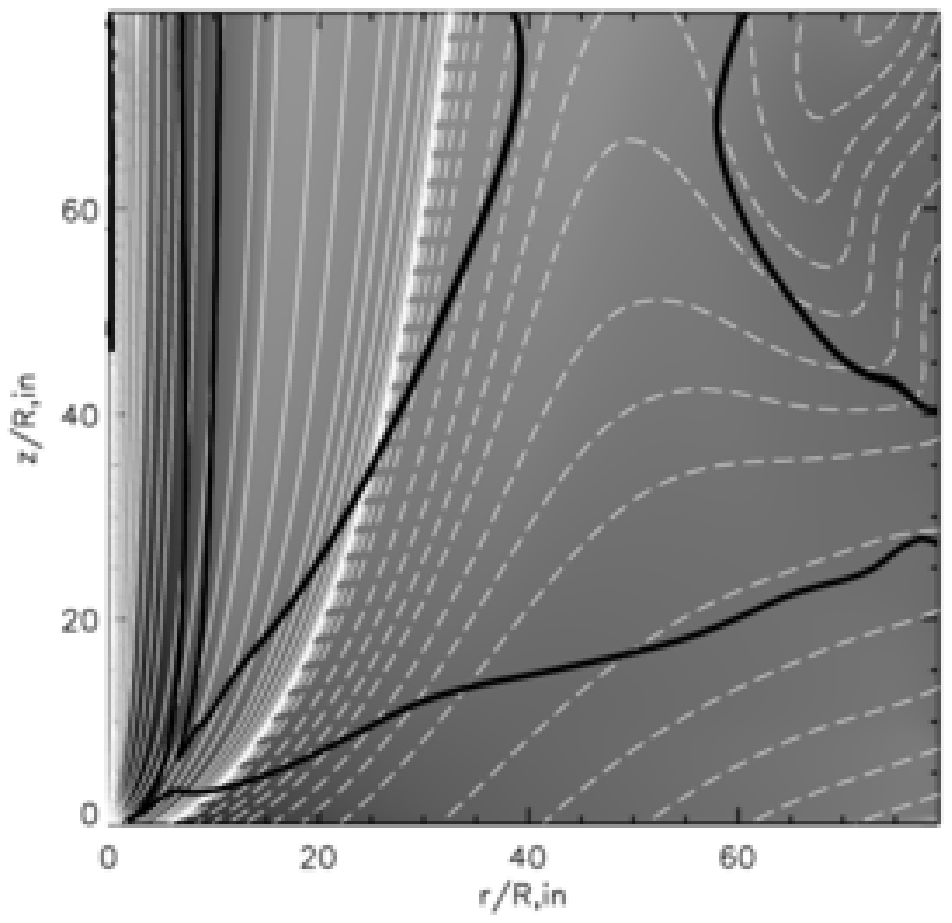}

\includegraphics[width=7cm]{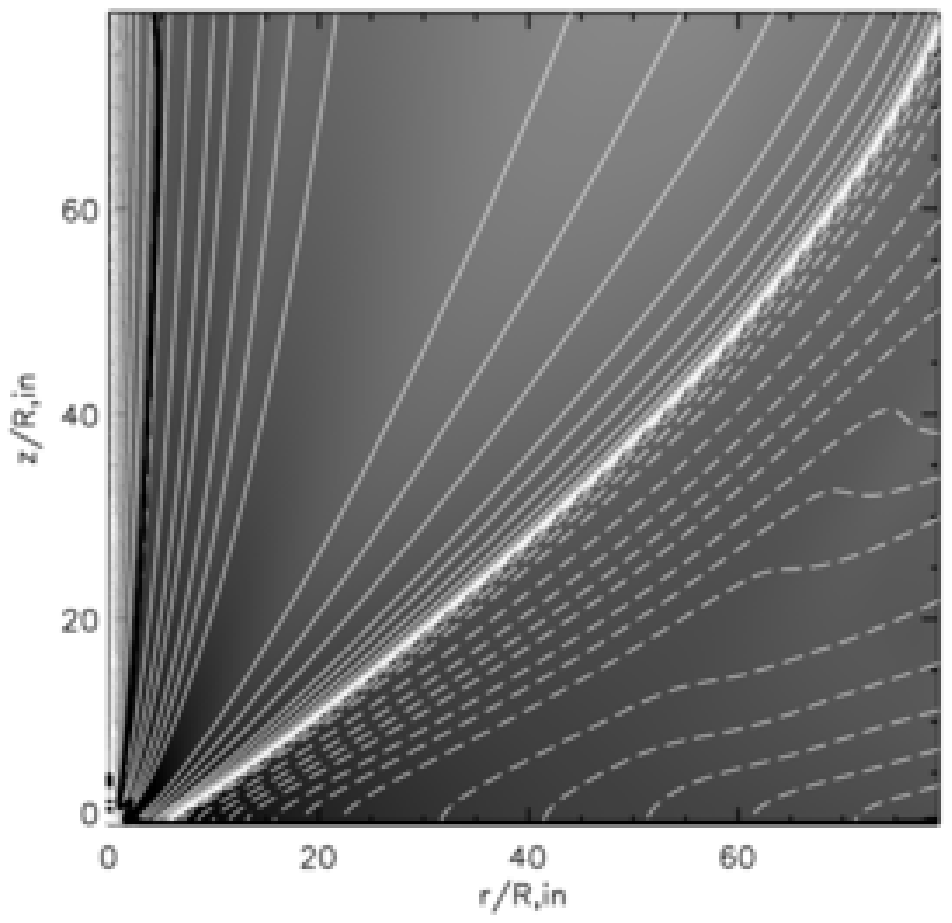}
\includegraphics[width=7cm]{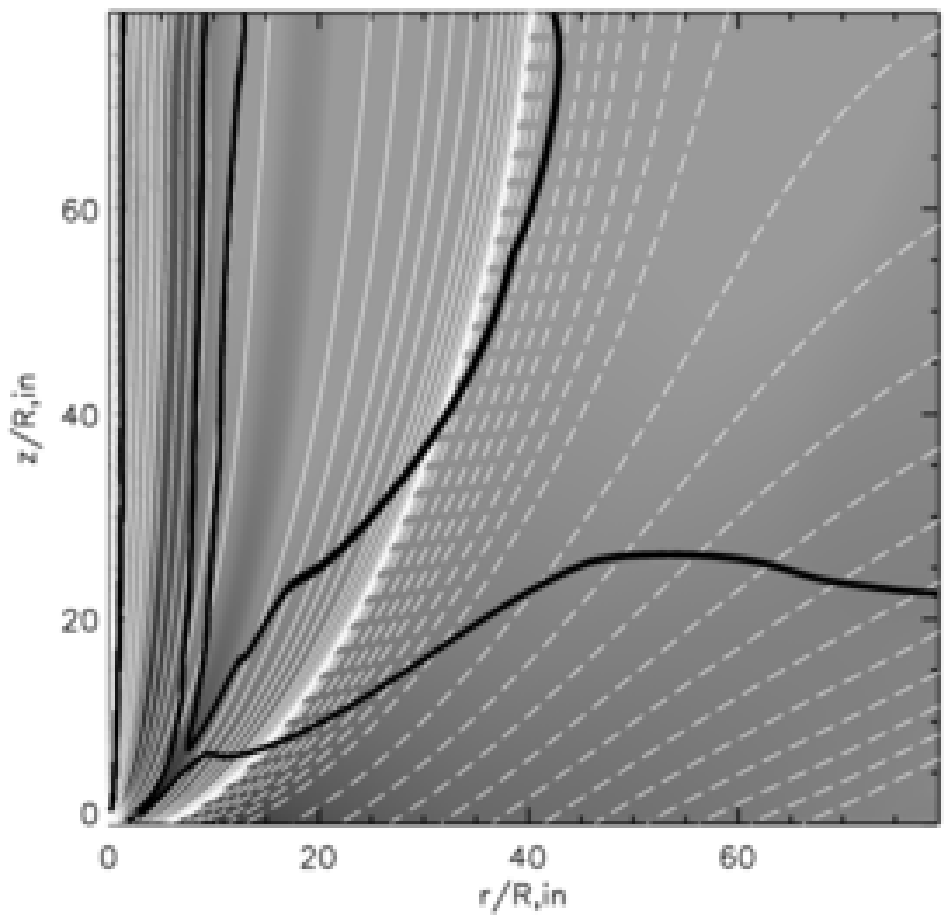}

\caption{Poloidal magnetic field distribution at the time when the simulation
         was stopped.
        Simulation 
A3 (t=500); A4a (t=3200); A10 (t=550); A13 (t=620); A12 (t=380); A14 (t=660) 
    (from top left to bottom right).
    Contour levels and grey scale as in Fig.~\ref{fig_evol1}.
\label{fig_final2}
}
\end{figure*}

This behavior is somewhat mirrored in the poloidal velocity profile. 
What is interesting for shock formation in the asymptotic jet, is a temporal change in 
jet velocity at certain jet radii (see below).
The maximum velocities in the outflow reach about two times the Keplerian speed
at the inner disk radius and vary by a factor of two.
Along the neutral field line the outflow velocity is fast-magnetosonic
(see Figs.~\ref{fig_small1} and \ref{fig_final2} where the fast surface is 
indicated).

As an estimate for the reconnection time scale we may apply the Sweet-Parker
approach with $\tau_{\rm SP} \simeq \sqrt{\tau_{\rm A} \tau_{\rm diff}}$ with the
dynamical (Alfv\'en) time scale $\tau_{\rm A} = l/v_{\rm A}$ and the diffusive
time scale $\tau_{\rm diff} = l^2 / \eta$.
With our model of turbulent magnetic diffusivity $\eta = \eta_0\rho^{1/3}$ we
find $\tau_{\rm SP} \simeq l^{3/2} (4\pi)^{1/4} \eta_0^{-1/2} \rho^{1/12}$.
For $\eta_0 = 0.01$, and $\rho \simeq 0.02$ at the reconnection 
area with $l\simeq 2$ (for simulation A16) that $\tau_{\rm SP} \simeq 30$. 
This is similar to the duration of the reconnection flare we observe between
$t=600$ and $t=650$.

\begin{figure*}
\centering

\includegraphics[width=7cm]{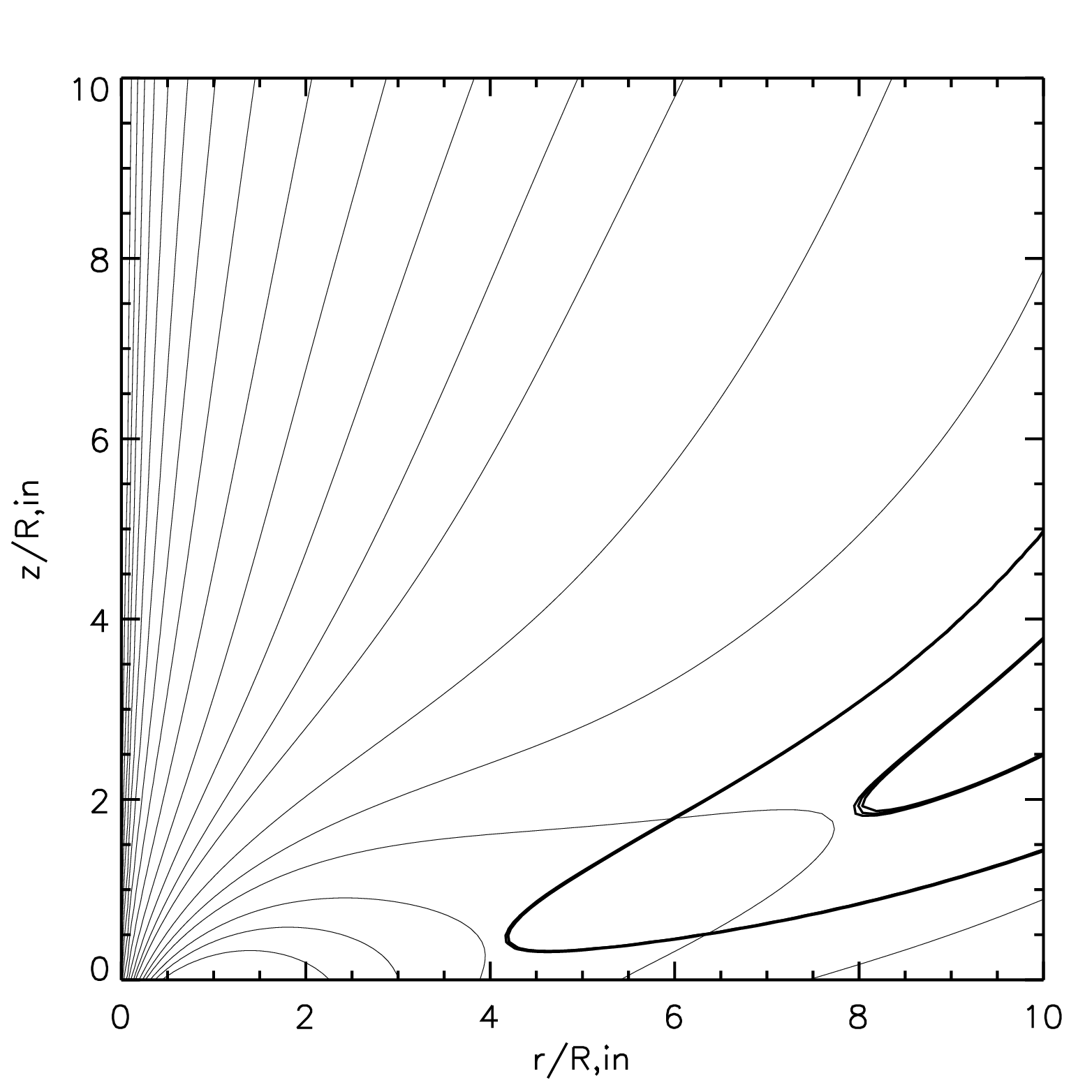}
\includegraphics[width=7cm]{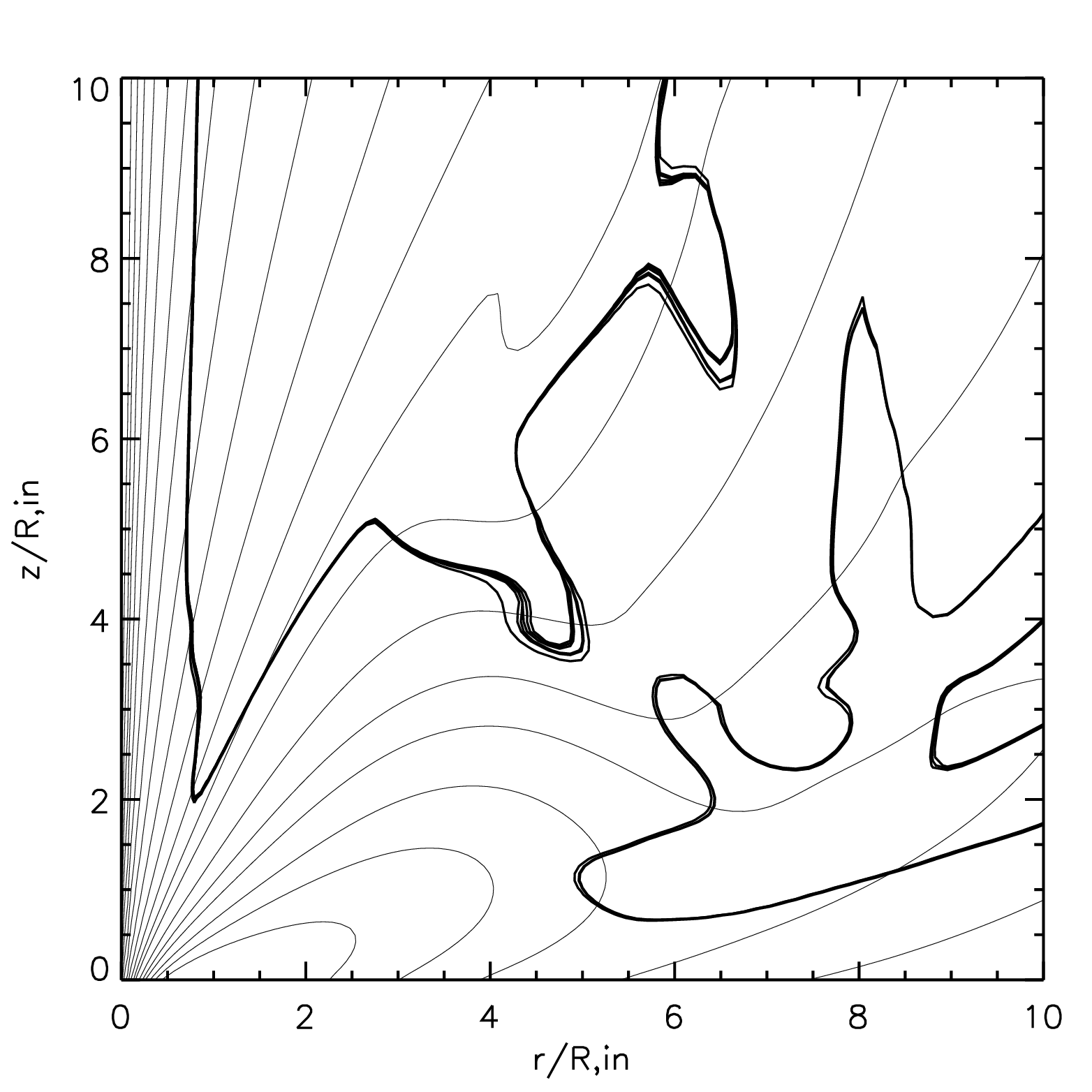}

\includegraphics[width=7cm]{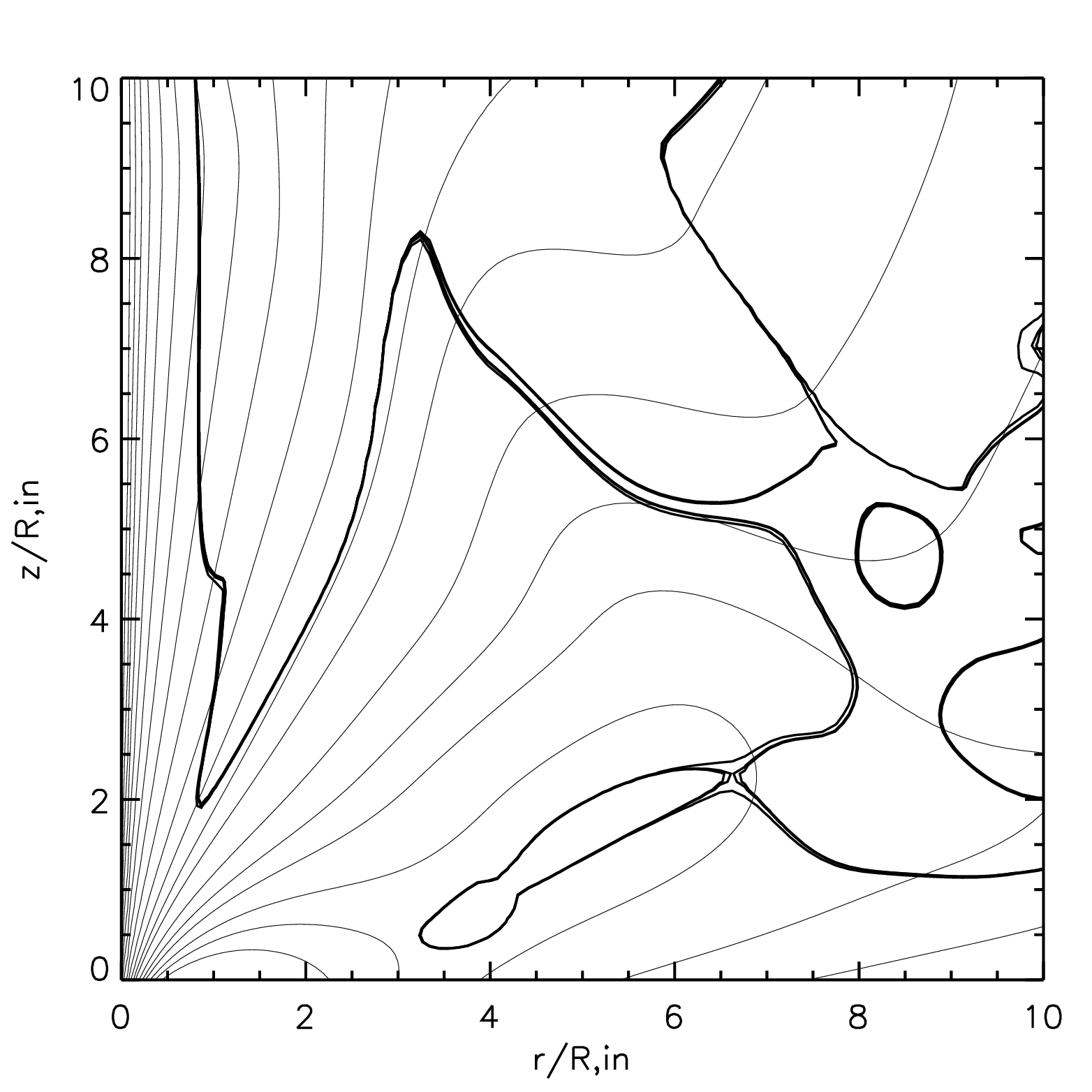}
\includegraphics[width=7cm]{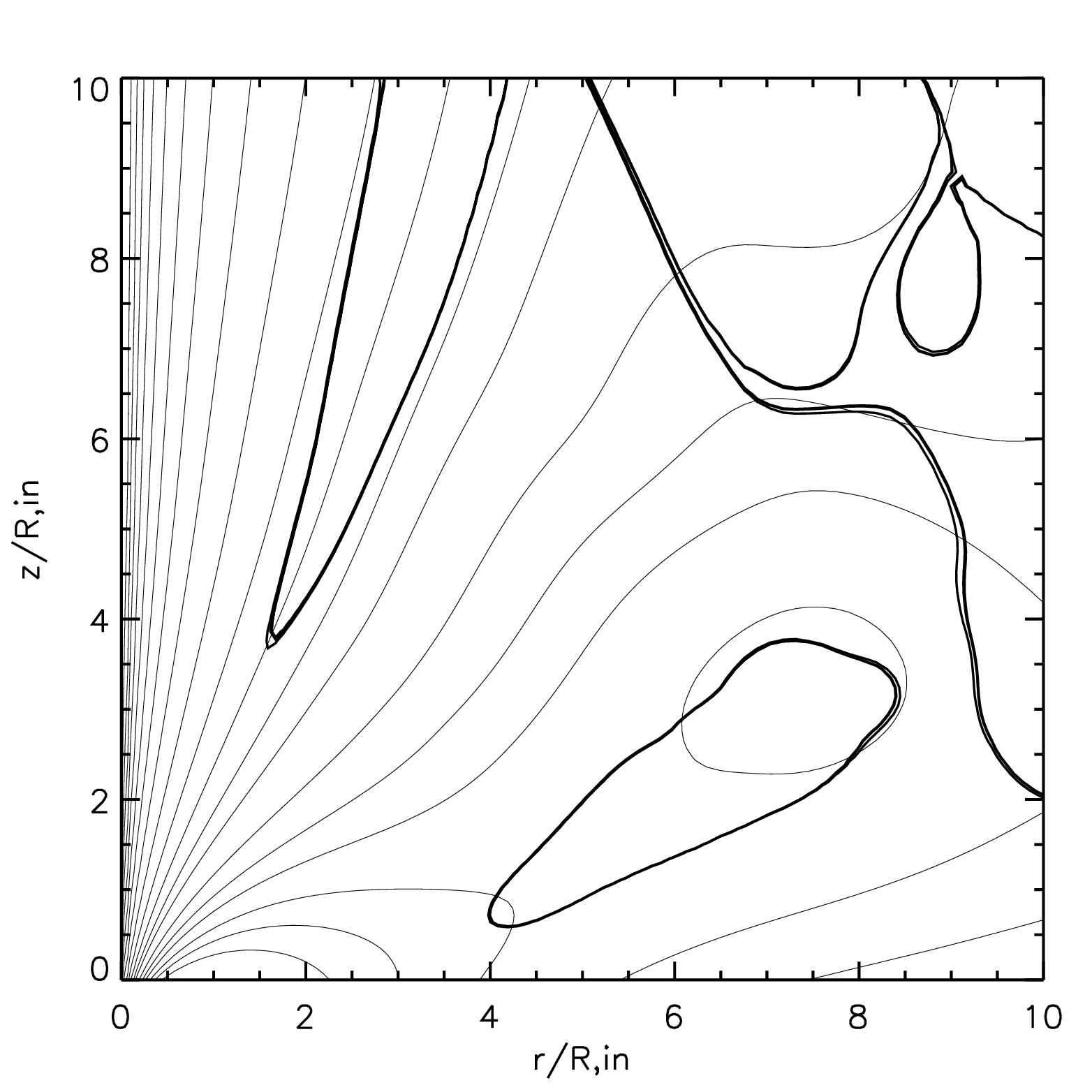}

\includegraphics[width=7cm]{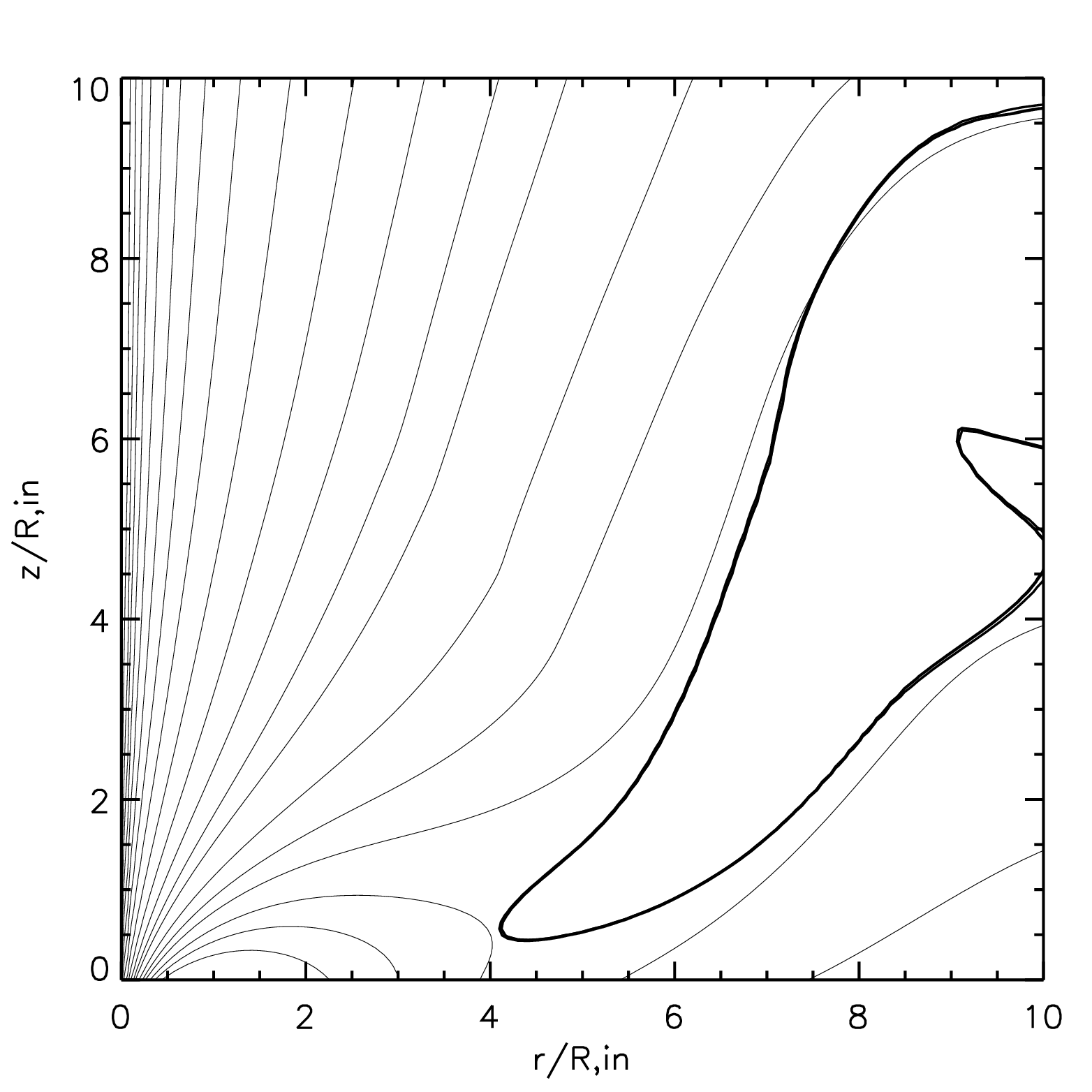}
\includegraphics[width=7cm]{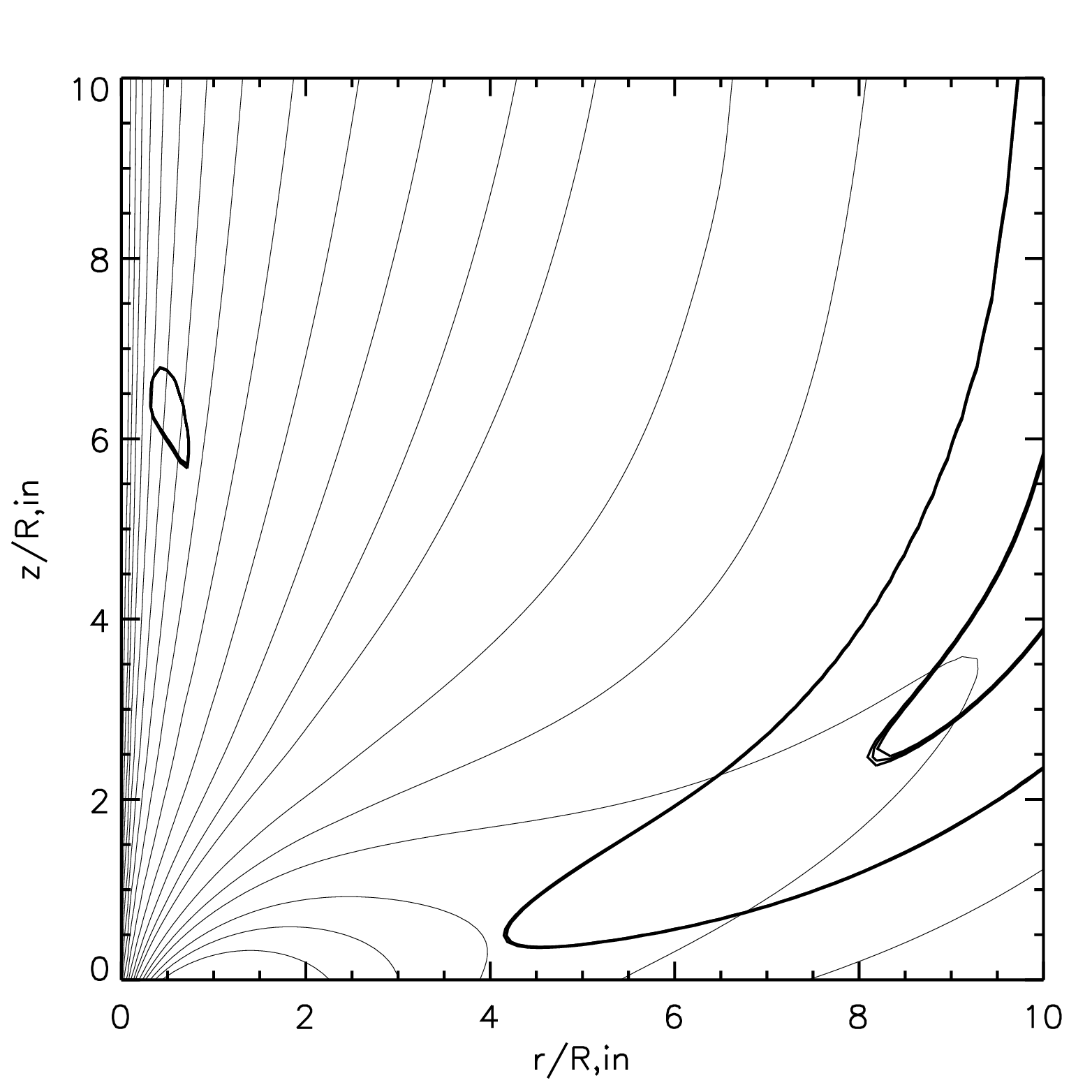}

\caption{Poloidal magnetic field distribution ({\em thin lines})
         during flare evolution (run A16).
    Contour levels as in Fig.~\ref{fig_evol1}.
        Simulation time steps
    t=  600, 620, 630, 640, 670, 760.
    (from top left to bottom right). Alfv\'en and fast surface are
    shown by {\em thick lines}. 
    Note the super-Alfvenic /super-fast stellar wind.
\label{fig_small2}
}
\end{figure*}

Note an interesting feature in both of the general model setups (aligned and 
anti-aligned field geometry) concerning the reconnection geometry.
Figures \ref{fig_small1} and \ref{fig_small2} show the inner structure of the
star-disk magnetosphere. 
These simulations (A04a, A16) were launched from differently aligned field
geometries (aligned/anti-aligned disk-stellar field).
In the anti-aligned case the initial X-point is along the axis.
However, as the magnetic field evolves, the dipole expands and sweeps off
the initial field along the axis. 
A new X-point evolves by disruption of some of the closed dipolar field loops.
The new X-point is located above the disk close to the inner disk radius (see above).
This situation is not very different from the the aligned case for which the 
initial X-point, located along the equatorial plane, moves up to a certain 
(small) height above the disk.
In summary, our simulations show that no magnetic X-point remains along the 
equatorial plane as e.g. assumed in the Shu et al. X-wind model. 
Instead we find from both initial configurations an X-point above the disk
(sometimes also called Y-point, \citep{ferr06}), 
located at time-averaged distances 
($z\simeq4, r\simeq2$) or ($z\simeq8, r\simeq2$), respectively.

\subsection{Flaring events as jet knot generator?}
The generation of knots in protostellar jets is a long-standing puzzle. 
In particular, it is unclear whether the shocks/knots are triggered by 
the internal engine or due to interaction with the ambient medium.
A strong argument for the first possibility is the existence of some
perfectly symmetric jets as HH\,212 \citep{zinn98},
however, the majority of jet sources looks asymmetric.
Internal shocks along the jet flow may be caused e.g. by a time-dependent 
velocity variation of the material injected into the jet (e.g.~\citet{raga07}).
Even if we have observational support for the idea of the jet knots 
being triggered by the central engine, we do not know {\em how} the 
central engine does that.
The time scale derived from typical knot separations $d_{\rm knot}$
and velocities $v_{\rm knot}$ correspond to
\begin{equation}
t_{\rm knot} = 10\,{\rm yrs} \left(\frac{v_{\rm knot}}{300\,{\rm km/s}}\right)
                            \left(\frac{d_{\rm knot}}{100\,{\rm AU}}\right)
\end{equation}
which is clearly different from the Keplerian time scale close to the inner
disk edge from where the jets are launched.

Considering the ejection of large-scale flares and the follow-up 
re-arrangement of outflow density and velocity distribution in our 
simulations,
one is tempted to hypothesize that the origin of knots is triggered by such
flaring events.

In our simulations (A4a, A16) the time scale of flare generation is about 
500-1000 rotational periods of the inner disk, corresponding 
to about 30-60 years (assuming an inner disk rotation period of, say two 
times the co-rotation period of a typical T\,Tauri star of about 10 days).
The variation in the hydrodynamic parameters lasts for about 30-40 
inner disk rotation times 
(respectively about $<400$ days for a 10\,day stellar rotation period).
The further evolution and generation of further flaring events is of
course not known as it is beyond our computation time.
The essential point, however, is that we detect a {\em long} time scale,
substantially longer than the typical dynamical time scales at the jet
formation area.
That time scale is surprisingly similar to the time scale defined by the 
observed knot separation and velocity.
Of course, it is to early to draw firm conclusions from 
such a tentative agreement.
The reconnection time-scale is governed by the magnetic diffusivity
for which we have applied our self-consistent model of turbulent 
magnetic diffusivity \citep{fend02}.

It is left to further simulations of the asymptotic collimated 
jet beam to check whether the detected variation in outflow speed
and mass flux 
(Figs.~\ref{fig_mflux_i},\ref{fig_mflux_mv},\ref{fig_mflux_v})
is sufficient in order to generate strong internal shocks comparable
to the observed knots.

\subsection{Collimation degree, mass loss rate and field alignment}
In this section we compare the overall collimation behavior of
the different simulation runs.
In previous studies \citep{fend02, fend06} we quantified the degree of 
outflow collimation $\zeta$ by comparing the mass flow rates in axial and 
lateral direction,
\begin{equation}
\zeta \equiv \frac{\dot{M}_{\rm z}}{\dot{M}_{\rm r}}
= {
{  2\pi \int_{0}^{r_{\rm max}} r \rho v_z dr }
 \over
{ 2\pi r_{\rm max} \int_{0}^{z_{\rm max}} \rho v_r dz }.
}
\label{eqn_zeta}
\end{equation}
In Tab.~\ref{tab_all} we provide mass flow rate within four differently 
sized volumes, $\dot{M}_{{\rm z},i}, \dot{M}_{{\rm r},i}$,
considering different sub-grids of the whole computational domain,
i.e.cylinders of radius and height
$(r_{\rm max} \times z_{\rm max}) =
(12\!\times\!12), (24\!\times\!24), (43\!\times\!43)$, and $(76\!\times\!76)$.
We calculate the mass fluxes {\em ratio} normalized to the size of the
area threaded by the mass flows in r- and z-direction,
$\hat{\zeta}_1, \hat{\zeta}_2, \hat{\zeta}_3, \hat{\zeta}_4$.
The ratio of grid extension $(r_{\rm max}/ z_{\rm max})$ as displayed
in our figures converts into a ratio of cylinder surface areas of
$ A_{\rm z}/A_{\rm r} \sim 0.5 (r_{\rm max}/ z_{\rm max})$.
As a general measure of collimation degree, we give an average value
$< \zeta >$ which may consider also the dynamical state of the simulation
run.

Compared to our previous work on pure disk wind collimation, 
where the relative mass flux of the collimating disk wind is a sufficient
quantitative measure of collimation the situation in the present setup 
is not exactly the same.
Naturally, the stellar wind mass flux emerges close to the outflow
axis and will tend to stay close to it.
Thus, a high stellar mass flux will naturally cause a more collimated
mass flow.

In order to compare the collimation degree within the large volume, 
it is essential to check the evolutionary state of the flow.
In all of our simulation runs the initial corona has completely swept out 
of the grid. In this case, all of the mass flux we measure has
been launched by the disk wind.
Exceptions are simulations A10, A13, and A12, where a relict of the 
initial bow shock are still visible in the outer 
layers. The mass loss rates in these examples have to taken with
care, at least the values from the larger volumes.
Run A16 is evolving very slowly in the outer part and it was not possible
within reasonable CPU time to evolve the disk wind to larger distances
from the disk surface.

\subsubsection{Collimation along the outflow}
The degree of collimation derived from the simulations is in general
different for the the different volumes,
$\hat{\zeta}_1, \hat{\zeta}_2, \hat{\zeta}_3, \hat{\zeta}_3$.

In most cases, the degree of mass flux collimation increases
along the flow 
which is exactly the signature of MHD self-collimation.
The outflow starts as un-collimated disk/stellar wind, and reaches the
outer grid boundaries as collimated beam.

A good example is simulation A7, where the collimation of mass flow
changes from ration $\sim 0.3$ for the inner region to $\sim 7$ in
the outer parts.
The same arguments hold for simulations A8, A15, A14.
A similar behavior is seen e.g. in simulations A10, A13, A14, however,
the collimation in mass flux around the largest volume is low.
In these cases the outermost flow structure has not yet evolved into
a steady state and either parts of the initial bow shock or the
initial corona are still in the computational domain.

\begin{figure}
\centering

\includegraphics[width=8cm]{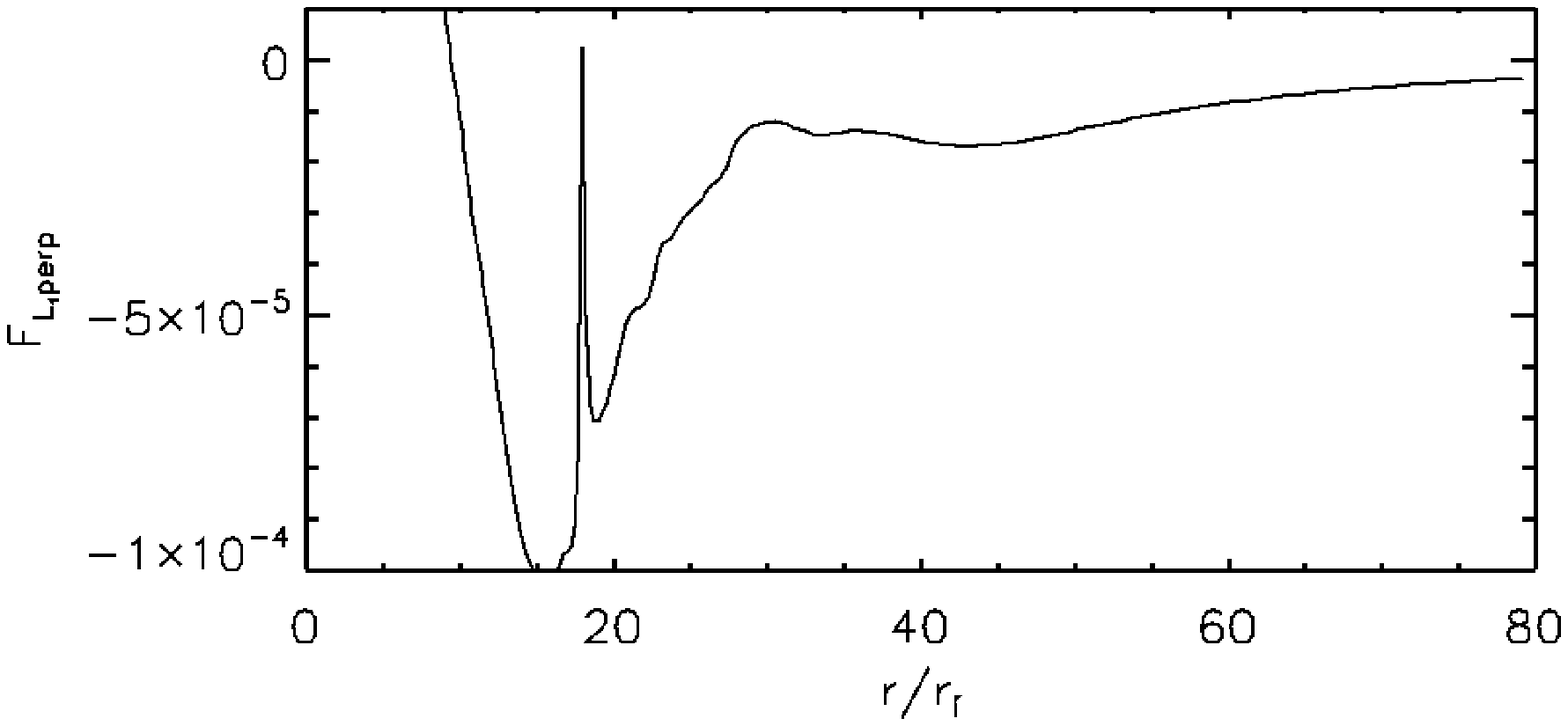}
\includegraphics[width=8cm]{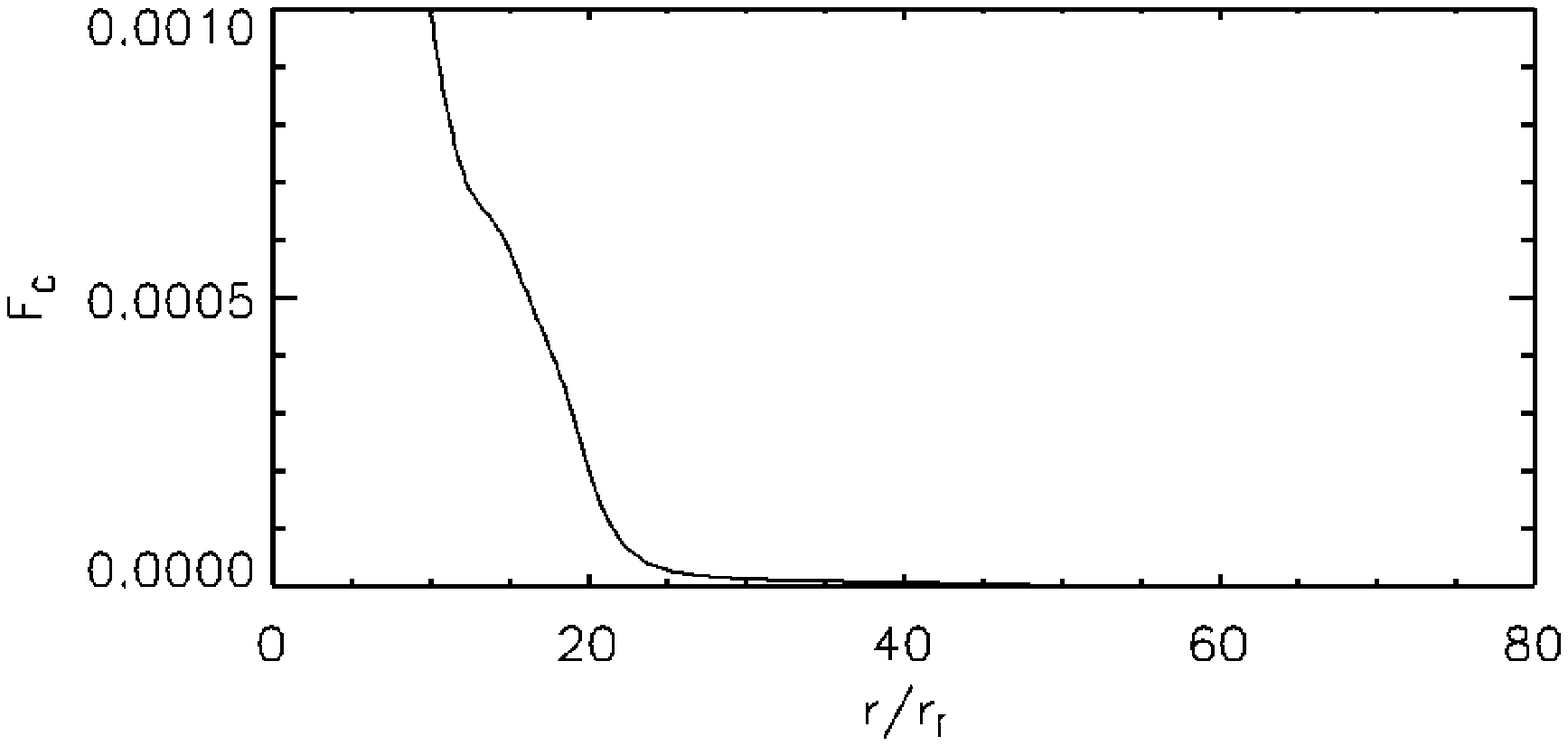}

\caption{Perpendicular Lorentz force (top) and centrifugal force (bottom)
across simulation A12 at height $z=50$ at time $t=380$.
\label{fig_lorentz}
}
\end{figure}

A counter example is simulation A12, which stays un-collimated in
mass flux, although the magnetic field distribution looks collimated.
Simulation A12 is exceptional for its high mass flux launched from 
the disk surface. Thus, the disk wind starts super-Alfv\'enic and 
super-seed the magnetosonic speed quickly.
The standard Blandford-Payne magneto-centrifugal acceleration is
not very efficient in this case, which also results in only little
induction of toroidal magnetic field and, thus, Lorentz force, see
Fig,\ref{fig_lorentz}. The radial component of the perpendicular 
Lorentz force component is negative, i.e. directed inwards, however,
too weak to balance the strong inner centrifugal force, resulting in
a weak collimation of the outer part.

Table \ref{tab_all} shows also the time evolution of the collimation 
degree.
Example A15 demonstrates how the collimation degree progresses
in time. From earlier times (t=550) to later times ($t=2100$)
collimation increases as the outflow evolves into a new dynamical
state over the whole numerical grid. The same hold for example A14
and others. Examples A7, A8, A9 are calculated for a high stellar wind
mass load, thus the time-evolution of the outer disk wind does
not play a big role concerning collimation. This outflows reach
a high degree of collimation early in time.

\subsubsection{Collimation and mass loss rates}
The mass loss rates from star and disk are prescribed as boundary condition.
A higher mass load in the stellar wind component would naturally
result in a higher degree of collimation in the mass flux as this mass
is just launched closer to the outflow axis.

\begin{figure*}
\centering

\includegraphics[width=7cm]{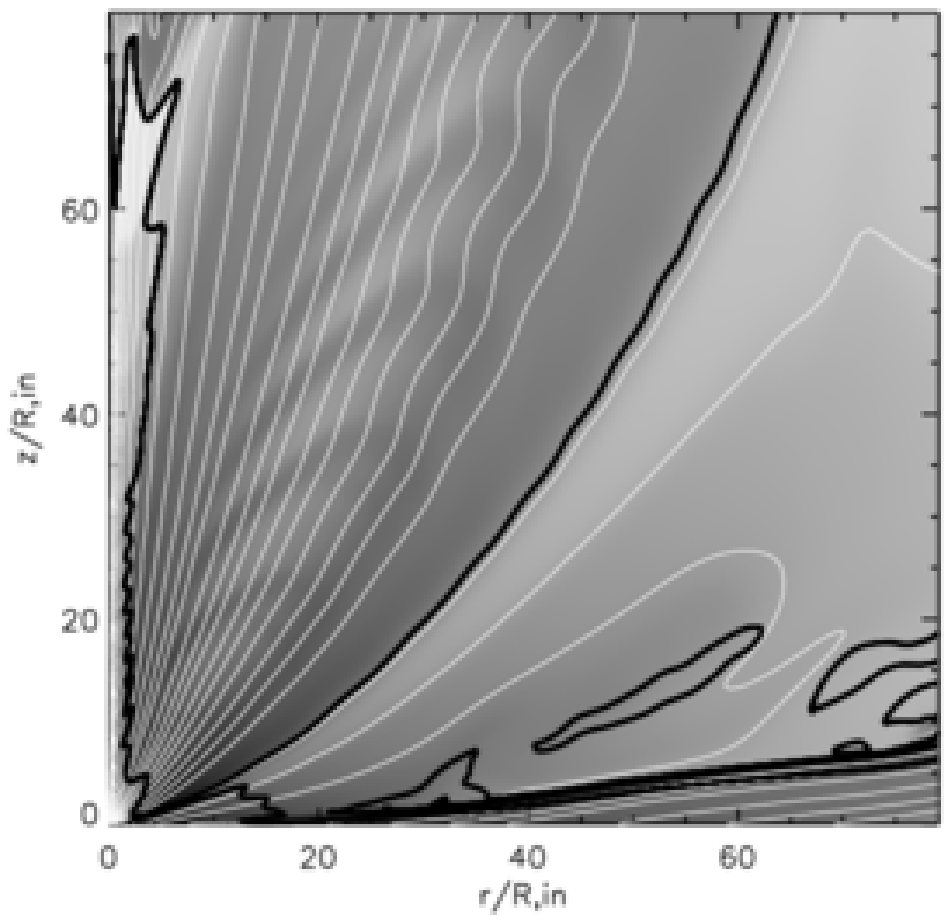}
\includegraphics[width=7cm]{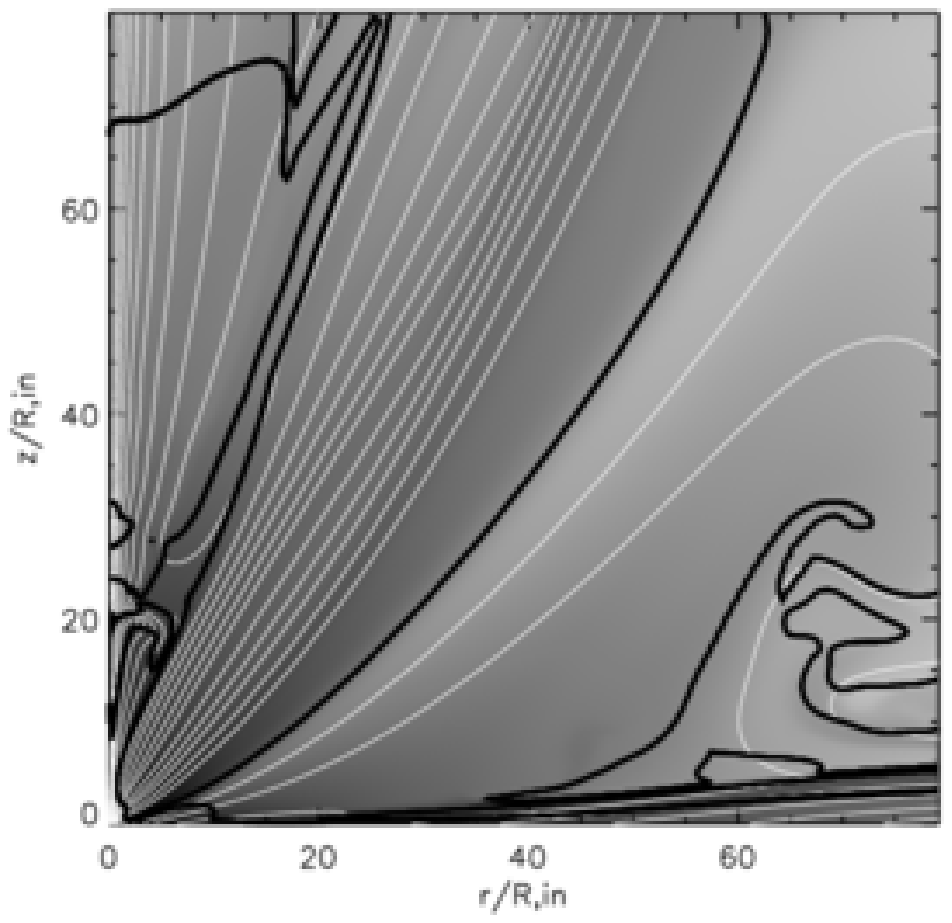}

\includegraphics[width=7cm]{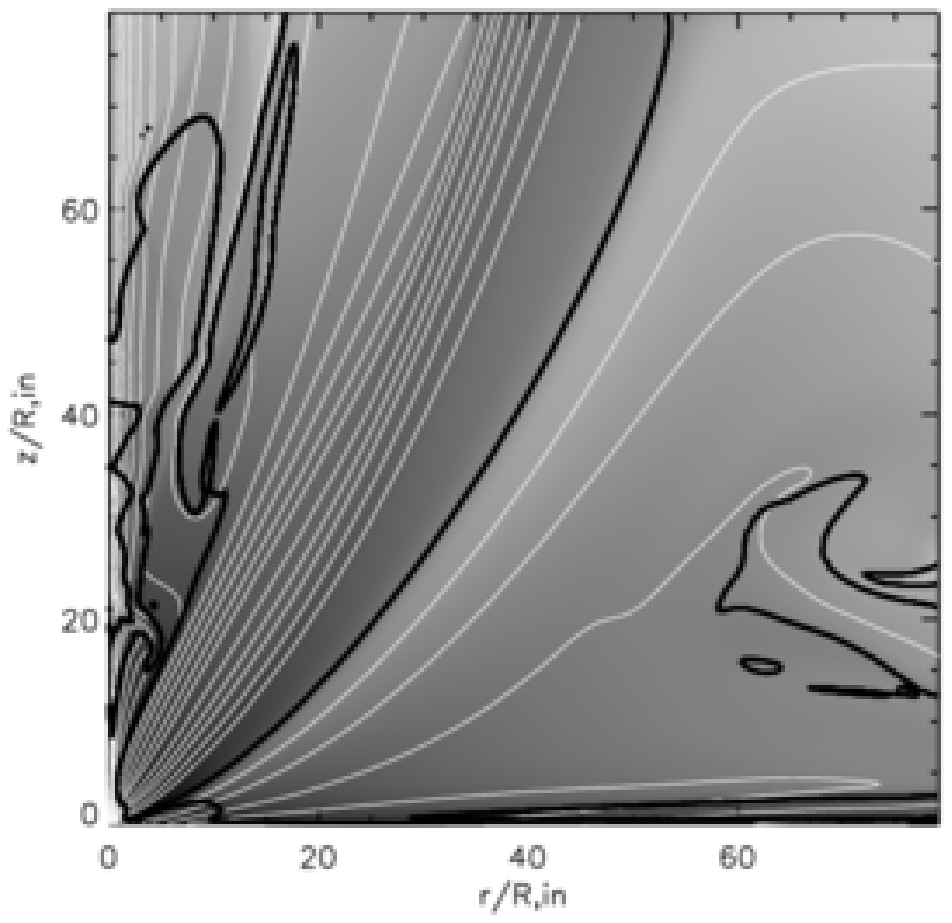}
\includegraphics[width=7cm]{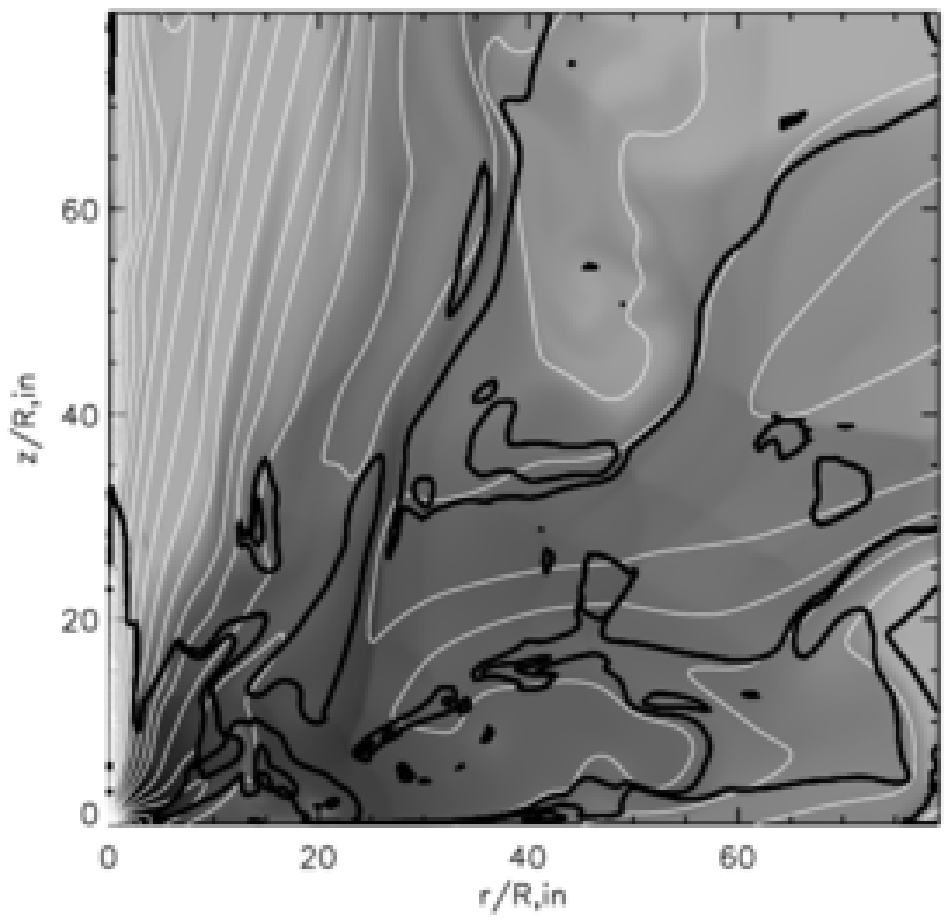}

\includegraphics[width=7cm]{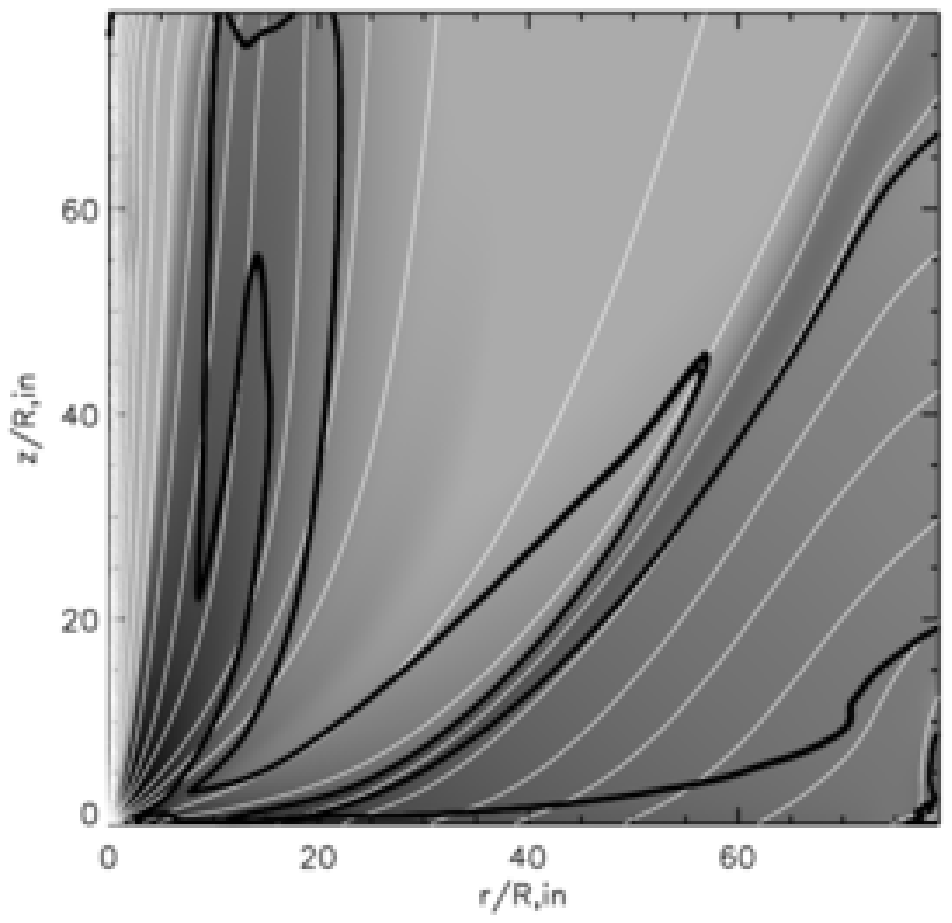}
\includegraphics[width=7cm]{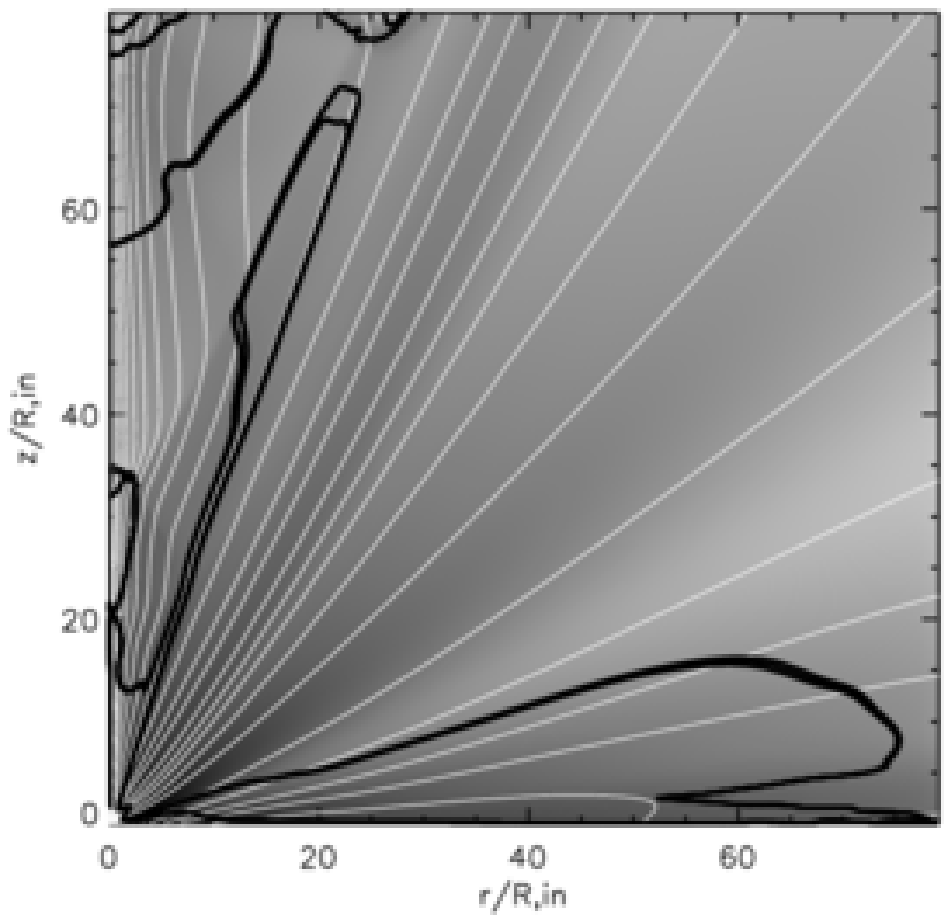}

\caption{Poloidal magnetic field lines at the time when the simulation
         was stopped.
        Simulation 
A9 (t=740); A8 (t=1000); A7 (t=1400); A16 (t=1350) A15 (t=2100); A2 (t=700)
    (from top left to bottom right).
Thick and medium thick lines indicate the location of the Alfv\'en and the 
magnetosonic surface.
    Contour levels and grey scale as in Fig.~\ref{fig_evol1}.
\label{fig_final1}
}
\end{figure*}

Simulations A2 and A3 are good examples. The stellar wind mass loss is 
2-3 times the disk wind mass flux. The resulting mass flux ratio is
similar, thus this outflow is rather well collimated hydrodynamically.
However, in spite the quite good collimation in mass flux, 
the magnetic field structure is not collimated with almost 
(spherical) radially expanding poloidal field lines 
(see Fig.~\ref{fig_final1}).
The field structure (and thus the poloidal velocity field) has a conical 
shape and is thus un-collimated.
Similar as for simulation A12 (see above) the disk wind is too weak in order 
to collimate the stellar wind to a high degree.

Simulations A12 and A13 have the same star-disk magnetic field profile,
however with a different mass load by a factor of ten. Run A12 reaches the
same dynamical evolutionary state as A13, but earlier at $t=300$ instead of
$t=620$. The less magnetized outflow A13 is stronger collimated.
Run A14 has the same mass load as A13, but the double magnetic flux
resulting in similar collimation degree.

Simulations A10 and A13 differ only by the stellar wind mass load
(factor two), however this mass load is low and does not result
in a variation of the overall degree of collimation.
Note that for both simulations, the outer area of the outflow has
not yet evolved in a steady state and a relict from the initial 
bow shock are still visible.

\subsubsection{Collimation and magnetic field alignment}
\label{sect_final}
From the technical point, the simulation of the case where magnetic dipole
and disk magnetic field are anti-aligned (A3, A4, A10, A13, A12, A14)
are computationally less expensive and were therefore able to evolve
to considerably longer time.

For the aligned cases (A9, A8, A7, A16, A15, A2) only the less magnetized 
evolve into a quasi stationary state in the outer (disk) outflow in reasonable 
time.
We mention in particular run A15 with a reasonably strong disk magnetic flux
which enables a collimating disk wind (compare to A16 and A7). 

Simulations A9, A8 and A7 demonstrate that a low mass flux disk wind mass flux 
cannot evolve into a collimated disk wind even if the magnetic flux is
relatively high if it is dominated by a strong central stellar jet.

The most promising model setup in order to explain strong stellar jets from
a star-disk magnetosphere are those with relatively strong disk wind and
disk magnetic flux.
The stellar wind dominated simulations may give a high degree of collimation,
however they collimate to too small radii.
Stellar magnetic flux dominated simulations tend to stay un-collimated.

%
\begin{table*}
\scriptsize
\begin{center}
\caption{Summary of simulation runs ID.
Shown are disk and stellar mass loss rates, 
${\dot{M}_{\rm d}}, {\dot{M}_{\star}}$, 
corresponding magnetic fluxes, $\Psi_{0,\rm d}, \Psi_{0,\star}$,
and the physical time step $\tau$, when the collimation degree
has been calculated.
Mass loss rates $\dot{M}_{{\rm z}i}, \dot{M}_{{\rm r}i}$ in 
$z$- and $r$-direction are integrated along three sub-grids 
$(r_{i,{\rm max}} \times z_{i,{\rm max}}) =
(11.6 \times 11.6),(23.6 \times 23.6), (43.4 \times 43.4), (76.2 \times 76.2),$
(for the $(80.0 \times 80.0)$ grid).
The average degree of collimation $<\!\!\zeta\!\!>$ is defined by the 
relative mass fluxes in $z$ and $r$-direction, normalized by the area threaded.
Simulations labeled with a '*' were run on a smaller, higher resolution 
$(40.0 \times 40.0)$ grid.
\label{tab_all}
}
\begin{tabular}{clcccllllllll}
\tableline\tableline
\noalign{\smallskip} ID & ${\dot{M}_{\rm D}}\over{\dot{M}_{\star}}$
 & $\Psi_{0,disk}$ & $\Psi_{0,\star}$ & $\tau $ &
 $\dot{M}_{z1},\dot{M}_{r1}$ & $\dot{M}_{z2},\dot{M}_{r2}$ & 
 $\dot{M}_{z3},\dot{M}_{r3}$ & $\dot{M}_{z4},\dot{M}_{r4}$ &
 $\hat{\zeta}_1, \hat{\zeta}_2, \hat{\zeta}_3, \hat{\zeta}_4 $ & 
 $<\!\!\zeta\!\!>$ \\
\noalign{\smallskip}
\tableline
\noalign{\medskip}

A9  & ${1.55}\over{5.62}$ & 0.1 & 10.0 & 740 &
  2.37, 5.82 & 4.72, 3.87 & 6.42, 1.64 & 7.44, 1.09 & 0.41, 1.2, 3.9, 6.8 &  \\
     &  &     &      & 700 &
  2.90, 5.99 & 4.73, 4.10 & 6.18, 1.87 & 7.23, 1.11 & 0.48, 1.2, 3.3, 6.5 & 6.7 \\
     &  &     &      & 600 &
  2.93, 5.12 & 5.87, 3.23 & 6.99, 1.48 & 7.06, 1.23 & 0.57, 1.8, 4.7, 5.7 &  \\
\noalign{\medskip}

A8  & ${1.57}\over{5.62}$ & 0.04 & 5.0 & 1010 &
  2.67, 5.40 & 4.64, 3.76 & 6.95, 1.49 & 7.41, 1.18 & 0.49, 1.2, 4.7, 6.3 &   \\
     &  &  & &  900 &
  2.68, 5.38 & 4.72, 3.66 & 6.97, 1.46 & 7.43, 1.11 & 0.50, 1.3, 4.8, 6.7 & 6.5 \\
     &  &  & &  800 &
  2.66, 5.39 & 4.58, 3.77 & 6.90, 1.45 & 7.48, 1.08 & 0.49, 1.2, 4.8, 6.9 &  \\
\noalign{\medskip}

A7  & ${1.65}\over{5.61}$ & 0.02 & 5.0 & 1450 &
  1.88, 6.37 & 2.28, 5.51 & 5.20, 1.31 & 6.77, 1.48 & 0.30, 0.41, 4.0, 4.6 &      \\
     &  &  & & 1350 &
  3.02, 5.26 & 5.99, 2.46 & 7.41, 1.07 & 7.51, 1.05 & 0.57, 2.4, 6.9, 7.2 & 7.0  \\
     &  &  & & 700 &
  2.80, 5.44 & 4.60, 3.86 & 6.92, 1.51 & 7.42, 1.02 & 0.52, 1.2, 4.6, 7.3 &   \\
\noalign{\smallskip}

A16  & ${1.65}\over{0.28}$ & 0.02 & 5.0 & 1300 &
  0.23, 0.25 & 0.52, -0.02 & 0.63, 0.30 & 0.81, 0.11 &     &      \\
     &  &      &     &  1200 &
  0.19, 0.27 & 0.55, 0.04 & 0.79, -0.14 & 0.38, 0.13 &     &      \\
     &  &      &     &  1000 &
  0.37, 0.02 & 0.52, -0.5 & 0.49, -0.54 & 0.59, -0.20 &     &      \\

\noalign{\smallskip}
A15  & ${1.51}\over{0.28}$ & 0.04 & 3.0 & 2100 &
   0.22, 0.83 & 0.40, 0.91 & 0.71, 0.93 & 1.40, 0.76 &   0.27, 0.44, 0.76, 1.8   &   \\
     &  &      &     &  1900 &
   0.23, 0.82 & 0.41, 0.90 & 0.72, 0.90 & 1.41, 0.87 & 0.28, 0.46, 0.8, 1.6  & 2.0  \\
     &  &      &     &  550 &
   0.28, 0.80 & 0.65, 0.72 & 1.56, 0.95 & 1.04, 2.39 &  0.35, 0.9, 1.6, 0.44  &      \\

\noalign{\smallskip}
A2  & ${1.73}\over{5.61}$ & 0.01  & 5.0 & 700 &
  2.34, 5.93 & 3.21, 5.34 & 4.04, 4.79 & 4.95, 4.30 & 0.40, 0.60, 0.84, 1.2 & 1.2 \\
     &  &      &     &  650 &
  2.35, 5.92 & 3.24, 5.33 & 4.10, 4.75 & 5.14, 4.20 & 0.40, 0.61, 0.86, 1.2 &   \\
\noalign{\medskip}
\tableline
\noalign{\medskip}

A3  & ${2.2}\over{5.6}$  & -0.01 & 5.0 & 500 &
   2.60, 5.73 & 3.59, 5.25 & 5.05, 4.33 & 6.45, 3.25 & 0.45, 0.68, 1.2, 2.0 & \\
     &  &      &     &  480 &
   2.59, 5.73 & 3.61, 5.24 & 5.10, 4.32 & 6.50, 3.23 & 0.45, 0.69, 1.2, 2.0 & 2.0 \\
     &  &      &     &  450 &
   2.58, 5.73 & 3.61, 5.23 & 5.18, 4.12 & 6.63, 3.05 & 0.45, 0.69, 1.3, 2.2 & \\
\noalign{\medskip}
A4a & ${2.08}\over{5.56}$   & -0.1  & 3.0 & 3600 &
  6.64, 1.75 & 7.54, 1.07 & 8.14, 0.79 & 8.41, 0.54 & 3.8, 7.1, 10.3, 15.6 &  \\
     &  &      &     & 3300 &
  6.31, 2.12 & 7.46, 1.07 & 7.94, 0.69 & 8.14, 0.60 & 3.0, 7.0, 11.5, 13.6 & 12 \\
     &  &      &     & 2000 &
  6.85, 1.46 & 7.33, 1.18 & 7.72, 1.10 & 8.39, 0.35 & 4.7, 6.2, 7.0, 24.0  &  \\
     &  &       &    &   80 &
  5.71, 2.86 &  8.04, 2.83 & 15.0, 9.92 & .0004, .02 & 2.0, 2.8, 1.5, -    & \\
\noalign{\medskip}
A4b$^{\star}$ & ${1.77}\over{5.40}$  & -0.1 & 3.0 & 80 &
  5.96,  1.34 & 7.17,  1.77 & 16.5, 12.7 & - , - & 4.4, 4.1 , 0.73, -      &     \\
\noalign{\medskip}

A10  & $ {2.08}\over{0.56} $ & -0.1 & 3.0 & 550 &
  1.11, 0.80 & 1.57, 0.70 & 2.08, 0.78 & 3.10, 3.16 & 1.4, 2.2, 2.7, 0.98  &  \\
     &  &     &      & 500 &
  1.11, 0.80 & 1.59, 0.69 & 2.11, 0.81 & 3.17, 3.96 & 1.4, 2.3, 2.6, 0.80  & 3.0 \\
     &  &     &      & 400 &
  1.13, 0.79 & 1.64, 0.66 & 2.26, 0.92 & 3.63, 7.33 & 1.5, 2.5, 2.5, 0.50  &  \\
\noalign{\medskip}
A13  & $ {2.08}\over{0.28} $ & -0.1 & 3.0 & 620 &
  0.81, 0.74 & 1.29, 0.64 & 1.76, 0.69 & 2.81, 2.37 & 1.10, 2.02, 2.55, 1.19 &  \\
     &  &     &      & 600 &
  0.82, 0.74 & 1.30, 0.63 & 1.78, 0.70 & 2.78, 2.54 & 1.11, 2.06, 2.54, 1.10 & 2.5 \\
     &  &     &      & 350 &
  0.84, 0.71 & 1.41, 0.58 & 2.09, 1.11 & 3.92, 9.19 & 1.18, 2.43, 1.88, 0.43 &  \\
\noalign{\medskip}
A12  & $ {20.8}\over{2.78} $ & -0.1 & 3.0 & 380 &
  6.25, 9.10 & 7.66, 11.1 & 8.86, 12.9 & 11.1, 15.0 & 0.69, 0.69, 0.69, 0.74 &            \\
     &  &     &      & 350 &
  6.26, 9.10 & 7.66, 11.1 & 8.87, 12.9 & 11.8, 15.4 & 0.69, 0.69, 0.69, 0.77  & 0.8  \\ 
     &  &     &      & 300 &
  6.26, 9.10 & 7.66, 11.1 & 8.91, 13.1 & 13.5, 14.9 & 0.69, 0.69, 0.68, 0.91 &  \\
\noalign{\medskip}
A14  & $ {2.08}\over{0.28} $ & -0.2 & 6.0 & 660 &
  0.79, 0.81 & 1.15, 0.81 & 1.60, 0.77 & 2.31, 1.72 & 0.98, 1.4, 2.1, 1.34   &  \\
     &  &     &      & 500 &
  0.75, 0.80 & 1.16, 0.76 & 1.65, 0.72 & 2.32, 2.51 &  0.94, 1.53, 2.3, 0.9   & 2.0  \\
     &  &     &      & 350 &
  0.78, 0.78 & 1.28, 0.71 & 1.74, 1.00 & 3.62, 7.14 & 1.0, 1.8, 1.7, 0.5      &  \\
\noalign{\medskip}
A$6^{\star}$  & ${1.77}\over{5.40}$ & -0.2 & 6.0 & 80 &
  5.86, 1.46 & 6.83, 1.44 & 13.0, 18.0 &   - ,    - & 4.0, 4.7, 0.72, -    &  \\
\noalign{\medskip}
\tableline
\end{tabular}
\end{center}
\end{table*}

\section{Summary}
We have performed axisymmetric MHD simulations of jet formation from accretion 
disks surrounding a magnetised star.
The simulations start from an equilibrium state of the star-disk corona
(hydrostatic density distribution, force-free field).
Our physical grid size is $(80\times 80)$ inner disk radii corresponding to about
$(14\times 14)\,$AU for $\ri \simeq 10\,\rsun$.
Disk and stellar surface are taken as a time-independent boundary condition for 
the outflow mass loss rates and the magnetic flux profile.

The major goal of this paper was to investigate the long-term interrelation between 
a stellar dipolar field and a disk field and how that affects the outflow collimation.
Certain combinations of stellar versus disk magnetic flux and field directions
were considered namely the cases of an aligned resp. anti-aligned magnetic field 
direction. 
 
The stellar magnetic field has an important impact on the jet formation process
by providing 
 additional magnetic flux, 
 additional central (magnetic) pressure component,
 excess angular momentum in the jet launching region.
It might disturb the outflow axisymmetry, and/or trigger a time-variation in 
outflow rate in the case of an inclined  central stellar magnetosphere.

Our results of MHD simulations of a superposed stellar \& disk magnetosphere
in general demonstrate the de-collimation of the disk jet by the central stellar
magnetosphere, respectively the collimation of the
stellar wind by the surrounding disk jet.

The interplay between disk wind and stellar wind, may result in flares
like coronal mass ejections, generating
a variation in the mass loading and velocity of the asymptotic jet. We
find variations in mass load by a factor of four and in velocity by a factor
of two. 
The time scale for such flaring events is {\em long}, i.e. several hundred
inner disk rotations and of the order of 10.000 days.
We therefore hypothesized whether such flaring events and the corresponding
hydrodynamic variations may be responsible for generating  internal shocks
in the asymptotic jet visible as knots.

In summary we conclude that strong protostellar jets must be generated
by collimating disk winds with reasonable mass load.
If the two-component system is dominated by the stellar wind, collimation
is either too weak (for low mass load) or too high (for high mass load).
It is only a disk jet which can provide a collimated outflow over a 
reasonable range in radius.

\acknowledgements
I thank the LCA team and M.~Norman and D.~Clarke for the possibility to use 
the ZEUS code.
Comments and suggestions by an unknown referee which helped to improve the
presentation of this work are acknowledged.


\end{document}